\def\editmode{0}

\def\bibfilenames{bibman_refs,refs_SK}

\def\spsformat{0}

\if\spsformat1

\documentclass{article}
\usepackage[utf8]{inputenc}
\usepackage{spconf}

\else
\documentclass[11pt,final,onecolumn]{IEEEtran}
\usepackage[left=1in,top=1in,right=1in,bottom=1in]{geometry}

\fi

\usepackage{savesym} %% Needed for compatibility with IEEEtran
\savesymbol{labelindent}
\usepackage{enumitem} % do not include the package enumerate. See below.
\if\editmode1  %%%%%%%%%%%%%%%%%%%%%%%%%%%%%%%%%%%%%%%%%%%%%%%%%%%%%%%%%%%
\usepackage[backend=bibtex,style=alphabetic,sorting=debug,maxbibnames=99]{biblatex}
\DeclareFieldFormat{labelalpha}{\thefield{entrykey}}
\DeclareFieldFormat{extraalpha}{}
\bibliography{\bibfilenames}
\newcommand{\cmt}[1]{\noindent\textcolor{lightgreen}{\underline{[#1]}}} % comment
\newcommand{\hc}[1]{\textcolor{blue}{#1}} % highlight command --> to
\setlistdepth{9}
\newlist{bulletlist}{enumerate}{9}
\setlist[bulletlist,1]{label=$\bullet$}
\setlist[bulletlist,2]{label=$\diamond$}
\setlist[bulletlist,3]{label=$\rightarrow$}
\setlist[bulletlist,4]{label=$\circ$}
\setlist[bulletlist,5]{label=$-$}
\setlist[bulletlist,6]{label=$\square$}
\setlist[bulletlist,7]{label=$\star$}
\setlist[bulletlist,8]{label=$\checkmark$}
\setlist[bulletlist,9]{label=$\Delta$}
\newenvironment{bullets}{\begin{bulletlist}}{\end{bulletlist}}

\newcommand{\blt}[1][noargpassed]{% add a label if an optional argument is passed
  \item%
  \ifthenelse{\equal{#1}{noargpassed}}{}{\cmt{#1}}%
}

\else
\usepackage{cite}
\newcommand{\cmt}[1]{} % comment
\newcommand{\hc}[1]{\textcolor{black}{#1}} % highlight command -->
\newenvironment{bullets}{}{}
\newcommand{\blt}[1][noargpassed]{\ignorespaces}

\fi
\newcommand{\printmybibliography}{
\if\editmode1 
\printbibliography
\else
\bibliography{\bibfilenames}
\fi
}

\usepackage{fixltx2e}

\usepackage{graphicx}

\usepackage{subcaption}              %subfigures with \mbox and \subfigure[]

\usepackage[utf8]{inputenc}

\usepackage{amsfonts}
\usepackage{amsmath}
\usepackage{mathtools}
\usepackage{amssymb}

\usepackage{bm}

\usepackage{color,verbatim}
\usepackage{multirow}
\usepackage{accents}

\usepackage{theoremref}
\usepackage{url}

\newcounter{rulecounter}
\newcommand{\resetrule}{ \setcounter{rulecounter}{0}}
\resetrule

\newtheorem{myauxproblem}{Problem}
\newtheorem{myauxoptionalproblem}{Optional Problem}
\newsavebox{\selvestebox}
\newenvironment{colbox}[1]
  {\newcommand\colboxcolor{#1}%
   \begin{lrbox}{\selvestebox}%
   \begin{minipage}{\dimexpr\columnwidth-2\fboxsep\relax}}
  {\end{minipage}\end{lrbox}%
   \begin{center}
   \colorbox{\colboxcolor}{\usebox{\selvestebox}}
   \end{center}}

\definecolor{orange}{rgb}{1,0.8,0}
\definecolor{gray}{rgb}{.9,0.9,0.9}
\definecolor{darkgray}{rgb}{.3,0.3,0.3}
\definecolor{darkblue}{rgb}{.1,0.0,0.3}
\definecolor{lightblue}{rgb}{0.7,0.7,1}
\definecolor{lightred}{rgb}{1,0.7,.7}
\definecolor{purple}{RGB}{204,153,255}
\definecolor{lightgray}{rgb}{.95,0.95,0.95}
\definecolor{lightgreen}{rgb}{0.3,0.5,0.3}
\definecolor{darkgreen}{rgb}{0.05,0.3,0.05}

\newcommand{\ra}{$\rightarrow$~}

\newcommand{\bbm}[1]{{\bar{\bm #1}}}

\newcommand{\hbm}[1]{{\hat{\bm #1}}}
\newcommand{\inv}{^{-1}}

\newcommand{\rfield}{\mathbb{R}}

\newcommand{\cov}{\mathop{\rm Cov}}

\newcommand{\diag}{\mathop{\rm diag}}

\newcommand{\rank}{\mathop{\rm rank}}

\newcommand{\transpose}{^\top}
 \newcommand{\define}{:=}
\newcommand{\expected}{\mathop{\textrm{E}}\nolimits}

\newcommand{\minimize}{\mathop{\text{minimize}}}

\DeclareMathOperator*{\argmin}{arg\,min}

\newtheorem{myproposition}{Proposition}
\newtheorem{myremark}{Remark}
\newtheorem{myproblemstatement}{Problem Statement}
\newtheorem{mylemma}{Lemma}
\newtheorem{mytheorem}{Theorem}
\newtheorem{mydefinition}{Definition}
\newtheorem{mycorollary}{Corollary}

\renewcommand{\hc}[1]{{#1}}

\usepackage{graphicx}
\usepackage[export]{adjustbox}

\usepackage{comment}
\usepackage{ifthen}

\excludecomment{blackp} % then the environment disappears

\newcommand{\includefigandtrim}[1]{\includegraphics[trim=20 0 50 30,clip,width=0.59\textwidth]{figs/#1}}

\newcommand{\includefig}[1]{\includegraphics[trim=0 0 0 0,clip,width=0.59\textwidth]{figs/#1}}
\newcommand{\ncite}[2][]{}

\renewcommand{\cov}{\text{Cov}} 

\newcommand{\region}{\hc{\mathcal{X}}} % geographic region of interest
\newcommand{\freqregion}{\hc{\mathcal{F}}} % frequency region of interest

\newcommand{\ch}{\hc{h}} % channel
\newcommand{\chest}{\hc{\hat h}} % channel estimate
\newcommand{\pow}{\hc{p}} % power
\newcommand{\powest}{\hc{\hat p}} % power estimate
\newcommand{\txpow}{\hc{p}^\text{TX}}
\newcommand{\rxpow}{\hc{p}^\text{RX}}
\newcommand{\powmean}{\hc{\mu_p}} % power mean
\newcommand{\pattern}{\hc{\Gamma}} % antenna pattern

\newcommand{\freq}{{\hc{f}}} % frequency
\newcommand{\freqind}{{\hc{n}}} % frequency index
\newcommand{\freqnum}{{\hc{N}_\freq}} % number of frequencies
\newcommand{\outage}{\hc{q}} % outage prob
\newcommand{\serv}{\hc{s}} % service
\newcommand{\servthresh}{\hc{\gamma}} % service threshold
\newcommand{\locs}{\hc{ x}} % scalar location
\newcommand{\loc}{{\hc{\bm x}}} % location
\newcommand{\locp}{\hc{\bm x}'} % location prime
\newcommand{\locvar}{{\hc{\bar{\bm x}}}} % location auxiliary variable
\newcommand{\txloc}{\hc{\bm x}^\text{TX}} % location
\newcommand{\rxloc}{\hc{\bm x}^\text{RX}} % location

\newcommand{\txind}{{\hc{s}}} % transmitter index
\newcommand{\txnum}{{\hc{S}}} % number of transmitters

\newcommand{\meas}[1]{{\hc{m}_{#1}}} % measurement (index)
\newcommand{\measvec}[1]{{\hc{\bm m}_{#1}}} % measurement vector (index)
\newcommand{\allmeasvec}{{\hc{\bm m}}} % vector with all measurements
\newcommand{\measpow}{{\hc{\tilde p}}} % measured power
\newcommand{\noise}[1]{{\hc{z}_{#1}}} % measurement noise (index)
\newcommand{\noisevec}{{\hc{\bm z}}} % measurement noise vector
\newcommand{\noisevar}{{\hc{\sigma}^2_{z}}} % measurement noise variance
\newcommand{\measind}{{\hc{n}}} % measurement index
\newcommand{\measnum}{{\hc{N}}} % number of measurements

\newcommand{\spbasisfun}{{\hc{\psi}}} % basis function
\newcommand{\spbasismat}{{\hc{\bm \Psi}}} % basis function
\newcommand{\spbasisnum}{{\hc{B}}} % number of basis functions
\newcommand{\spbasisind}{{\hc{b}}} % index of the basis functions
\newcommand{\spcoef}{{\hc{\alpha}}} % basis function coefficient
\newcommand{\spcoefvec}{{\hc{\bm\alpha}}} % vector of basis function
\newcommand{\spcoefvecest}{{\hc{\hbm\alpha}}} % estimate of the vector of
\newcommand{\fbasisfun}{{\hc{\phi}}} % basis function
\newcommand{\fbasisfreq}{{\hc{\tilde f}}} % basis function central frequency
\newcommand{\fbasisnum}{{\hc{ C}}} % number of basis functions
\newcommand{\fbasisind}{{\hc{ c}}} % index of the basis functions
\newcommand{\fcoef}{{\hc{\beta}}} % basis function coefficient
\newcommand{\rkhs}{{\hc{\mathcal{G}}}} %
\newcommand{\rkhsfun}{{\hc{g}}} % function in an RKHS
\newcommand{\kernel}{{\hc{\kappa}}} %
\newcommand{\kernelmat}{{\hc{\bm K}}} %
\newcommand{\loss}{{\hc{\mathcal{L}}}} %
\newcommand{\regpar}{{\hc{\lambda}}} %

\newcommand{\nnfun}{{\hc{g}}} % neural net function
\newcommand{\nnparvec}{{\hc{\bm w}}} %  neural net parameters

\newcommand{\slf}{{\hc{F}}} % spatial loss field
\newcommand{\pathloss}{{\hc{h}}^\text{PL}} %
\newcommand{\shad}{{\hc{a}}^\text{SF}} % shadow fading attenuation
\newcommand{\shadest}{{\hc{\hat a}}^\text{SF}} % shadow fading attenuation estimate
\newcommand{\shadmean}{{\hc{\mu}}^\text{SF}} % shadow fading attenuation mean
\newcommand{\shadmeas}{{\hc{\breve a}}^\text{SF}} % shadow fading measurement
\newcommand{\shadmeasvec}{{\hc{\breve{ \bm a}}}^\text{SF}} % shadow fading measurement
\newcommand{\shadmeasset}{{\hc{\breve{ \mathcal A}}}^\text{SF}} % shadow fading measurement set
\newcommand{\shadvar}{{\hc{\sigma}^2_\text{SF}}} % shadow fading attenuation variance
\newcommand{\shaddist}{{\hc{d}^\text{SF}}} % shadow fading attenuation 1/2 distance
\newcommand{\fad}{{\hc{a}}^\text{FF}} % fast fading attenuation
\newcommand{\fadmean}{{\hc{\mu}^\text{FF}}} % fast fading attenuation mean
\newcommand{\fadvar}{{\hc{\sigma}^2_\text{FF}}} % fast fading attenuation mean
\newcommand{\freqresp}{{\hc{\Gamma}}} % freq. response of a filter
\newcommand{\filtpow}{{\hc{\bar p}}} % power at the filter output
\newcommand{\fbasisvec}{{\hc{\bm \fbasisfun}}} %  vector with basis
\newcommand{\powlow}{{\hc{a}}} %  power interval left endpoint
\newcommand{\powhigh}{{\hc{b}}} %  power interval right endpoint
\newcommand{\branchind}{{\hc{l}}} % branch index in a filter bank
\newcommand{\branchnum}{{\hc{L}}} % branch number in a filter bank

\usepackage{float}
\makeatletter
\newcommand\floatc@simplerule[2]{{\@fs@cfont #1 #2}\par}
\newcommand\fs@simplerule{\def\@fs@cfont{\bfseries}\let\@fs@capt\floatc@simplerule
  \def\@fs@pre{\hrule height.8pt depth0pt \kern4pt}%
  \def\@fs@post{\kern4pt\hrule height.8pt depth0pt \kern4pt \relax}%
  \def\@fs@mid{\kern8pt}%
  \let\@fs@iftopcapt\iftrue}

\floatstyle{simplerule}
\newfloat{floatbox}{t}{lob}
\floatname{floatbox}{Box}

\newcommand{\changed}[1]{\textcolor{black}{#1}}
\newcommand{\changedr}{\color{black}} % changed region

\newcommand{\signalstrength}{\changed{signal strength} }
\newcommand{\Signalstrength}{\changed{Signal strength} }
\newcommand{\SignalStrength}{\changed{Signal Strength} }

\renewcommand{\expected}{{\mathbb{E}}}

\begin{document}

%%%%%%%%%%%%%%%%%%%%%%%%%%%%%%%%%%%%%%%%%%%%%%%%%%%%%%%%%%%%%%%%%%%%%%%%%%%%%%%%% 
\title{Radio Map Estimation:\\ A Data-Driven Approach to \\Spectrum Cartography}
%%%%%%%%%%%%%%%%%%%%%%%%%%%%%%%%%%%%%%%%%%%%%%%%%%%%%%%%%%%%%%%%%%%%%%%%%%%%%%%%% 
\if\spsformat1 \name{Author(s) Name(s)\thanks{Thanks to XYZ agency for
    funding.}}  \address{Author Affiliation(s)} \else \author{Daniel
  Romero and Seung-Jun Kim\thanks{D. Romero is with the Dept. of
    Information and Communication Technology, University of Agder, Jon
    Lilletunsvei 9, 4879 Grimstad, Norway. Email:
    daniel.romero@uia.no. S.-J. Kim is with the Dept. of Computer
    Science and Electrical Engineering, University of Maryland,
    Baltimore County, 1000 Hilltop Circle, Baltimore, MD 21250, USA. Email: sjkim@umbc.edu.}
    \thanks{This research has been funded in part by the Research Council of Norway under IKTPLUSS grant 311994.}
    }
\fi

\maketitle
%%%%%%%%%%%%%%%%%%%%%%%%%%%%%%%%%%%%%%%%%%%%%%%%%%%%%%%%%%%%%%%%%%%%% 
\begin{abstract}
Radio maps characterize quantities of interest in radio communication environments, such as the received signal strength and channel attenuation, at every point of a geographical region. Radio map estimation typically entails interpolative inference based on spatially distributed measurements. In this tutorial article, after presenting some representative applications of radio maps, the most prominent radio map estimation methods are discussed. Starting from simple regression, the exposition gradually delves into more sophisticated algorithms, eventually touching upon state-of-the-art techniques. To gain insight into this versatile toolkit, illustrative toy examples will also be presented.
\end{abstract}
%%%%%%%%%%%%%%%%%%%%%%%%%%%%%%%%%%%%%%%%%%%%%%%%%%%%%%%%%%%%%%%%%%%%%


%%%%%%%%%%%%%%%%%%%%%%%%%%%%%%%%%%
\section{Introduction}
%%%%%%%%%%%%%%%%%%%%%%%%%%%%%%%%%%
\label{sec:intro}

Spectrum cartography comprises a collection of techniques to construct
and maintain radio maps, which provide useful information on the RF
landscape, such as the received signal power, interference power,
power spectral density (PSD), electromagnetic absorption, and channel
gain across a geographic area; see
e.g.~\cite{alayafeki2008cartography,bazerque2010sparsity,yilmaz2013radio}\ncite{jayawickrama2013compressive}. \changed{A quick overview on the most representative types of radio map  is provided in
  Table~\ref{table:mapsummary}.}  

Radio maps find a myriad of applications in wireless communications and networking, such as network planning, interference coordination and mitigation, power control, resource allocation, handoff management, multi-hop routing, dynamic spectrum access, and cognitive radio networking tasks; see~\cite{romero2017spectrummaps,kim2011kriged} and the references therein. Radio maps are also useful for localization~\cite{bazerque2010sparsity} and tomography~\cite{romero2018blind}.

{\changedr
Arguably, spectrum cartography can be traced back to the application of Maxwell's equations to characterize the propagation of radio waves across space. However, due to insufficient computational capacity, this approach has been traditionally confined to problems involving relatively simple geometries, such as determining the electromagnetic field radiated by a dipole. To analyze more complex environments, numerous empirical models have been developed, such as the well-known P-recommendations from the International Telecommunication Union - Radiocommunication Sector~(ITU-R). Unfortunately, this kind of models often fail to provide estimates that are accurate enough for a given application~\cite{phillips2012bounding}. 

With the advent \changed{of  modern computational resources}, finite-element analysis and ray-tracing techniques paved the way for effectively approximating the solutions of Maxwell's equations in complex environments. However, besides their high computational complexity, their main limitation is that an accurate description of the propagation environment is required through  3D models of all objects and obstacles along with their electromagnetic properties.

To mitigate such limitations, \changed{radio map estimation (RME)} was proposed, originally in the context of cognitive radios~\cite{alayafeki2008cartography}. In RME, a collection of measurements acquired by spatially distributed sensors is used together with their locations to construct a map of the relevant RF descriptors, typically by applying some form of interpolation techniques. As this approach does not require physical modeling of the propagation environment, it constitutes a \emph{data-driven} alternative to the model-based techniques mentioned earlier. Since its conception, a sizable body of literature has emerged on the estimation of a variety of kinds of radio maps for a wide range of application scenarios; see e.g.~\cite{bazerque2010sparsity,dallanese2012gslasso,bazerque2011splines,romero2017spectrummaps,kim2011kriged}\ncite{jayawickrama2013compressive} and the references therein. Recently, the work in this area has intensified thanks to the boom of deep learning~\cite{parera2020tiltdependent,imai2019radiopredictioncnn,  iwasaki2020transferbasedpower,levie2019radiounet}.%\ncite{saito2019twosteppathlos,hayashi2020studyradiopropagation,levie2020pathlossprediction}.
}

This article provides an introduction to RME by guiding the readers on the foundations and applications of RME as well as on recent advances in this rapidly growing research area. \changed{To this end, the most common types of radio maps are first described. Afterwards, RME methods for signal strength and propagation maps are expounded in a tutorial fashion. Practical considerations and future directions are also discussed.}

\newcommand{\figintable}[1]{\raisebox{-\totalheight}{ \includegraphics[width=4.5cm]{figs/#1}}}

\begin{table}[bhtp] % table as image
  \centering

  \begin{center}
{ \color{black}    
    \begin{tabular}{|p{1.5cm} | p{4.6cm} | p{2.5cm} | p{2.5cm} | p{2cm} |}
      \hline
      Type of Map & Illustration in a 1D Scenario & Example \-Applications\- & Construction
                                                                   \-Complexity\- & Changes
                                                                                if\ldots \\
      \hline
      \hline
      Coverage Map & \figintable{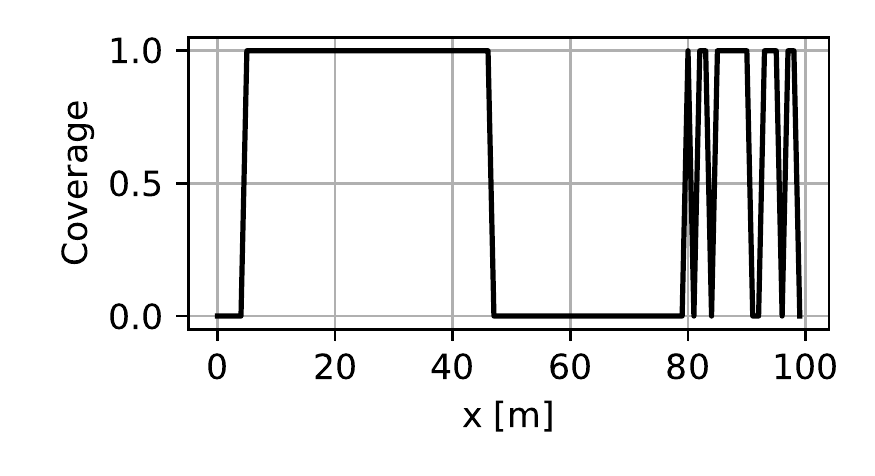} &
                                                          Find
                                                          coverage
                                                          holes.
                                                                 &\multirow{4}{2cm}{
                                                                   \hspace{-.56cm}
 \begin{tabular}{p{1.9cm}}The
                                                                   base
                                                                   station
                                                                   only
                                                                   needs
                                                                   to
                                                                   know
                                                                   if
                                                                   the
                                                                   mobile
                                                                   user
                                                                   can
                                                                   receive
                                                                     data.
                                                                     \end{tabular}
                                                                   }&
                                                                      \multirow{4}{2cm}{
                                                                      \hspace{-.56cm}
                                                                      \begin{tabular}{p{1.9cm}}
                                                                        \begin{itemize}[leftmargin=*]
                                                                          \item
                                                                          the
                                                                          environment changes.
                                                                          \item
                                                                          the
                                                                          transmission
                                                                          activity
                                                                          changes.
                                                                          \item
                                                                          the
                                                                          transmitter
                                                                          location
                                                                          or
                                                                          orientation
                                                                          changes. 
                                                                        \end{itemize}
                                                                      \end{tabular}
                                                                      }

\\
      \cline{1-3}
       Outage Probability Map
                   &
                     \figintable{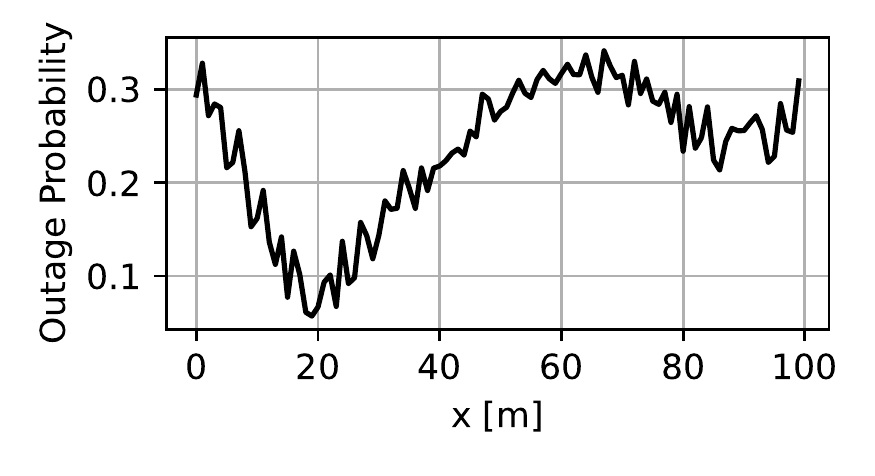}
                                            & 
                     Improve the reliability of a cellular network. 
                                              
                                                                         & &
      \\
      \cline{1-4}
      Power Map
                  &
                    \figintable{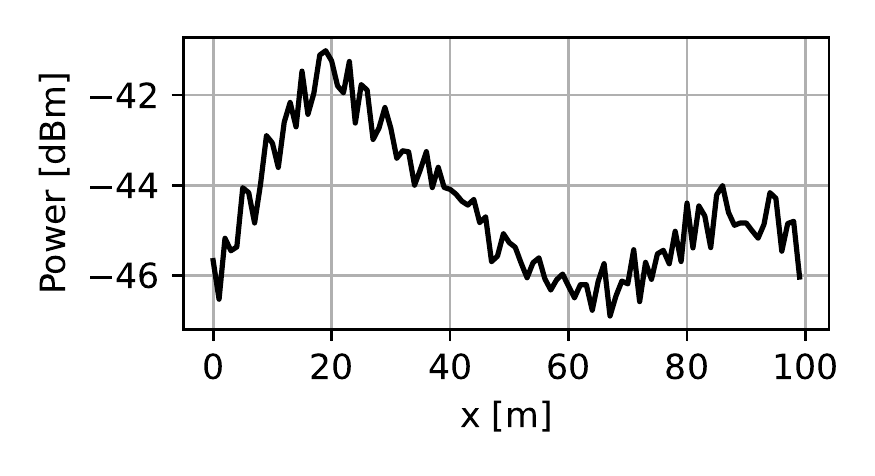}
                                                  &
                                                    Unveil regions of
                                                    high
                                                    interference. Determine
                                                    appropriate locations
                                                    for new base stations.
                                                                         &
                                                                           The
                                                                           mobile
                                                                           user
                                                                           reports
                                                                           power
                                                                           measurements.
                                                                              &
      \\
      \cline{1-4}
      PSD Map &
                %\hspace{-1cm}
                \raisebox{-\totalheight}{ \includegraphics[trim=25
                    0 35 0, width=4.5cm, clip]{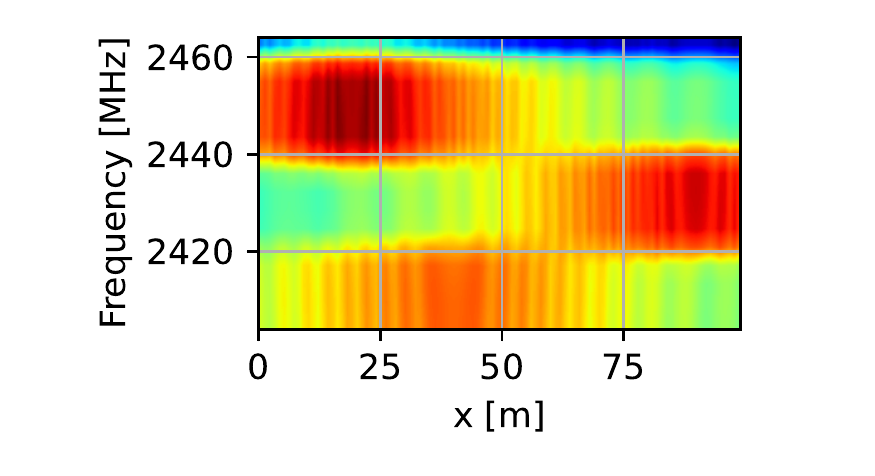}}
                                                  &       Maximize
                                                    frequency reuse.
                                                                 & The mobile user  reports power (density) measurements
                                                                   for
                                                                   each
                                                                   frequency
                                                                   (e.g. a
                                                                   periodogram).& 
      \\
      \hline
      Channel gain map
                  &\hspace{0cm}
                    \raisebox{-\totalheight}{ \includegraphics[trim=50
                    0 50 0, width=4.5cm, clip]{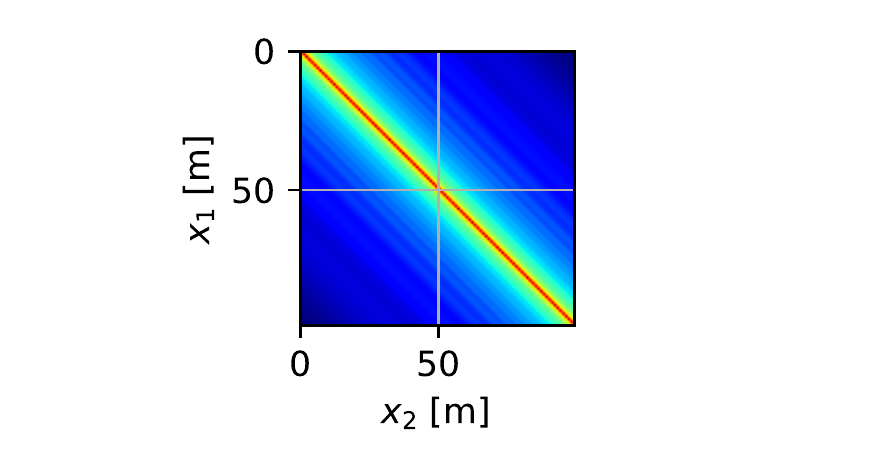}}
                                                  &
                                                    Resource
                                                    allocation for
                                                    device-to-device
                                                    communications.
                                                    & User at position
                                                      $x_1$ sends a
                                                      pilot
                                                      sequence. User
                                                      at $x_2$ sends an
                                                      estimate of the
                                                      received
                                                      power to the
                                                      base station after
                                                      normalizing by
                                                      the transmitted
                                                      power. &
                                                                          the
                                                               environment
                                                               changes.
                                                                            \\   
      \hline
    \end{tabular}
}
  \end{center}
  \caption{\changed{Illustration of the  prominent types of radio maps. Although
    radio maps find applications in many domains, 
    this table exemplifies their applicability in cellular
    communications for specificity. The $x$-coordinate indexes a point on a road or
    railway.  }}
  \label{table:mapsummary}
\end{table}

%%%%%%%%%%%%%%%%%%%%%%%%%%%%%%%%%%%%%%%%%%%%%%%%%%%%%%%%%%
\section{Radio Maps and Their Applications}
\label{sec:types}
%%%%%%%%%%%%%%%%%%%%%%%%%%%%%%%%%%%%%%%%%%%%%%%%%%%%%%%%%%
% We will then introduce different kinds of radio maps by organizing them into two categories, namely \emph{\signalstrength maps} and \emph{propagation maps}.  The taxonomy is geared to facilitate the readers' understanding of the diverse and often inconsistent definitions of radio maps proposed in the literature.  Along with each kind of map, its representative applications will be discussed in more detail. The relations among the various types of maps will also be elucidated along with the trade-offs involved in their usage.

%The role of radio maps is to characterize the RF spectrum environment over space. 
The signal received at a certain location is determined by i) the transmitted signal; and ii) the communication channel between the transmitter and the receiver. Depending on whether the focus is on the combined effect of the two, or rather on the effect of the propagation channel itself, two families of radio maps can be considered: \emph{\signalstrength maps} and \emph{propagation maps}. %Both families are described next along with their major properties and motivating applications.

For simplicity, unless stated otherwise, it will be assumed that the maps do not change significantly within the time interval under consideration. In practice, the length of the interval for which this assumption remains valid depends not only on the speed of variation but also specific applications.
  
%%%%%%%%%%%%%%%%%%%%%%%%%%%%%%%%%%%
\subsection{\SignalStrength Maps}
\label{sec:occupancy}
%%%%%%%%%%%%%%%%%%%%%%%%%%%%%%%%%%%
\Signalstrength maps focus on metrics of the received signal, which are determined by the
aggregate effects of the channel upon the signals transmitted by all
active sources. This is the case, for instance, if the goal is to map
interference power levels. Constructing such maps does not require 
knowledge of the number, locations, and power of the transmitters,
which is appealing in scenarios involving a large number of mobile
transmitters, as in device-to-device communications or cellular
uplink channels. % The term ``occupancy'' is brought here from the cognitive
% radio literature, where it refers to the aggregate contribution of all
% transmissions at the same time, frequency, and spatial
% location~\cite{agarwal2018spectrum}. 
Different kinds of \signalstrength maps are presented next with the increasing
level of detail they capture.

%%%%%%%%%%%%%%%%%%%%%%%%%%%%%%%%
\subsubsection{Coverage Maps}
%%%%%%%%%%%%%%%%%%%%%%%%%%%%%%%%
The coarsest characterization of the radio environment can be provided by a map that takes only binary values for coverage indication. Specifically, let $\pow(\loc)$ denote the signal power that a radio \changed{with an isotropic antenna\footnote{The case of non-isotropic antenna patterns is discussed later.}} receives at a spatial location $\loc\in \region$, where $\region$ represents a geographical region of interest, typically a subset of $\rfield$, $\rfield^2$, or $\rfield^3$. A coverage map is a function $\serv:\region\rightarrow \{0,1\}$ that takes the value $\serv(\loc)=1$ if $\pow(\loc)\geq \servthresh$ and 0 otherwise, where $\servthresh$ is a given threshold. This threshold may correspond to the minimum signal power necessary to guarantee a prescribed communication rate. Coverage maps may also be constructed by  replacing $\pow(\loc)$ in the above definition with the \emph{signal-to-noise-power ratio} (SNR) or the \emph{signal-to-interference-plus-noise-power ratio} (SINR).

Coverage maps are often  used by cellular and TV broadcast network operators to find areas of weak coverage, which allows them to determine suitable sites for deploying new base stations and relay antennas. A more recent application is mission planning for autonomous mobile robots or vehicles that require network connectivity, where coverage maps may assist in, e.g., minimizing the time and distance traversed without connectivity.

%%%%%%%%%%%%%%%%%%%%%%%%%%%%%%%%%%%%%%%%%
\subsubsection{Outage Probability Maps}
%%%%%%%%%%%%%%%%%%%%%%%%%%%%%%%%%%%%%%%%%
A soft version of  coverage maps can be constructed by adopting a probabilistic perspective, as the effects of the channel, such as fading and shadowing, are often modeled as random. An outage probability map $\outage(\loc)$ \changed{is a function $\outage:\region \rightarrow [0,1]$ that} provides the probability that $\pow(\loc)< \servthresh$. Since outage probability maps capture more detailed information than coverage maps, the former can be readily employed in the applications of the latter. However, the additional information provided by outage probability maps allows more sophisticated decision making, as in route planning~\cite{romero2019noncooperative}.

%%%%%%%%%%%%%%%%%%%%%%%%%%%%%
\subsubsection{Power Maps}
\label{sec:powermaps}
%%%%%%%%%%%%%%%%%%%%%%%%%%%%%
A substantially finer characterization of the  \signalstrength is obtained by a power map, \changed{defined as a function $\pow:\region\rightarrow \rfield$, which returns} the received power $\pow(\loc)$  at every spatial location $\loc \in \region$. 
%Being more detailed than coverage or outage probability maps, power maps can also be used for applications such as network planning or trajectory optimization. Besides, the increased degree of detail embodied in power maps renders them also suitable e.g. for source localization, where the goal is to locate a transmitter by measuring the power it radiates at a set of locations~\cite{bazerque2010sparsity}. Yet another application is fingerprinting-based localization: for instance, one may construct one power map per WiFi access point in a certain building. A mobile device that measures the received power from the access points within range can estimate its location as the spatial position that yields the greatest agreement between the measured values and those predicted by the maps. 
As the information contained in  power maps is richer than that in  coverage or  outage probability maps,  power maps can be used not only for  tasks such as network planning and trajectory optimization, but also for localizing  transmitters~\cite{bazerque2010sparsity}. Also, in  fingerprint-based localization, a mobile device can measure the received powers of nearby access points and determine its position by matching the measurements with the  values of the map.

%%%%%%%%%%%%%%%%%%%%%%%%%%%
\subsubsection{PSD Maps}
%%%%%%%%%%%%%%%%%%%%%%%%%%%
One is sometimes interested not only in the power distribution across space but also across the frequency domain. A PSD map is a function \changed{$\pow: \region \times \freqregion \rightarrow \rfield$ that provides the PSD $\pow(\loc, \freq)$ of the received} signal at each location $\loc \in \region$. \changed{Here, $\freq \in \freqregion$ is the frequency variable and the set $\freqregion\subset \rfield$ contains the frequencies of interest. If the latter is discretized as $\freqregion=\{\freq_1,\ldots,\freq_\freqnum\}$, }constructing a PSD map is tantamount to constructing a collection of power maps proportional to $\pow(\loc,\freq_1),\ldots,\pow(\loc, \freq_\freqnum)$.

In addition to the applications mentioned for the previous kinds of \signalstrength maps, PSD maps enable additional use cases. For example, they can be used for speeding up handoff procedures in cellular networks by providing the quality of the relevant channels at a given location, obviating the need for time-consuming channel measurement or feedback processes. PSD maps can also be utilized for interference coordination where concurrent transmissions are assigned to different frequency band channels based on the transceiver locations, promoting efficient spectrum reuse.  In cognitive radio networks, PSD maps can unveil underutilized ``white spaces’’ in the space/frequency/time domains, which can  be exploited opportunistically by  unlicensed users ~\cite{axell2010sensing}\ncite{yucek2009surveysensing}.

%%%%%%%%%%%%%%%%%%%%%%%%%%%%%%%
\subsection{Propagation Maps}
\label{sec:propagationmaps}
%%%%%%%%%%%%%%%%%%%%%%%%%%%%%%%
Whereas \signalstrength maps capture the aggregate effect of the transmitted signals and the channels, propagation maps focus exclusively on the channel. Each  parameter of interest gives rise to a different kind of propagation map. As described next, \emph{channel gain maps} constitute the simplest kind. Suppose that $\rxpow$ denotes the power received at location $\rxloc$ due to a transmitter with power $\txpow$ at location $\txloc$. \changed{A \emph{channel gain map} is a function $\ch: \region \times \region \rightarrow \rfield$ of the transmitter and receiver locations
that
provides the channel gain\footnote{\changed{More sophisticated propagation maps arise by accounting for frequency selectivity. For example, the power gain that each subcarrier sees in an orthogonal frequency division multiplexing (OFDM) system can be mapped. For simplicity, this article focuses on  channel gain maps, which provide the overall gain that affects a single narrow frequency band.}}  $\ch(\txloc, \rxloc)=\rxpow/\txpow$}.

Clearly, given a channel gain map $\ch(\txloc, \rxloc)$ together with the locations $\txloc_1,\ldots,\txloc_\txnum$ and transmit powers $\txpow_1,\ldots,\txpow_\txnum$ of $\txnum$ sources in a region, one can obtain the power map as $\pow(\loc)=\sum_\txind \ch(\txloc_\txind,\loc)\txpow_\txind$, provided that the signals transmitted by different sources are uncorrelated, \changed{as generally occurs in practice, except, e.g., in \emph{single-frequency networks} such as the ones utilized by \emph{digital television broadcast}.} Thus, propagation maps can be readily used in the applications of \signalstrength maps provided that the locations and  transmit-powers of the sources are known. On the other hand, propagation maps offer more versatile information than \signalstrength maps: whereas a \signalstrength map may provide the total interference at each location, a propagation map  reveals the contribution of each source. This enhanced flexibility is instrumental for tasks such as interference coordination or network planning.  

Observe that changes in the locations and transmit-powers of the sources give rise to changes in \signalstrength maps, whereas propagation maps remain unaffected. On the other hand, alterations in the scattering environment, such as the construction of new buildings or seasonal changes of foliage, affect both propagation and \signalstrength maps. Thus, the time scale of variations of \signalstrength maps is never greater than that of  propagation maps. Hence, propagation maps can be used to construct \signalstrength maps in highly dynamic setups, such as the  uplinks of cellular networks, where mobile users  rapidly change their positions and activity patterns. %To this end, information on the locations and transmit power of the users would need to be collected. 

Propagation maps can also help address the classical problem of predicting the potential interference inflicted to passive receivers, which arises in the context of cognitive radios~\cite{axell2010sensing}. For example, when reusing the TV spectrum, the challenge is to carry out unlicensed transmissions without introducing detrimental interference to TV receivers. With a propagation map, one can ensure that no receivers in a certain area will be negatively affected without the need to know their precise locations~\cite{dallanese2011kriging}.

Yet another application is the problem of \emph{aerial base station placement}, where a propagation map of the air-to-ground channels can be constructed to determine the best set of locations to deploy unmanned aerial vehicle (UAV)-mounted base stations to serve  ground users~\cite{romero2022aerial}.

%%%%%%%%%%%%%%%%%%%%%%%%%%%%%%%%%%%%%%%%%%%%%%%%%
\section{Estimation of  \SignalStrength  Maps}
%%%%%%%%%%%%%%%%%%%%%%%%%%%%%%%%%%%%%%%%%%%%%%%%%
\label{sec:eststrength}

{\changedr

In a typical RME formulation, the goal is to construct a radio map using a set of measurements acquired by spatially dispersed sensors together with their locations. For \signalstrength maps, consider $N$ measurements, where the $\measind$-th measurement $\meas{\measind}$ is acquired by a sensor at location $\loc_\measind$. In the case of power maps, $\meas{\measind}$ may be the average power measured in a certain band within a given time interval, which can be modeled as  $\meas{\measind}= \pow(\loc_\measind) + \noise{\measind}$. Here, $\noise{\measind}$ denotes  measurement noise, which is caused, e.g., by the finite length of the averaging time interval. For estimating PSD maps, $\meas{\measind}$ can contain power spectrum measurements such as periodograms. The RME problem becomes constructing the desired \signalstrength map given the pairs $\{(\loc_\measind,\meas{\measind})\}_{\measind=1}^{\measnum}$.

It is worth noting that each sensor may collect measurements at multiple locations provided that they are taken within a time window whose length is small relative to the scale of variations of the target map. Thus, the number of sensors may be much smaller than $\measnum$. \blt[Surveying]In fact, the RME formulation can be extended to accommodate the decision on where to acquire the measurements sequentially, as discussed in Sec.~\ref{sec:ssurv}. Furthermore, a sensor need not be a special-purpose device. For example, a user terminal in a cellular network may function as a sensing device.

The rest of the section presents the main approaches for constructing \signalstrength maps.
}

  \newenvironment{inputlist}{
    \hspace{-.4cm}
    \begin{tabular}{p{3.5cm}}
      \begin{itemize}[leftmargin=*]
      }{\end{itemize}
    \end{tabular}
  }

  \newenvironment{strengthlist}{\hspace{-.4cm}
    \begin{tabular}{p{5.2cm}}
      \begin{itemize}[leftmargin=*]
      }{
      \end{itemize}
    \end{tabular}
  }

  \newenvironment{limitationlist}{\begin{strengthlist}
      }{
      \end{strengthlist}
  }

\begin{table}[bhtp] % table as image
  \centering

  \begin{center}
{ \color{black}    
  \begin{tabular}{|p{1.5cm} | p{3.6cm} | p{5.3cm} | p{5.3cm} |}
    \hline
    Method & Input (besides measurements) & Strengths & Limitations \\
    \hline
    \hline
    Linear Parametric RME &
                            \begin{inputlist}
                              \item
                              Transmitter
                              locations $\txloc_1,\ldots,\txloc_\txnum$
                              \item
                              Path loss
                              law;
                              e.g. $\spbasisfun_\txind(\loc):=1/\|\loc-\txloc_\txind\|^2$
                            \end{inputlist}
                                          &
                                            \begin{strengthlist}
                                              \item
                                              Simplicity
                                              \item Closed form
                                              \item
                                              Accuracy
                                              in
                                              line-of-sight (LOS)
                                              conditions
                                              \item
                                              Can
                                              easily
                                              accommodate knowledge of
                                              transmit
                                              antenna patterns
                                            \end{strengthlist}                                            
                                                      &
                                                        \begin{limitationlist}
                                                          \item
                                                          Inaccurate
                                                          in non-LOS
                                                          (NLOS)
                                                          conditions

                                                          \item
                                                          Requires
                                                          transmitter locations
                                                        \end{limitationlist}
    \\
    \hline
    Kernel-based Learning&                           
                            \begin{inputlist}
                              \item Reproducing kernel
                              $\kernel(\loc,\locp)$
                              \item Loss $\loss $
                              \item Regularization parameter $\regpar$
                            \end{inputlist}
                                          &
                                            \begin{strengthlist}
                                              \item High flexibility
                                              \item Does not require
                                              transmitter locations
                                            \end{strengthlist}                                      
                                                      &
                                                        \begin{limitationlist}
                                                          \item
                                                          Sensitive to
                                                          the choice
                                                          of the
                                                          kernel
                                                          \item
                                                          Depending on
                                                          $\loss$, a
                                                          numerical
                                                          solver may
                                                          be necessary
                                                          \item
                                                          $\regpar$
                                                          must be
                                                          tuned,
                                                          e.g. via cross-validation
                                                        \end{limitationlist}\\
    \hline
    Kriging&
                           \begin{inputlist}
                              \item
                              Map's mean $\mu_p(\loc)$ and covariance $\cov[\pow(\loc),\pow(\locp)]$
                              \item
                              Measurement noise variance $\noisevar$
                             
                            \end{inputlist}
                                          &
                                            \begin{strengthlist}
                                              \item LMMSE optimality
                                              \item Closed form
                                              \item Naturally suited
                                              to the customary log-normal shadowing model
                                              \item Estimation error
                                              can be quantified
                                            \end{strengthlist}                                            
                                                      &
                                                        \begin{limitationlist}
                                                          \item
                                                          Accurate covariance structure
                                                          may be
                                                          hard to obtain
                                                          \item
                                                          Requires
                                                          user
                                                          locations
                                                          
                                                        \end{limitationlist}\\
      \hline
      Sparsity-based Methods&                           
                            \begin{inputlist}
                              \item Discrete grid
                              \item Regularization parameter $\lambda$
                            \end{inputlist}
                                          &
                                            \begin{strengthlist}
                                              \item Efficient algorithms available for solution
                                              \item Recovered sparse solution readily interpretable
                                            \end{strengthlist}                                            
                                                      &
                                                        \begin{limitationlist}
                                                          \item Prior knowledge on propagation characteristics needed
                                                          \item Errors due to grid mismatch
                                                        \end{limitationlist}\\
            \hline
      Matrix Completion&                           
                            \begin{inputlist}
                              \item Regular grid
                              \item Regularization parameter $\lambda$
                            \end{inputlist}
                                          &
                                            \begin{strengthlist}
                                              \item Agnostic to propagation characteristics
                                              \item Spatial correlation structures exploited
                                            \end{strengthlist}
                                                      &
                                                        \begin{limitationlist}
                                                          \item Low-rank condition is critical
                                                          \item Sufficient number of measurements required for stable interpolation
                                                        \end{limitationlist}\\
      \hline
%       \end{tabular}
%  }
%  \end{center}
% % \caption{\changed{Comparison of the main methods discussed in this
% %     tutorial for estimating power maps (Part 1/2).}}
% % \label{table:pmestsummary1}
% \end{table}

% \begin{table}[bhtp] % table as image
%  \centering
%  \begin{center}
% { \color{black}    
%  \begin{tabular}{|p{1.5cm} | p{3.6cm} | p{5.3cm} | p{5.3cm} |}
%     \hline
%     Method & Input (besides measurements) & Strengths & Limitations \\
%     \hline
%     \hline  
      Dictionary Learning&                            
                            \begin{inputlist}
                              \item Dictionary size $Q$
                              \item Regularization parameters $\lambda_s$, $\lambda_L$
                            \end{inputlist}
                                          &
                                            \begin{strengthlist}
                                              \item  Powerful union-of-subspace structure for spatial patterns
                                              \item Can accommodate high temporal dynamics
                                            \end{strengthlist}
                                                      &
                                                        \begin{limitationlist}
                                                          \item  Nonconvex optimization
                                                          \item Hyperparameter tuning is necessary
                                                        \end{limitationlist}\\
      \hline
      Deep Learning&                          % It depends on the specific method, but it may include
                            \begin{inputlist}
                              \item Terrain maps
                              \item Vegetation maps
                              \item Building height maps
                              \item Network architecture
                              \item Training parameters
                              \item etc.
                            \end{inputlist}
                                          &
                                            \begin{strengthlist}
                                              \item Can learn
                                              propagation patterns
                                              from a data set
                                              \item More accurate than
                                              other methods if
                                              sufficient data is available~\cite{teganya2020rme}
                                            \end{strengthlist}                                            
                                                      &
                                                        \begin{limitationlist}
                                                          \item Large
                                                          amount of
                                                          data is
                                                          required
                                                          \item
                                                          Training is
                                                          computationally intensive
                                                        \end{limitationlist}\\
      \hline
    \end{tabular}
  }
  \end{center}
  \caption{\changed{Comparison of the power map estimation methods discussed in this
      tutorial.}}
  \label{table:pmestsummary1}
\end{table}

%%%%%%%%%%%%%%%%%%%%%%%%%%%%%%%%%%%%%%%%  
\subsection{Estimation of Power Maps}
%%%%%%%%%%%%%%%%%%%%%%%%%%%%%%%%%%%%%%%%

%%%%%%%%%%%%%%%%%%%%%%%%%%%%%%%%%%%%%%%%%%%
\subsubsection{Linear Parametric RME}
\label{sssec:linear}
%%%%%%%%%%%%%%%%%%%%%%%%%%%%%%%%%%%%%%%%%%%
Let us start from the simple yet illustrative scenario where there is a single transmitter with known location $\txloc_1$ in free space. As per Friis' transmission equation, the received power at location $\loc$ is inversely proportional to the squared distance $\|\loc-\txloc_1\|^2$. In other words, $\pow(\loc)$ can be written as $\pow(\loc)=\spcoef_1\spbasisfun_1(\loc)$, where $\spbasisfun_1(\loc) :=1/\|\loc-\txloc_1\|^2$ and $\spcoef_1$ depends on the (unknown) transmit power. Therefore, to estimate $\pow(\loc)$ everywhere, it suffices to obtain $\spcoef_1$. Clearly, this could be accomplished from a single noiseless measurement $\meas{1}= \pow(\loc_1)$ at $\loc_1$ by setting $\spcoef_1 = \meas{1}/\spbasisfun_1(\loc_1)$.

Similarly, if $\txnum$ transmitters with known locations $\txloc_1,\ldots,\txloc_\txnum$ are active in a certain region, one can let $\spbasisfun_\txind(\loc):=1/\|\loc-\txloc_\txind\|^2$ to write $\pow(\loc)$ as
    \begin{align}
      \label{eq:pardec}
      \pow(\loc)=\spcoef_1\spbasisfun_1(\loc) + \ldots +
      \spcoef_\txnum\spbasisfun_\txnum(\loc),
    \end{align}
\changed{so long as the transmitted waveforms are uncorrelated. } Based on \eqref{eq:pardec}, one can  typically  estimate the $\txnum$ coefficients $\{\alpha_s\}$ from $\txnum$ noiseless measurements by solving the system of equations
    \begin{align}
      \meas{1}& = \spcoef_1\spbasisfun_1(\loc_1) + \ldots +
                \spcoef_\txnum\spbasisfun_\txnum(\loc_1)\nonumber\\
      &\hspace{6em} \vdots \label{eq:S_eqs}\\
      \meas{\txnum}& = \spcoef_1\spbasisfun_1(\loc_\txnum) + \ldots +
                     \spcoef_\txnum\spbasisfun_\txnum(\loc_\txnum).\nonumber
    \end{align}
In practice, however, the measurements are noisy and one may use more than $\txnum$ of them to estimate the coefficients. Upon defining $\spcoefvec:=[\spcoef_1,\ldots,\spcoef_\txnum]\transpose$, $\allmeasvec:=[\meas{1},\ldots,\meas{\measnum}]\transpose$, $(\spbasismat)_{\measind,\txind}:=\spbasisfun_\txind(\loc_\measind)$ and $\noisevec := [\noise{1},\ldots,\noise{\measnum}]\transpose$, \eqref{eq:S_eqs} can be extended to the case with $N > S$ measurements as $\allmeasvec=\spbasismat\spcoefvec+\noisevec$. The least squares (LS) estimate of $\spcoefvec$ is therefore $\spcoefvecest=\argmin_\spcoefvec\|\allmeasvec-\spbasismat\spcoefvec\|^2$. Because the number $\txnum$ of parameters to be estimated does not depend on the number $\measnum$ of measurements, this approach is termed \emph{parametric}. Further parametric and \emph{non-parametric} estimators are discussed in the rest of this section.
    
\begin{figure}[!t]
  \centering
  \includefigandtrim{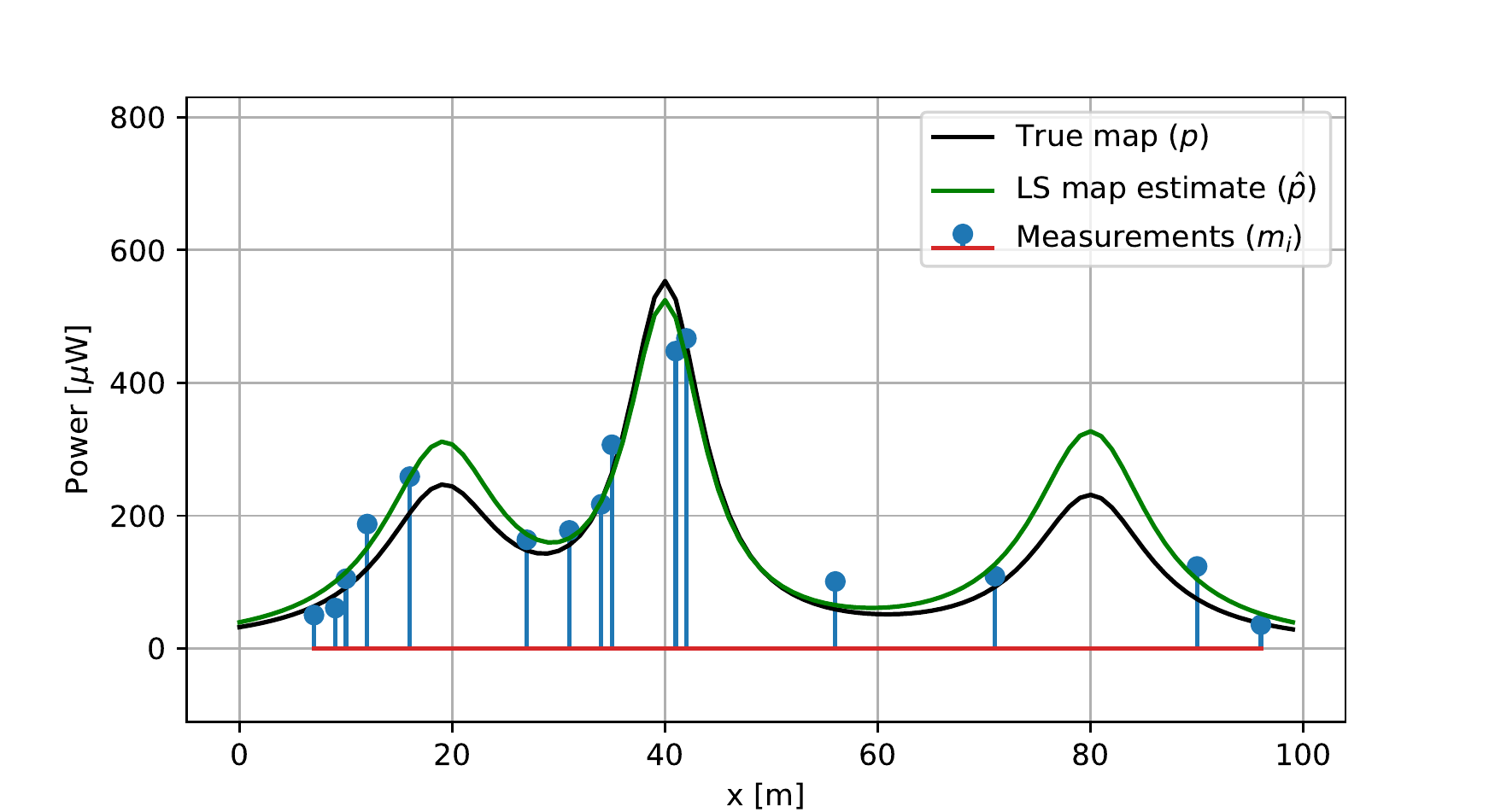}
  \caption{Example of map estimation in 1D using a parametric
    estimator that knows the transmitter locations. The estimate is
    reasonably accurate despite the low number of measurements. }
  \label{fig:locaware}
\end{figure}

Fig.~\ref{fig:locaware} illustrates a setup where a map needs to be estimated on a line, i.e. the region of interest is given by $\region\subset \rfield$, which may correspond e.g. to a road or a railway. The true map and the estimated map obtained by substituting $\spcoefvecest$ into the right-hand side (RHS) of \eqref{eq:pardec} are compared. The estimated map is seen to be reasonably accurate and can be shown to converge to the true map for $\measnum\rightarrow \infty$ under mild conditions.  
    
%%%%%%%%%%%%%%%%%%%%%%%%%%%%%%%%%%%%%%%%%%%%%
%%%%%%%%%%%%%%%%%%%%%%%%%%%%%%%%%%%%%%%%%%%%%
 
%%%%%%%%%%%%%%%%%%%%%%%%%%%%%%%%%%%%%%%%%%%%%
\begin{figure}[!t]
  \centering
  \includefigandtrim{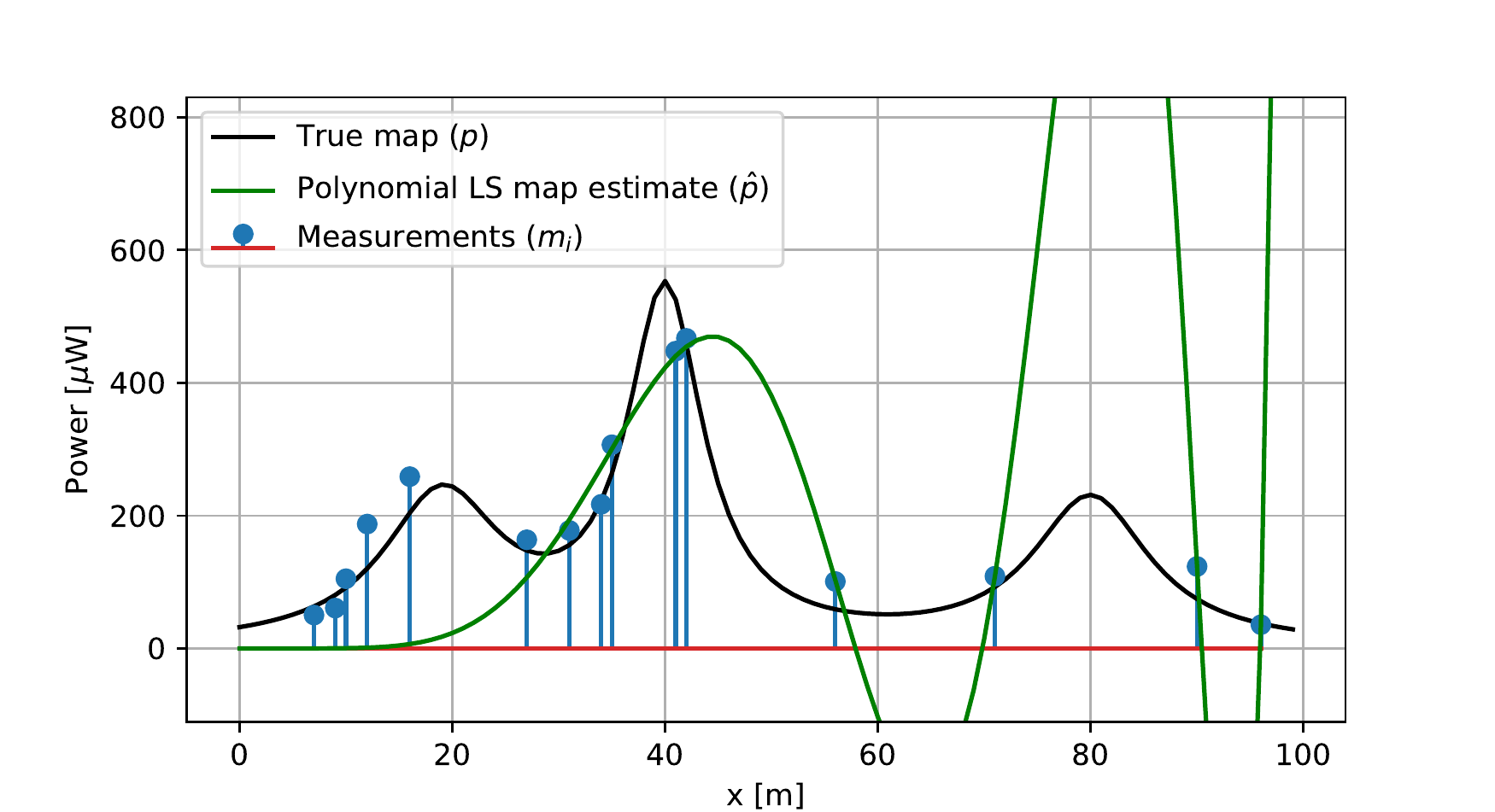}
  \caption{Example of map estimation by fitting a polynomial of degree
    13 via LS. The estimate is clearly unsatisfactory despite the fact
    that the estimate fits accurately most of the measurements.  }
  \label{fig:poly}
\end{figure}
%%%%%%%%%%%%%%%%%%%%%%%%%%%%%%%%%%%%%%%%%%%%%

So far, it was assumed that propagation takes place in free space. If this is not the case, then the basis functions $\spbasisfun_\txind(\loc)=1/\|\loc-\txloc_\txind\|^2$ may not yield a satisfactory fit. Although one can in principle adopt other families of basis functions, such as those determined by the well-known Okumura-Hata model\ncite[Sec. 9.1.2]{jeruchim2000},  the flexibility of such an approach is rather limited. Besides, the location of the sources is required, which may not be a realistic assumption in some applications. These observations suggest generalizing~\eqref{eq:pardec} to 
    \begin{align}
      \label{eq:pardec2}
      \pow(\loc)=\spcoef_1 \tilde \spbasisfun_1(\loc) + \ldots +
      \spcoef_\spbasisnum \tilde\spbasisfun_\spbasisnum(\loc),
    \end{align}
where $\tilde \spbasisfun_\spbasisind(\loc)$ can take an arbitrary form and need not even be linked to any particular transmitter.  For example, in the case where a map needs to be constructed on a line, $\{\tilde \spbasisfun_\spbasisind(\loc)\}_\spbasisind $ could form a polynomial basis by setting $\tilde  \spbasisfun_\spbasisind(\loc)=\tilde \spbasisfun_\spbasisind(\locs)=\locs^{\spbasisind-1}$. The coefficients $\{\spcoef_b\}$ can again be found by LS estimation. However, despite the appealing simplicity of this approach, the quality of the estimates is often poor. As illustrated by Fig.~\ref{fig:poly} for the same setup as in Fig.~\ref{fig:locaware}, this kind of regression methods may be sensitive to the choice of the basis functions.

%%%%%%%%%%%%%%%%%%%%%%%%%%%%%%%%%%%%%%%%
\subsubsection{Kernel-Based Learning}
\label{sec:kbl}
%%%%%%%%%%%%%%%%%%%%%%%%%%%%%%%%%%%%%%%%
\changed{The main challenge faced by the parametric methods described in the previous section lies in the difficulty to select suitable basis functions. This difficulty is further exacerbated  in higher dimensions, such as when $\region =\rfield^2$ or $\rfield^3$.} Kernel-based learning can sidestep this issue while enjoying simplicity, universality, and good performance~\cite{scholkopf2001}.

%%%%%%%%%%%%%%%%%%%%%%%%%%%%%%%%%%%%%%%%%%%%%
\begin{figure}[!t]
  \centering
  \includefigandtrim{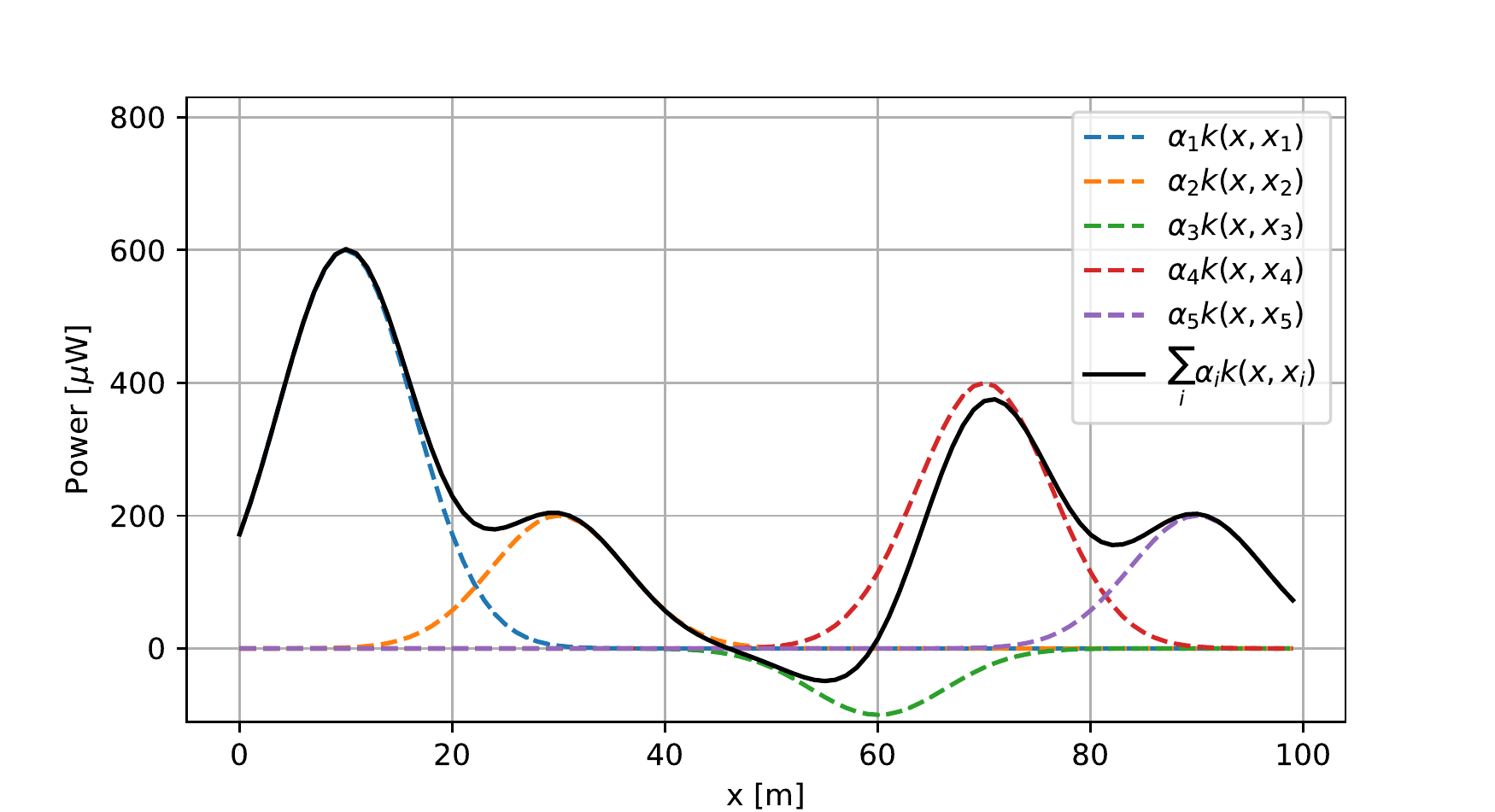}
  \caption{Example of a function in an RKHS obtained with the
    expansion in \eqref{eq:rkhs} with only 5 terms. 
  }
  \label{fig:rkhs}
\end{figure}
%%%%%%%%%%%%%%%%%%%%%%%%%%%%%%%%%%%%%%%%%%%%%

Upon postulating a family of functions $\rkhs$, the goal is to select, based on the data $\{(\loc_\measind,\meas{\measind})\}_{\measind=1}^{\measnum}$, a function $\powest$ in $\rkhs$ that satisfies $\powest(\loc)\approx\pow(\loc)~\forall \loc$. In kernel-based learning, $\rkhs$ is a special class of functions termed \emph{reproducing-kernel Hilbert space} (RKHS), given by
\begin{align} \label{eq:rkhs}
    \rkhs:=\left\lbrace \rkhsfun: \rkhsfun(\loc)=\sum_{i=1}^ \infty \spcoef_i
        \kernel(\loc,\loc'_i),~\loc'_i \in \region,~ \spcoef_i \in
        \mathbb{R}~ \changed{\forall i}\right\rbrace.
\end{align}
Here, $\kernel:\region \times \region \rightarrow \mathbb{R}$ is a \emph{reproducing kernel}~\cite[Ch. 2]{scholkopf2001}, which is a function that is (i) symmetric, i.e., $\kernel(\loc,\locp)=\kernel(\locp,\loc)~\forall\locp,\loc$; and (ii) positive-definite, meaning that the matrix $\bar \kernelmat$ with entries $(\bar \kernelmat)_{i,j}=\kernel(\loc_i,\loc_j)$ is positive-definite for any set of points $\{\loc_1,\ldots,\loc_\measnum\}$. A common choice is the so-called Gaussian \emph{radial basis function}~(RBF) $\kernel(\loc,\loc'):=\exp\left( - \| \loc-\loc'\|^2/{2\sigma^2}\right) $, where $\sigma>0$ is a prescribed parameter. Seen as a function of $\loc$, $\kernel(\loc,\loc'_i)$ is a bell-shaped surface centered at $\loc'_i$. Thus, it can be observed from \eqref{eq:rkhs} that a function in $\rkhs$ is a superposition of (a possibly infinite number of) Gaussian bells with different centers and amplitudes, as illustrated in Fig.~\ref{fig:rkhs}.

In view of \eqref{eq:rkhs}, finding a suitable estimate $\powest$ in $\rkhs$ amounts to determining a set of coefficients $\{\alpha_i\}$ and centroids $\{\loc_i'\}$. To this end, a typical approach is to solve 
\begin{align} \label{eq:pregr}
        \powest= \argmin_{{\rkhsfun}\in
        \rkhs}\frac{1}{\measnum}\sum_{\measind=1}^\measnum
        \loss \left(\meas{\measind}, {\rkhsfun}(\loc_\measind)\right) +  \regpar \| {\rkhsfun}  \| _{\rkhs}^2,
\end{align}
where $\regpar>0$ is a pre-determined regularization parameter and $\loss$ is a loss function quantifying the deviation between the observations $\{\meas{\measind}\}_{\measind=1}^\measnum$ and the predictions $\{\rkhsfun(\loc_\measind)\}_{\measind=1}^\measnum$ produced by a candidate $\rkhsfun$. If the \emph{square loss} $\loss(\meas{\measind}, {\rkhsfun}(\loc_\measind)) = (\meas{\measind}- {\rkhsfun}(\loc_\measind))^2 $ is adopted, \eqref{eq:pregr} becomes \emph{kernel ridge regression} (KRR)~\cite[Ch. 4]{scholkopf2001}. The RKHS norm\footnote{\changed{The term $\| {\rkhsfun} \| _{\rkhs}^2$ in \eqref{eq:pregr}  can be replaced by other increasing functions of $\| {\rkhsfun} \| _{\rkhs}$.}} of $\rkhsfun(\loc)=\sum_{i=1}^ \infty \spcoef_i \kernel(\loc,\loc'_i)$ is given by 
\begin{align}
\label{eq:rkhsnorm}
  \left \| \rkhsfun \right \| _{\rkhs} := \sqrt{\sum_{i=1}^{\infty} \sum_{j=1}^{\infty}\spcoef_i \spcoef_j \kernel(\loc'_i,\loc'_j)}.
\end{align}
  
\changed{To understand the role of the regularization term $\regpar \| {\rkhsfun}  \| _{\rkhs}^2$ in \eqref{eq:pregr}, first note that $\loss$ is typically designed so that its minimum is attained when $\rkhsfun(\loc_\measind) = \meas{\measind}$. Thus, in the absence of the regularization term, owing to the infinite degrees of freedom of $\rkhsfun$ (cf.~\eqref{eq:rkhs}), the solution $\powest$ to \eqref{eq:pregr} would achieve a perfect fit for {\em all} measurements. However, such a $\powest$ would typically be  highly irregular since it would fit even the noise component of the measurements and, thus, likely differ significantly from $\pow$ at the locations where no measurements were  taken. The regularization term helps avoid such {\em overfitting} by promoting smoothness in $\powest$. The reason is that, since $\kappa$ is positive-definite, $\left \| \rkhsfun \right \| _{\rkhs}^{2}$ penalizes large values of $\{\spcoef_i\}$, which tend to occur in  overfitted solutions. Parameter $\lambda$ is adjusted to achieve the ``sweet spot'' between data fitting and regularization.}

\begin{figure}[!t]
  \centering
  \includefigandtrim{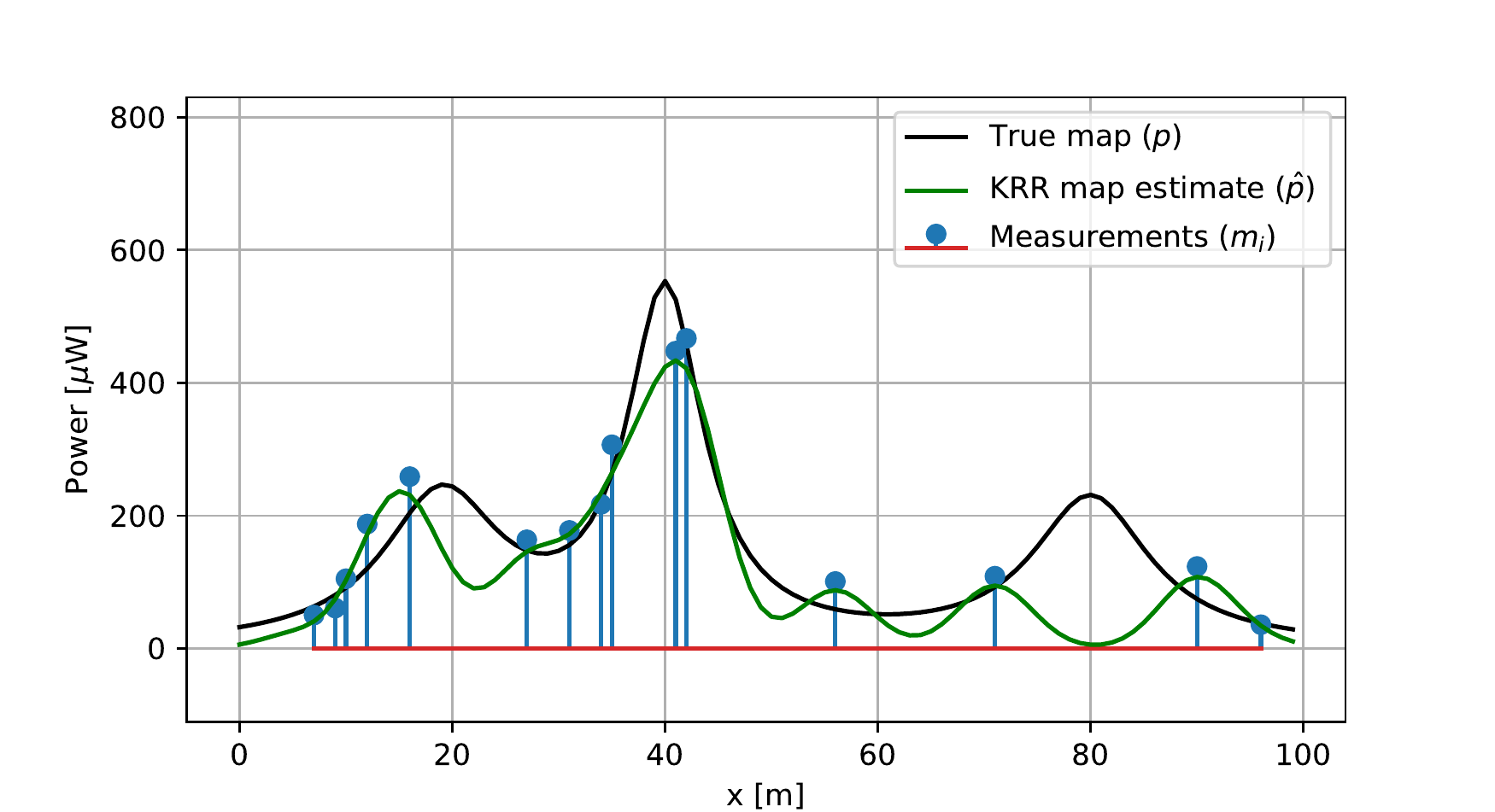}
  \caption{Example of KRR estimate. As expected, the quality of the
    fit is higher in regions with higher  measurement density.
  }
  \label{fig:krrlow}

  \centering
  \includefigandtrim{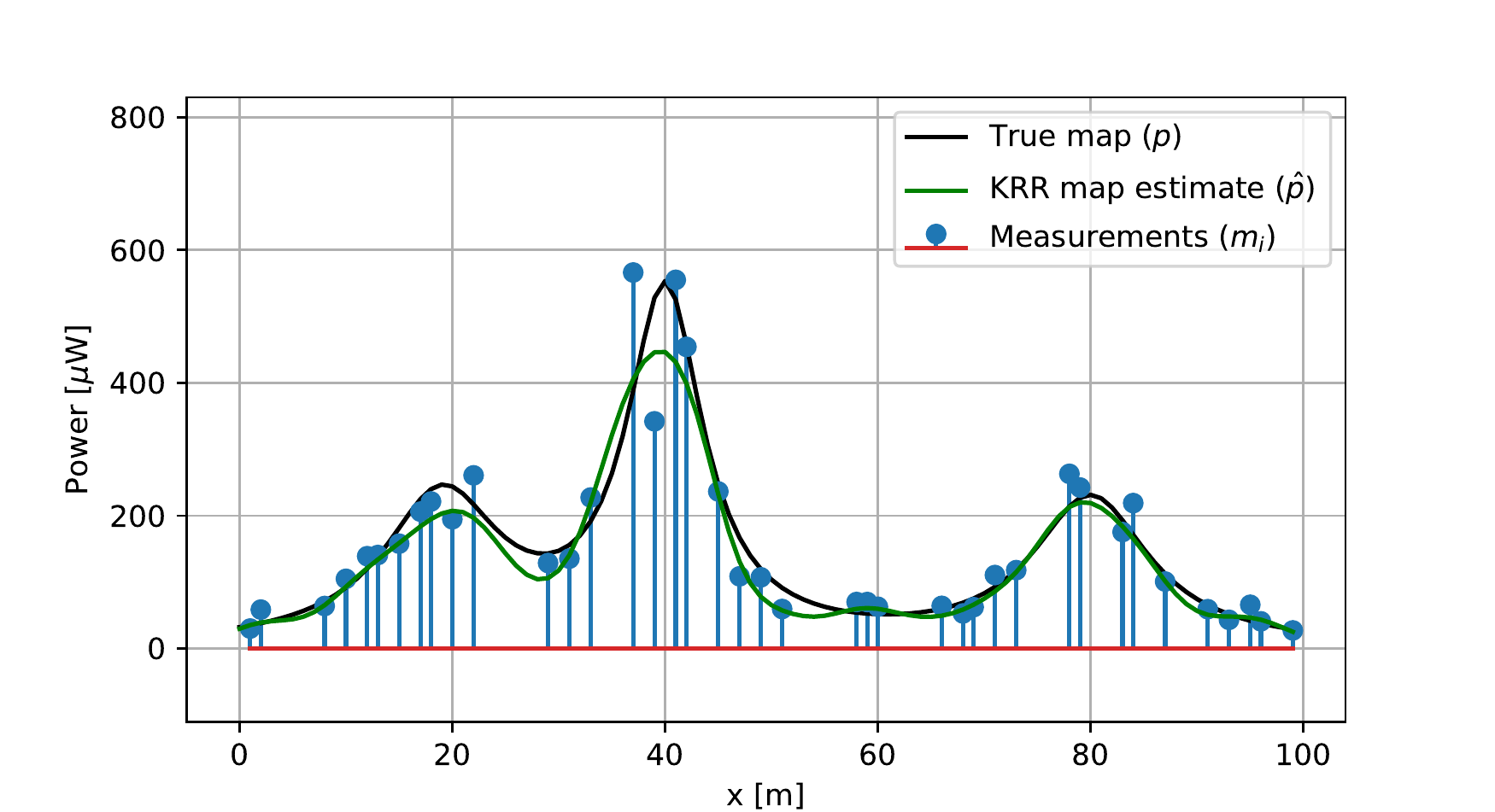}
  \caption{Example of KRR estimate with more measurements than in
    Fig.~\ref{fig:krrlow}. The fit is considerably better. 
  }
  \label{fig:krrhigh}
\end{figure}

To solve \eqref{eq:pregr}, one could initially think of substituting the expansion \eqref{eq:rkhs} into \eqref{eq:pregr} and optimizing over the infinitely many coefficients $\{\spcoef_i\}$ and centroids $\{\loc_i'\}$. However, this approach is obviously intractable. Instead, the so-called \emph{representer theorem} can be invoked~\cite[Th. 4.2]{scholkopf2001}, which states that the solution to \eqref{eq:pregr} must be of the form 
\begin{align} \label{eq:estp}
      \powest(\loc)=\sum_{\measind=1}^\measnum \spcoef_{\measind}\kernel(\loc,\loc_{\measind})
\end{align}
for some $\{ \spcoef_{\measind}\}_{\measind=1}^\measnum$. Observe that the centroids in \eqref{eq:estp} are precisely the measurement locations. This effectively reduces an optimization problem with infinitely many variables to a problem with just the $\measnum$ variables $\spcoef_1,\ldots,\spcoef_\measnum$. For example, if one adopts the square loss, substituting \eqref{eq:estp} into \eqref{eq:pregr} yields
\begin{align}
        \label{eq:optalpha} \spcoefvecest =
        \argmin_{{\spcoefvec}} \frac{1}{\measnum}\left \| \allmeasvec
        -\kernelmat{\spcoefvec} \right \| ^2+ \regpar
        \spcoefvec^{\top}\kernelmat\spcoefvec,
\end{align}
where (with some abuse of notation) $\spcoefvec:=[\spcoef_1,...,\spcoef_\measnum]^{\top}$ and $\kernelmat$ is an $N$-by-$N$ matrix with $(\kernelmat)_{i,j} = \kernel(\loc_{i},\loc_{j})$. It should be noted that now the number of parameters to be determined depends on the number of measurements $\measnum$, which is why this kind of  methods are called \emph{non-parametric}. Problem~\eqref{eq:optalpha} admits the closed-form solution 
\begin{align}
    \label{eq:krrcoef}
          \spcoefvecest= (\kernelmat+\regpar \measnum \textbf{I}_{\measnum})^{-1} \allmeasvec,
\end{align}
from which $\powest$ can be obtained via \eqref{eq:estp}.  Figs.~\ref{fig:krrlow} and~\ref{fig:krrhigh} show the KRR-based map estimates in the same setup as in Figs.~\ref{fig:locaware} and~\ref{fig:poly}. It can be seen that as the number of measurements increases, the estimated map becomes closer to the true map. 
  
%%%%%%%%%%%%%%%%%%%%%%%%%%%%%%%%%%%%%%%%%%%
%%%%%%%%%%%%%%%%%%%%%%%%%%%%%%%%%%%%%%%%%%%

{\changedr It is worth mentioning that RME based on  kernel methods is best suited for scenarios where no prior knowledge on the propagation environment is available. % due to their ability to approximate spatial fields with arbitrarily high accuracy\ncite{carmeli2010vector}. A brief discussion on the extensions to KRR is in order. The remainder of this section  briefly discusses extensions to the kernel ridge regression estimator discussed here.
\changed{When some prior information, such as the transmitter locations or the path loss exponent, is indeed available, it is also possible to combine the flexibility of non-parametric kernel methods with the ability of  parametric methods to} capture prior information by means of appropriate basis functions. To this end, one can postulate that $\pow$ can be represented as the sum of a function in the form of \eqref{eq:pardec2} and a function in an RKHS~\cite{romero2015onlinesemiparametric}. Such an approach also generalizes the so-called \emph{thin-plate spline regression}, which has well-documented merits in RME~\cite{bazerque2011splines,romero2017spectrummaps}.
}

\changed{Another limitation of kernel-based methods is the need for choosing the kernel (including its parameters), which may affect estimation performance significantly. This difficulty may be alleviated through \emph{multikernel} learning, where a dictionary of kernels can be specified and a suitably designed algorithm uses the measurements to construct a kernel by combining the kernels in the dictionary; see references in \cite{teganya2020rme}}\ncite{bazerque2013basispursuit}.

%%%%%%%%%%%%%%%%%%%%%%%%%%%%
\subsubsection{Kriging}
\label{sec:kriging}
%%%%%%%%%%%%%%%%%%%%%%%%%%%%
RME can also be formulated in a statistical framework, where $\pow(\loc)$ is treated as a random process. A popular approach is kriging, which is a linear spatial interpolator  based on  the linear minimum mean square error (LMMSE) criterion~\cite{alayafeki2008cartography,agarwal2018spectrum,shrestha2022surveying}\ncite{beers2004kriging,boccolini2012wireless}. In {\em simple kriging}, the mean and the covariance of $\pow(\loc)$ are assumed to be known. That is, $\mu_p(\loc) := \expected[\pow(\loc)]$ and $\cov[p(\loc),p(\loc')]$ are given  for all $\loc$ and $\loc'$. How to obtain these functions is discussed later.

Under the measurement model $m_n = \pow(\loc_n) + z_n$, $n=1,2,\ldots,N$, assume that $z_n$ is zero-mean with variance $\sigma_z^2$ and uncorrelated with $z_{n'}$ for all $n'\neq n$ and with $\pow(\loc)$ for all $\loc$. Thus, the mean and  covariance of the measurements are respectively  $\expected[m_n] = \mu_p(\loc_n)$ and $\cov[m_n,m_{n'}] = \cov[p(\loc_n),p(\loc_{n'})] + \sigma_z^2 \delta_{n,n'}$, where $\delta_{n,n'}$  equals  $1$ if $n=n'$ and $0$ otherwise. It can also be verified that $\cov[p(\loc),m_n] = \cov[p(\loc),p(\loc_n)]$. Then, it can be shown that the  LMMSE estimator of $\pow(\loc)$ based on the measurements $\allmeasvec\define[\meas{1},\ldots,\meas{\measnum}]\transpose$ is given by 
\begin{align}
\label{eq:lmmse}
      \powest(\loc) = \powmean(\loc) + \cov[\pow(\loc),\allmeasvec] \cov\inv[\allmeasvec,\allmeasvec](\allmeasvec - \expected[\allmeasvec]),
\end{align}
\changed{
where $\cov[\allmeasvec,\allmeasvec]$ is the $\measnum \times \measnum$ matrix whose $(\measind,\measind')$-th entry is $\cov[\meas{\measind}, \meas{\measind'}]$ and  $\cov[\pow(\loc),\allmeasvec]$ is the $\measnum\times 1$ vector with $\measind$-th entry equal to $\cov[\pow(\loc),\meas{\measind}]$.}

It is worth comparing \eqref{eq:lmmse} with \eqref{eq:estp} and \eqref{eq:krrcoef}. It can be easily seen that, except for the mean terms in \eqref{eq:lmmse}, the estimators provided by \eqref{eq:lmmse} and \eqref{eq:estp} coincide if one sets $\kernel(\loc,\loc')=\cov[\pow(\loc),\pow(\loc')]$ and $\regpar$ is adjusted properly. This is a manifestation of the well-known fact that a reproducing kernel can be thought of as a generalization of covariance. \changed{As a result, some of the practical issues and corresponding mitigation strategies for kernel-based learning apply to kriging as well.}

To obtain the mean $\mu_p(\loc)$ and the covariance $\cov[p(\loc),p(\loc')]$ of the map $\pow(\loc)$ to be estimated, one can rely on historic measurement data. Given the covariance function, {\em universal kriging} also provides a framework to estimate $\mu_p(\loc)$ as a part of the kriging estimator. 

Next,  a simple example with a single transmitter at location $\txloc$ transmitting with power $\txpow$ will be used to illustrate how the mean and  covariance can be derived from common propagation models; a more sophisticated example involving the idea of universal kriging and incorporating temporal variations as well will be presented in Sec.~\ref{sec:nontomo}.
%
  %Another approach to radio map estimation arises by adopting a probabilistic model. \emph{kriging} may be one of the simplest and most popular approaches~\cite{alayafeki2008cartography,
  %agarwal2018spectrum, shrestha2022surveying}\ncite{beers2004kriging,boccolini2012wireless}. There are several ways to arrive at the kriging estimator. This section pursues one of them that models $\pow(\loc)$ as a random variable by relying on customary channel assumptions.
%Assume for simplicity that there is a single transmitter at location $\txloc$ that transmits with power $\txpow$.
%
\def \dB {{\mathrm{dB}}}
  \begin{bullets}
    \blt[shadowing + fading + noise]To this end, note that the received power in  logarithmic scale can be written as $\pow_\dB(\loc) = \txpow_\dB + \ch_\dB(\txloc, \loc)$, where $\txpow_\dB$ and $\ch_\dB(\txloc, \loc)$ are expressed in dB. A common decomposition for the latter is
    $\ch_\dB(\txloc, \loc) = \pathloss(\loc) - \shad(\loc) - \fad(\loc)$, where 
    \begin{bullets}
      \blt $\pathloss(\loc)$ is the path loss, 
      \blt $\shad(\loc)$ is the attenuation due to shadow fading, and
      \blt $\fad(\loc)$ is the attenuation due to fast fading.
    \end{bullets} The dependence on $\txloc$ and the subscript $\dB$ on the RHS have been omitted for brevity.
    Recall that shadow fading is produced by obstructions in the line of sight between the transmitter and the receiver, whereas fast fading is due to the constructive and destructive interference between the different multipath components arriving at the receiver. 
    
With the above decomposition, it is common to model $\pathloss(\loc)$ as a deterministic function of $\loc$. Furthermore, $\shad(\loc)$ and $\fad(\loc')$ can be assumed to be uncorrelated for all $\loc$ and $\loc'$ and to have means $\shadmean$ and $\fadmean$, respectively. The spatial structure of $\shad(\loc)$ is often captured by a simple correlation model, such as the Gudmundson model~\cite{gudmundson1991correlation}, which prescribes that $\cov[\shad(\loc),\shad(\loc')] = \shadvar 2^{-\|\loc - \loc'\|/\shaddist}$. Here, $\shadvar$ is a constant and $\shaddist$ is the distance at which the correlation decays by 50\%. On the other hand, due to the rapid spatial variability of $\fad(\loc)$, it is reasonable to set $\cov[\fad(\loc),  \fad(\loc')] = \fadvar \delta_{\loc,\loc'}$. Then, we have $\powmean_{,\dB}(\loc) = \txpow_\dB + \pathloss(\loc) - \shadmean -\fadmean$ and $\cov[\pow_\dB(\loc),  \pow_\dB(\loc')] =
\shadvar 2^{-\|\loc - \loc'\|/\shaddist} + \fadvar \delta_{\loc,\loc'}$.

  \end{bullets}

\subsubsection{Leveraging Sparsity}
\label{sec:sparsity}
%%%%%%%%%%%%%%%%%%%%%%%%%%%%%%%%%%%%%%%%%

%\begin{bullets}
%  \blt[Power maps]
%  \begin{bullets}
%    \blt compressive sampling~\cite{jayawickrama2013compressive,bazerque2010sparsity},
%    \blt sparse Bayesian learning~\cite{huang2014sparsebayesian},
%  \end{bullets}
%\end{bullets}
%\input{sparsity.tex}

\def \bbm {{\boldsymbol{m}}}
\def \bbx {{\boldsymbol{x}}}
\def \bbz {{\boldsymbol{z}}}

\def \bbalpha {{\boldsymbol{\alpha}}}

\def \bbPsi {{\boldsymbol{\Psi}}}

\def \TX {{\text{TX}}}
\def \grid {{\text{grid}}}

In many practical RME problems,  estimation performance can be significantly improved by incorporating  prior information. The sparsity prior has played a critical role in compressive sensing (CS), in which framework  RME problems can often be formulated. Moreover, depending on the choice of the basis functions, the sparsity prior can be physically interpreted in terms of the spatial, temporal, and spectral scarceness of the RF energy distribution~\cite{bazerque2010sparsity,bazerque2011splines}\ncite{jayawickrama2013compressive,}. %, revealing the spectrum occupancy in those domains as well

Consider once more the linear parametric RME model \eqref{eq:pardec}, but rather than assuming that the number $S$ and  locations $\{\bbx_s^\TX\}$ of the transmitters are known, simply discretize the map area using $N_g$ grid points $\{\bbx_{n_g}^\grid\}_{n_g}\subset \region$ representing the possible locations of the transmitters. Then, upon defining $\tilde \bbalpha := [\tilde \alpha_1,\ldots,\tilde \alpha_{N_g}]^\top$ and $\tilde \bbPsi \in \mathbb{R}^{N \times N_g}$ with $(\tilde \bbPsi)_{n,n_g} = \psi_{n_g}(\bbx_n) := 1/\|\bbx_n - \bbx_{n_g}^\grid\|^2$ for $n = 1,\ldots,N$ and $n_g = 1,\ldots,N_g$, one has the model $\bbm = \tilde \bbPsi \tilde \bbalpha + \bbz$. In practical scenarios, it is expected that only a small subset of the grid points are actually occupied by transmitters, that is, $S\ll N_g$. Thus, one can impose the sparsity prior on $\tilde \bbalpha$. For example, a Lasso problem can be formulated as $\hat {\tilde \bbalpha} := \arg \min_{\tilde \bbalpha} \|\bbm - \tilde \bbPsi \tilde \bbalpha\|_2^2 + \lambda \|\tilde \bbalpha\|_1$, where $\lambda>0$ and the term $\|\tilde \bbalpha\|_1 := \sum_{n_g = 1}^{N_g} |\tilde \alpha_{n_g}|$ is known to promote sparsity in $\tilde \bbalpha$\ncite{Tib96}. The non-zero entries of the obtained $\hat{\tilde \bbalpha}$ reveal the (grid-based) locations $\{\bbx_s^\TX\}$ and the number $S$ of the transmitters. Then, one can reconstruct the desired power map $p(\bbx)$ using~\eqref{eq:pardec}.

% In using the basis functions $\psi_{n_g}(\bbx) = 1/\|\bbx - \bbx_{n_g}^\grid\|^2$, the propagation characteristic is assumed to be known deterministically~\cite{jayawickrama2013compressive,bazerque2010sparsity}. The restrictive assumption can be alleviated via a sparse total least-square (TLS) approach~\cite{dallanese2012gslasso}, kernel-based learning~\cite{bazerque2011splines}, and sparse Bayesian learning~\cite{huang2014sparsebayesian}. In~\cite{huang2014sparsebayesian}, the locations of the transmitters were directly estimated without relying on a set of grid points.

%D: I just rephrased to address R2.2. 

\changed{As in the linear parametric RME approach, the
adopted basis functions $\psi_{n_g}(\bbx) = 1/\|\bbx -
\bbx_{n_g}^\grid\|^2$ may not accurately capture the actual propagation
characteristics. Possible remedies for this issue include
sparse total least-squares (TLS)~\cite{dallanese2012gslasso},
kernel-based learning~\cite{bazerque2011splines}, and sparse Bayesian
learning techniques~\cite{huang2014sparsebayesian}. In particular, the basis mismatch issue due to the grid-based discretization of space can be mitigated in the atomic norm minimization framework.}

%\changed{On the other hand, discretizing the space into a grid may
%also introduce a certain error. This limitation is alleviated
%in~\cite{huang2014sparsebayesian}, where the locations of the
%transmitters are directly estimated without relying on a set of grid
%points.}

%%%%%%%%%%%%%%%%%%%%%%%%%%%%%%%%%%%%%%%%%
\subsubsection{Matrix Completion}
\label{sec:matrixcompletion}
%%%%%%%%%%%%%%%%%%%%%%%%%%%%%%%%%%%%%%%%%
%Power maps:~\cite{ding2015devicetodevice}.

%\input{lowrank.tex}

\def \bbx {{\boldsymbol{x}}}

\def \bbM {{\boldsymbol{M}}}
\def \bbP {{\boldsymbol{P}}}

\def \ccalO {{\mathcal{O}}}
\def \ccalR {{\mathcal{R}}}

\def \grid {{\text{grid}}}
\def \rank {{\textrm{rank}}}

Another useful framework for RME is low-rank matrix completion. For
instance, consider building a power map over a rectangular area
$\ccalR \subset \mathbb{R}^2$. By discretizing $\ccalR$ using a
regular grid $\{\bbx_{(i,j)}^\grid: i=1,\ldots,I,\ j=1,\ldots,J\}$,
one can obtain a power map matrix $\bbP \in \mathbb{R}^{I \times J}$
where $(\bbP)_{i,j} := p(\bbx_{(i,j)}^\grid)$. Of course, only a small
subset of the entries will be actually observed by the
sensors. However, when the grid is dense enough compared to the
spatial variability of the map, adjacent entries of $\bbP$ will be
similar, which will, in turn, manifest itself as an approximate
rank deficiency of $\bbP$, %\acom{daniel: this is a bit unclear to me:
  %if the entries of $\bbP$ are correlated, then they are random
  %variables, but the approach is totally deterministic. Also, the
  %reader may not understand why this implies approximately low rank}
that is, $\rank(\bbP) \ll \min\{I,J\}$. Matrix completion thus
tries to estimate the unobserved entries of $\bbP$ under a low-rank
prior. Since directly promoting low rank gives rise to non-convex problems,  tractable formulations are typically pursued by penalizing the nuclear norm of the estimate, which is
the sum of its singular values. Denote the set of indices of the
observed entries as
$\ccalO \subset \{1,\ldots,I\} \times \{1,\ldots,J\}$ and the nuclear
norm of $\bbP$ as $\|\bbP\|_*$. Also, let $\bbM$ be the matrix whose $(i,j)$-th element equals the sensor measurement at
$\bbx_{(i,j)}^\grid$ if $(i,j) \in \ccalO$ and $0$ otherwise. A matrix
completion problem for the power map can be posed
as~\ncite{ding2015devicetodevice}
\begin{align}
\minimize_\bbP \frac{1}{2} \sum_{(i,j) \in \ccalO} \left[ (\bbP)_{(i,j)} - (\bbM)_{(i,j)} \right]^2 + \lambda \|\bbP\|_*.
\end{align} %\acom{Explain lambda.} 
With a sufficient number of observed entries, which depends on the rank and the \emph{incoherence} of $\bbP$, the desired map can be reconstructed reliably\ncite{candes2009exact}.

\changed{When $\ccalR$ grows large, the rank of $\bbP$ may increase, as the power distribution may become more diverse. %\acom{daniel: same as before}. 
In this case, local matrix completion on submatrices of $\bbP$ may be a viable approach~\cite{KHG18}. The matrix completion idea can also be extended to tensors, when the maps in a 3-D space are desired~\cite{SCS19},  or when the time and frequency domains are considered together with space~\ncite{ZWP21}\cite{zhang2019spectrum}.}

%%%%%%%%%%%%%%%%%%%%%%%%%%%%%%%%%%%%%%%%%
\subsubsection{Dictionary Learning}
\label{sec:dictionarylearning}
%%%%%%%%%%%%%%%%%%%%%%%%%%%%%%%%%%%%%%%%%
\def \bba {{\boldsymbol{a}}}
\def \bbd {{\boldsymbol{d}}}
\def \bbm {{\boldsymbol{m}}}
\def \bbs {{\boldsymbol{s}}}
\def \bbv {{\boldsymbol{v}}}

\def \bbA {{\boldsymbol{A}}}
\def \bbD {{\boldsymbol{D}}}
\def \bbM {{\boldsymbol{M}}}
\def \bbL {{\boldsymbol{L}}}
\def \bbO {{\boldsymbol{O}}}
\def \bbS {{\boldsymbol{S}}}

\def \ccalD {{\mathcal{D}}}
\def \ccalM {{\mathcal{M}}}
\def \ccalN {{\mathcal{N}}}

\def \obs {{obs}}

When it is desired to capture the temporal variations of the power map, e.g., to exploit unused spectral resources over both time and space, it is useful to learn a {\em library} of power maps, from which the suitable one can be chosen to explain the power distribution at a given time. Dictionary learning is an unsupervised learning method that seeks a  possibly overcomplete basis, termed a {\em dictionary}, such that the data vectors can be expressed as linear combinations of a small number of  vectors in the dictionary. %Dictionary learning has been successfully applied to various problems in signal processing and machine learning. %Dictionary learning can also be viewed as matrix bi-factorization, where one of the factors is constrained to be a sparse matrix.

Denote the power measurements of the $N$ sensors at time $t$ as $\bbm(t) := [m_1(t),\ldots,m_N(t)]^\top$ for $t=1,\ldots,T$. Dictionary learning postulates that $\bbm(t)$ can be represented using a dictionary $\bbD \in \mathbb{R}^{N \times Q}$ as $\bbm(t) \approx \bbD \bbs(t)$, where $\bbs(t) \in \mathbb{R}^Q$ is a {\em sparse} vector of coefficients for the measurements at time $t$. The columns of $\bbD$ are called the {\em atoms}. Collecting the data samples into a matrix $\bbM := [\bbm(1),\ldots,\bbm(T)] \in \mathbb{R}_+^{N\times T}$, one can appreciate that finding such a dictionary can be viewed as a matrix factorization task since $\bbM \approx \bbD \bbS$, where $\bbS := [\bbs(1),\ldots,\bbs(T)]$ is a sparse matrix. There are various optimization formulations to learn $\bbD$ from $\bbM$~\cite{KSVD}. %\ncite{MBP10} and the references therein. 

In the present context of power map estimation, consider the case where the sensors do not report their measurements every time due to e.g. energy-saving sleep modes or congested signaling channels. Thus, the network controller must apply an appropriate interpolation technique to estimate the missing observations. A helpful piece of side information is the topology of the network of sensors, which is typically maintained for various network control tasks such as routing. To leverage this topology information, let $\bbA \in \{1,0\}^{N \times N}$ be the adjacency matrix of the network topology, i.e., the $(n,n')$-th entry $a_{n,n'}$ of $\bbA$ is equal to $1$ if nodes $n$ and $n'$ can communicate directly with each other and $0$ otherwise. The Laplacian matrix $\bbL$ is defined as $\bbL := \diag\{\bbA {\bf 1}\} - \bbA$, where ${\bf 1}$ is the all-one vector. As seen next, this matrix can be used to promote spatial smoothness in the sense that the power estimates at adjacent sensors are similar.

For training, at each time $t$, a subset $\ccalN^\obs(t) \subset \ccalN := \{1,\ldots,N\}$ of sensors acquire power measurements, which are stacked in vector $\bbm^\obs(t) \in \mathbb{R}_+^{|\ccalN^\obs(t)|}$. Also, let $\bbO(t)$ denote the matrix which contains the $n$-th row of the $N \times N$-identity matrix if and only if $n \in \ccalN^\obs(t)$. Then, upon defining 
\begin{align}
f(\bbs,\bbD; \bbm^\obs(t),\bbO(t)) := \frac{1}{2} \|  \bbm^\obs (t) -  \bbO(t) \bbD \bbs \|_2^2 + \lambda_s \|\bbs\|_1 + \frac{\lambda_L}{2} \bbs^\top \bbD^\top \bbL \bbD \bbs, \label{eq:dl_cost}
\end{align}
the dictionary can be learned via
\begin{align}
\hat \bbD := \argmin\limits_{\bbD \in \ccalD,\{\bbs(t)\}} \sum_{t=1}^T f(\bbs(t),\bbD; \bbm^\obs(t),\bbO(t)), \label{eq:dictl}
\end{align}
where $\ccalD := \{[\bbd_1,\ldots,\bbd_Q] \in \mathbb{R}^{N \times Q}: \|\bbd_q\|_2^2 \le 1, \ q=1,\ldots,Q\}$. The first term in \eqref{eq:dl_cost} promotes the fitness of the reconstruction to the training datum in a LS sense;  the second term, with an adjustable weight $\lambda_s > 0$, is an $\ell_1$-norm-based regularizer encouraging sparsity in $\bbs$; and the third term, with  weight $\lambda_L \ge 0$, captures the prior information that the power levels at the neighboring sensor nodes should be similar, since it holds that $\bbv^\top \bbL \bbv = \frac{1}{2} \sum_{n=1}^N \sum_{n'=1}^N a_{nn'} (v_n - v_{n'})^2$ for any $\bbv \define[ v_1,\ldots v_N]\transpose\in \mathbb{R}^N$.  Problem~\eqref{eq:dictl} can be solved efficiently via a block coordinate descent (BCD) algorithm~\cite{kim2013dictionary}. 

In the operational phase, once the dictionary $\hat \bbD$ is obtained from~\eqref{eq:dictl}, given a (new) set of measurements  $\bar \bbm^\obs$ and the corresponding observation matrix $\bar \bbO$ (corresponding to the observation set $\bar \ccalN^\obs$), one first finds the sparse coefficients by solving $\bar \bbs := \argmin_{\bbs} f(\bbs,\hat \bbD; \bar \bbm^\obs,\bar \bbO)$. Then, the missing power levels for sensors $n \in \bar \ccalN^{miss} := \ccalN \backslash \bar \ccalN^\obs$ can be obtained by first reconstructing the whole $\hat {\bar \bbm} = \hat \bbD \bar \bbs$ and extracting the entries $\{\hat {\bar m}_n\}_{n \in \bar \ccalN^{miss}}$. %from $\hat {\bar \bbm}$. 
\changed{A practical challenge is to implement the algorithm for online and distributed operation to handle large-scale real-time computation~\cite{kim2013dictionary,kim13camsap}. Additionally,  tuning the hyperparameters, such as the dictionary size, and the regularization parameters may require cross-validation based on historic measurements.}

%%%%%%%%%%%%%%%%%%%%%%%%%%%%%%%%%%%%%%%%
\subsubsection{Deep Learning}
\label{sec:deeplearning}
%%%%%%%%%%%%%%%%%%%%%%%%%%%%%%%%%%%%%%%%
\begin{bullets}
  \blt[intro dl]
  \begin{bullets}
    \blt[DNN] A deep neural network (DNN) is a function
    $\nnfun_\nnparvec$ that can be expressed as the composition of more
    elementary functions called \emph{layers}, which are parameterized by a vector
    $\nnparvec$. Training a DNN 
    involves finding
    $\nnparvec$ so that $\nnfun_\nnparvec$ fits the 
    given data set.
    \blt[features]DNNs feature a
    large learning capacity and can be efficient trained via
    stochastic optimization methods. Spatial structures in the data can be readily exploited utilizing \emph{convolutional layers}, in which
    case the DNN is called a \emph{convolutional neural network}
    (CNN). 
  \end{bullets}
  \blt[task formulation]Next, multiple
  approaches to use DNNs for \signalstrength map estimation are described.
\end{bullets}

\paragraph{Pointwise DNN Estimators}
\label{sec:pointwise}

\begin{bullets}
  \blt[descr]The simplest approach is to use a DNN to construct a function, where the input is the sensor location and the output is the signal strength at that location.
  \blt[example]This approach was pursued 
  in~\cite{parera2020tiltdependent}, where the input was encoded using
  a spherical coordinate
  system located at the (single) transmitter.
  \blt[benefits]%
  \begin{bullets}%
    \blt[small size]Since the dimensionality of the input is small,
    the network architecture can be kept simple
    \blt[curse of dim.] and the resulting estimator is not affected by
    the so-called \emph{curse of dimensionality}~\cite[Sec. 4.3]{scholkopf2001}. 
  \end{bullets}
  
  \blt[limitations]%
  \begin{bullets}%%
    \blt[not CNN]\changed{However, such an approach cannot easily capture the spatial structure of the map using CNNs.} %a limitation is that this input
      %encoding is not amenable to CNN architectures, which therefore
      %relinquishes their full potential to exploit spatial
      %information.}
    % 
    \blt[no learning from experience] \changed{Besides, the DNN needs to be
    re-trained for each specific RF environment. Therefore,  it cannot benefit from
    measurements previously collected in other scenarios, such as different
    cities. }
  \end{bullets}
\end{bullets}

\paragraph{Local DNN Estimators}
\label{sec:localdnn}
\begin{bullets}
  \blt[descr]To alleviate the aforementioned limitations of pointwise DNN estimators,
  the network input can be replaced with
  a collection of matrices that capture information about the local
  environment of the sensor. These matrices, typically stacked as
  slabs of a tensor, can be thought of as \emph{local maps}
  \changed{defined over a rectangular grid} centered at the sensor.
  \blt[examples]\changed{A transmitter (alternatively a sensor) distance map, for
    example, is a matrix whose $(i,j)$-th entry equals the distance
    from the $(i,j)$-th grid point to the transmitter
    (sensor)~\cite{imai2019radiopredictioncnn,iwasaki2020transferbasedpower}.}
  \blt[Metadata]
  \begin{bullets}
    \blt[Terrain maps]It is also possible to include a local terrain
    map that indicates the altitude of the terrain at each grid point.
    \blt[building maps]Further kinds of local maps include building
    indicator maps~\cite{teganya2020rme}, 
    \blt[height maps]building height maps~\cite{ratnam2020fadenet,krijestorac2020deeplearning},
    \blt[foliage maps]or foliage maps~\cite{ratnam2020fadenet}.
    \blt[picture]One can also use  aerial or satellite
    images of the surroundings of the sensor as a local
    map \changed{\cite{thrane2020model}}. Fig.~\ref{fig:localdnn} provides an illustration of this kind
    of setup.

\begin{figure}[!t]
  \centering
  \includegraphics[width=0.6\textwidth]{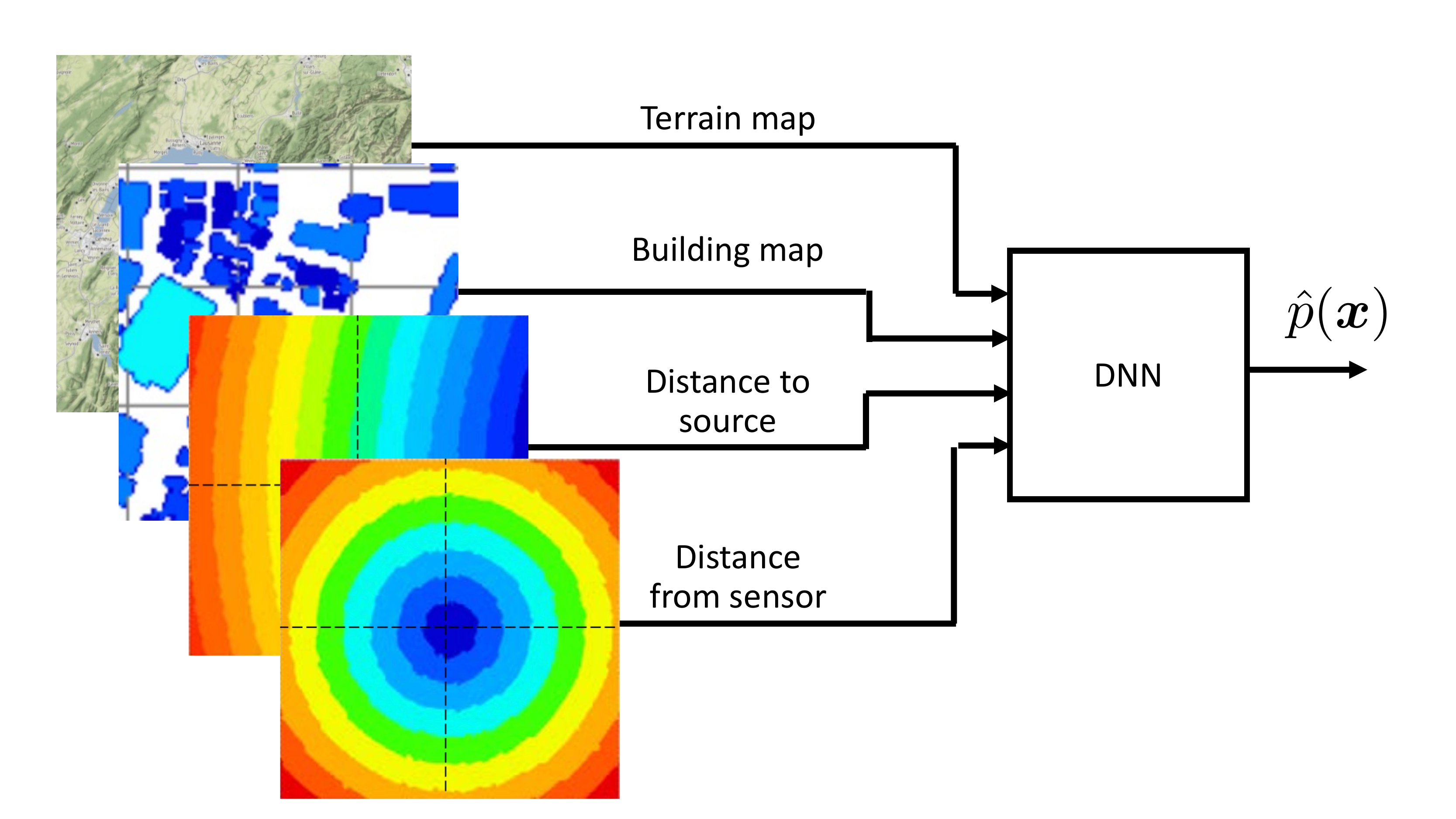}
  \caption{Illustration of a local DNN estimator, which provides $\powest(\loc)$ at a single $\loc$.
  }
  \label{fig:localdnn}
\end{figure}

  \end{bullets}

  \blt[benefits]
  \begin{bullets}
    \blt[conv layers]This input format lends itself to CNN
    architectures that leverage spatial information in the vicinity of
    the sensor to predict the received power.
    \blt[learn from experience]To learn across different environments
    where the transmitters possibly employ different transmit power, one
    can set the output of the network to be the gain between each
    transmitter and the sensor and work out the received power
    afterwards.  This effectively sets this approach halfway  between \signalstrength and propagation map estimation.
    % 
    %\blt[gridless]
  \end{bullets}
  
  \blt[limitations]
  \begin{bullets}
    \blt[]\changed{The practical limitation of this approach is that it requires
    knowledge of the locations (and the transmit powers if one
    wishes to estimate the gains) of all transmitters and the
    measurements must be obtained separately for each transmitter.  }
    \blt[exploits only rx. surroundings]\changed{Furthermore, it only exploits
    the information in the vicinity of the sensor but, in practice,
    obstacles or scatterers far away from the sensor may also affect
    the received power significantly.}
    \blt[fw pass per pt]\changed{In addition, networks designed in this way
    provide the received power (or channel gain) only at a single
    location per evaluation (also known as forward pass). To construct the entire map, the estimator
    needs to be evaluated repeatedly for each point on a grid, resulting in  significant computational complexity.}
  \end{bullets}
\end{bullets}

\paragraph{Global DNN Estimators}
\label{sec:globaldnn}

\begin{bullets}
  \blt[descr]To accommodate global, rather than local, environment
  information, one can create a regular grid across the region where
  the map needs to be constructed and formulate the RME problem as a
  matrix or tensor completion
  task~\cite{niu2018recnet,teganya2020rme,krijestorac2020deeplearning,ratnam2020fadenet,levie2019radiounet}.
  \blt[completion net]To this end, each measurement is associated with
  the nearest grid point and a matrix is constructed with an entry per grid
  point. If a single measurement is assigned to a grid point, the
  corresponding entry contains the measurement. If multiple
  measurements are assigned to a grid point, the corresponding entry may contain their average.  Those points with no associated
  measurements can simply be filled with  physically unlikely values~\cite{niu2018recnet,han2020power,krijestorac2020deeplearning}, or a separate binary mask matrix can be included in the input~\cite{teganya2020rme,shrestha2021deep}. 
  
  \blt[metadata]Other maps with side information, such as the ones used in local DNN estimators, can also be appended to %stacked to the
  %aforementioned measurement matrix and sampling mask to form 
  the input
  tensor to the network. %The same kind of maps as for local DNN
  %estimators (terrain, building, vegetation, etc) can be used, but
  However, note that now
  these maps must be \emph{global} in the sense that they capture the entire region
  of interest. 
  
\blt[architectures]The global input can naturally be processed by CNN architectures. The most common ones are 
  \begin{bullets}%%
    \blt[autoencoder]autoencoders~\cite{han2020power,teganya2020rme}
    \blt[unet]and
    UNets~\cite{krijestorac2020deeplearning,levie2019radiounet}. The motivation for the former is described in Box~\ref{box:manifold}.
  \end{bullets}
A global DNN estimator is illustrated in Fig.~\ref{fig:globaldnn}.

\begin{figure}[!t]
  \centering
  \includegraphics[width=0.6\textwidth]{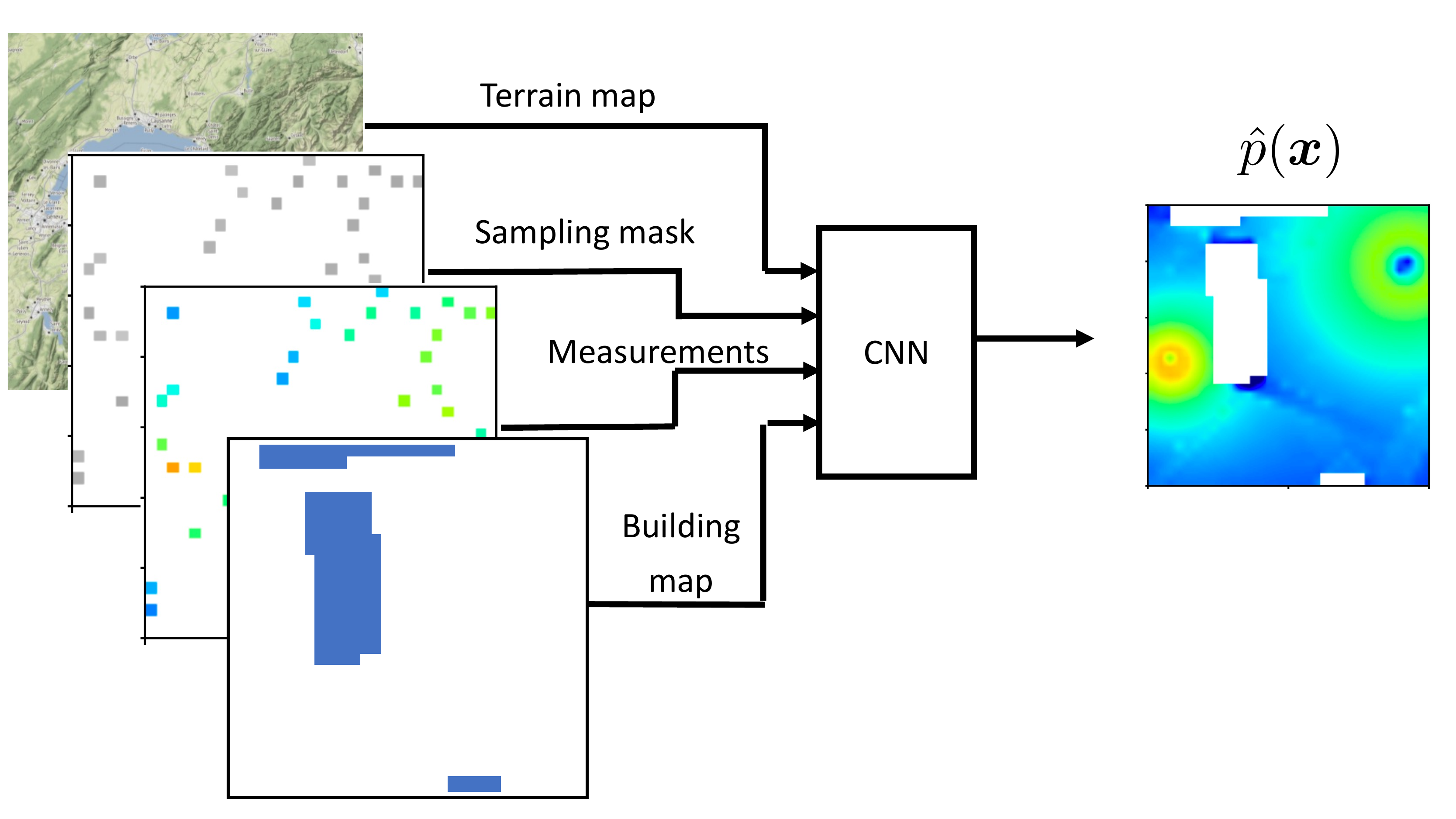}
  \caption{Illustration of a global DNN estimator, which provides $\powest(\loc)$ for all values of  $\loc$ on a grid.
  }
  \label{fig:globaldnn}
\end{figure}

  \begin{floatbox}{}{}
   
   Autoencoder networks are attuned to situations where the data lies on a low-dimensional manifold embedded in a high-dimensional space. To see that this is the case of radio maps, consider the values of a power map in 2D produced by 2 sources radiating with a fixed height and power in free space. A data set can be generated where each map is obtained by placing the sources over random locations on the horizontal plane. Each map is therefore uniquely identified by the 4 scalars corresponding to the locations of the sources. If the maps are defined on a 32 $\times$ 32 grid, they comprise $32^2=1024$ points, which means that these maps lie on a manifold of dimension 4 embedded in a space of dimension 1024. 
   
   This observation is corroborated in \cite{teganya2020rme} by training an autoencoder on the aforementioned data set. An autoencoder is the concatenation of an encoder and a decoder. In this case, the encoder takes a $32\times 32$ map and produces a code vector $\bm \lambda$ of length 4. The decoder takes this vector at its input and aims at reconstructing the original $32\times 32$ map. For properly trained encoder and decoder, the output of the decoder resembles closely the input of the encoder, which means that the code effectively condenses  the information of the map in just 4 numbers. 
   
   Each value of the code identifies a point in the manifold. The top panel of Fig.~\ref{fig:manifold} shows the output of the decoder when its input equals the average of the codes associated with each map in the data set. The rest of panels show the output of the decoder applied to the result of perturbing the entries of this average code indicated by index set $\mathcal{S}$  by an amount equal to the standard deviation of that entry across the data set. This procedure yields different points in the manifold. All panels approximately correspond to maps of the kind composing the data set, which supports the above manifold hypothesis. 
   
   If propagation does not take place in free space or if the power or height of the sources is variable, a longer code needs to be utilized to capture all information in the maps. Experiments with other data sets reveal that, in presence of propagation phenomena such as shadowing and fading, radio maps lie \emph{close} to a manifold of low dimension~\cite{teganya2020rme}. 
   
   \caption{\textbf{Manifold structure of power maps.}}
    \label{box:manifold}
  \end{floatbox}

  %%%%%
  \begin{figure*}[t!]
    \centering
    \includegraphics[width=7cm,height=11.5cm]{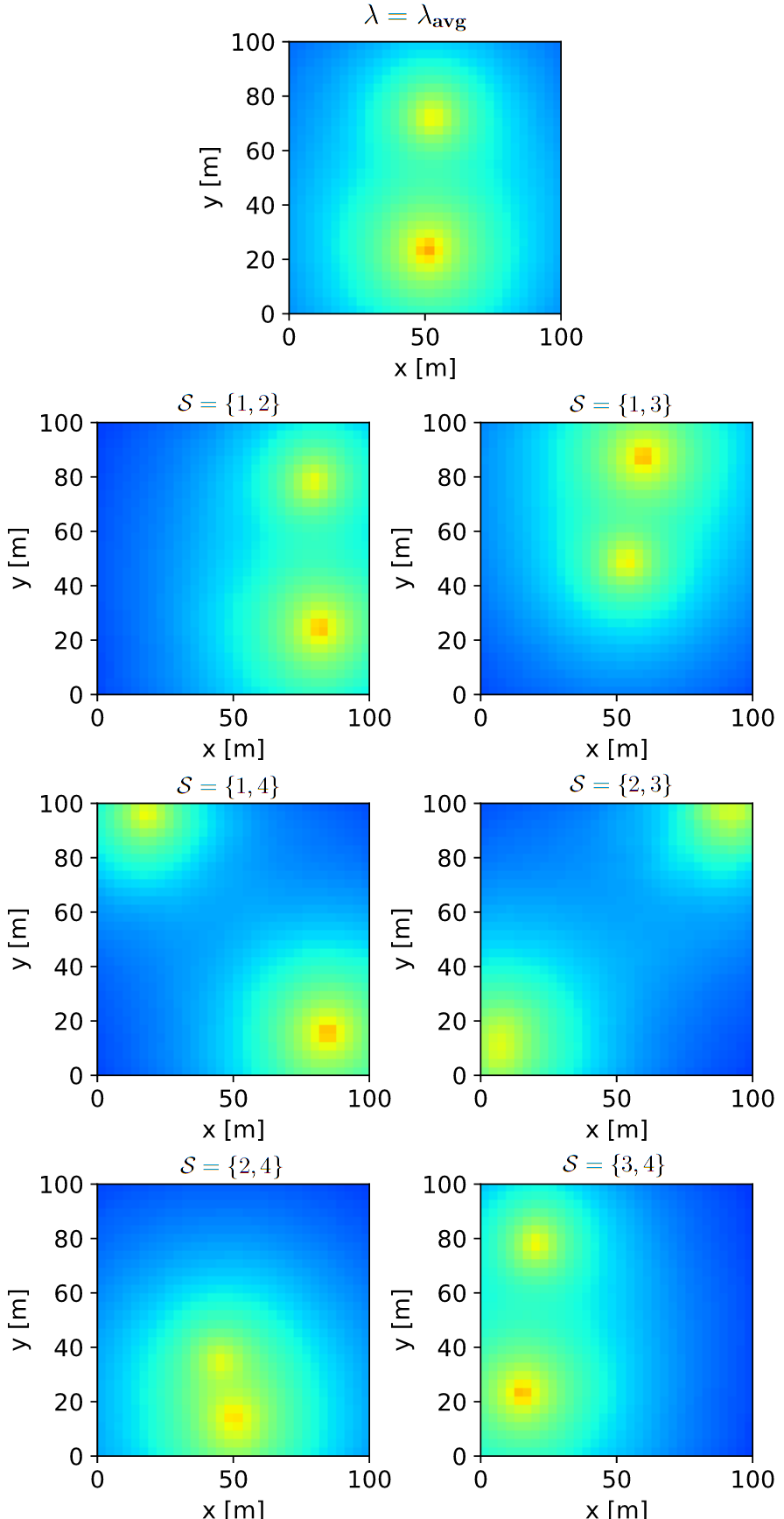}

    \caption{Decoder outputs for the average code and its perturbed versions obtained with an autoencoder with code length 4~\cite{teganya2020rme}.  }
    \label{fig:manifold}
  \end{figure*}
  
  %%%%%%
  
  \blt[Benefits]
  \begin{bullets}
    \blt[exploits global information]%Besides the fact that global DNN
    %estimators exploit information about the entire environment,
    %
    \blt[conv layers]%the formulation they adopt is amenable to CNN
    %architectures.
    % 
    \blt[single fw pass]Unlike local DNN estimators, a single forward
    pass of the DNN produces the entire map. Furthermore, using  map measurements collected from multiple environments, the architecture can readily learn across different RF environments.
    %since each training
    %instance will generally correspond to a different environment,
    %these approaches naturally learn from experience. 
  \end{bullets}
  \blt[limitations]
  \begin{bullets}
    \blt[tr. data]\changed{On the other hand, collecting a sufficiently large
    data set to train such a network may be challenging.} To alleviate
    this difficulty, one may resort to data augmentation or incorporate synthetic data from ray-tracing simulators~\cite{teganya2020rme}.
    \blt[grid discretization]\changed{Another limitation deals with the spatial
    resolution of the constructed maps. A high
    resolution map requires a dense grid,
     significantly increasing computational complexity.}
  \end{bullets}
\end{bullets}
%\acom{If needed, expand with training techniques}
%\acom{talk about fitting in dB?}

%%%%%%%%%%%%%%%%%%%%%%%%%%%%%%%%%%%%%%%
\subsection{Estimation of PSD Maps}
%%%%%%%%%%%%%%%%%%%%%%%%%%%%%%%%%%%%%%%
PSD maps describe how the power distributes not only across space but also across the frequency domain. To estimate a PSD map $\pow(\loc,\freq)$, most schemes assume that the sensors measure the power that they receive at a set of frequencies $\freq_1,\ldots,\freq_\freqnum$. The $\measind$-th measurement is therefore a vector $\measvec{\measind}=[\measpow(\loc_\measind,\freq_1),\ldots,\measpow(\loc_\measind,\freq_\freqnum)]\transpose$, where $\measpow(\loc,\freq)$ denotes the measured PSD at location $\loc$ and frequency $\freq$, possibly obtained by using a periodogram  or Welch's method. Relying on these  measurements, the goal is to obtain a PSD map estimate $\powest$ such that $\powest(\loc,\freq)$ is as close to the true $\pow(\loc,\freq)$ as possible. To this end, several alternatives are explored next.

%%%%%%%%%%%%%%%%%%%%%%%%%%%%%%%%%%%%%%%%%%%%%%%%%%%%
\subsubsection{Separate Estimation per Frequency}
\label{sssec:perfreq}
%%%%%%%%%%%%%%%%%%%%%%%%%%%%%%%%%%%%%%%%%%%%%%%%%%%%
The simplest approach is to consider each frequency separately and essentially decompose the problem of estimating a PSD map at $\freqnum$ frequencies as $\freqnum$ problems of estimating a single power map~\cite{han2020power}. More specifically, the $\freqind$-th power map is estimated from PSD measurements of $\pow(\loc_1,\freq_\freqind),\ldots,\pow(\loc_\measnum,\freq_\freqind)$ using the techniques  described earlier. \changed{The main limitation of this approach is that it disregards any structure in the frequency domain, making it more sensitive to measurement noise than other schemes explored later.} On the upside, these approaches are simple and do not require prior knowledge on the channel or transmit PSD characteristics. Moreover, a twofold benefit arises in terms of the sizes of the training set and the parameters for schemes such as  deep learning estimators. First, provided that the propagation environment affects all frequencies in a similar fashion, considering each frequency separately will increase the number of training examples by a factor of $\freqnum$. On the other hand, \changed{if the neural network takes per-frequency measurements as the input rather than processing all frequencies jointly,} the number of parameters to be learned can be significantly reduced~\cite[Sec.~III-C1]{teganya2020rme}.

%%%%%%%%%%%%%%%%%%%%%%%%%%%%%%%%%%%%%%%%%%%%%%%%%%%%
\subsubsection{Estimation in Narrowband Channels}
\label{sssec:nbch}
%%%%%%%%%%%%%%%%%%%%%%%%%%%%%%%%%%%%%%%%%%%%%%%%%%%%
When the width of the band of interest is small or moderate, it makes sense to assume that the channel is not frequency selective~\cite{zhang2019spectrum,shrestha2021deep}. This means that the true PSD map can be written as $\pow(\loc, \freq)=\sum_\txind \ch_\txind(\loc)\txpow_\txind(\freq)$, where $ \ch_\txind(\loc)=\ch(\txloc_\txind,\loc) $ is the channel gain from the $\txind$-th transmitter to location $\loc$. This implies that the measurements essentially provide $\freqnum$ noisy linear combinations of the $\txnum$ functions $ \ch_1(\loc),\ldots,\ch_\txnum(\loc)$. Therefore, when $\freqnum\gg \txnum$, one can effectively exploit the structure in the frequency domain, improving robustness to measurement noise. One of the main benefits of this approach is that no knowledge of the transmit PSD is required, as it can often be estimated using tools such as nonnegative matrix factorization without requiring any prior knowledge~\cite{shrestha2021deep}.

%%%%%%%%%%%%%%%%%%%%%%%%%%%%%%%%%
\begin{figure}[!t]
  \centering
  \includegraphics[width=0.6\textwidth]{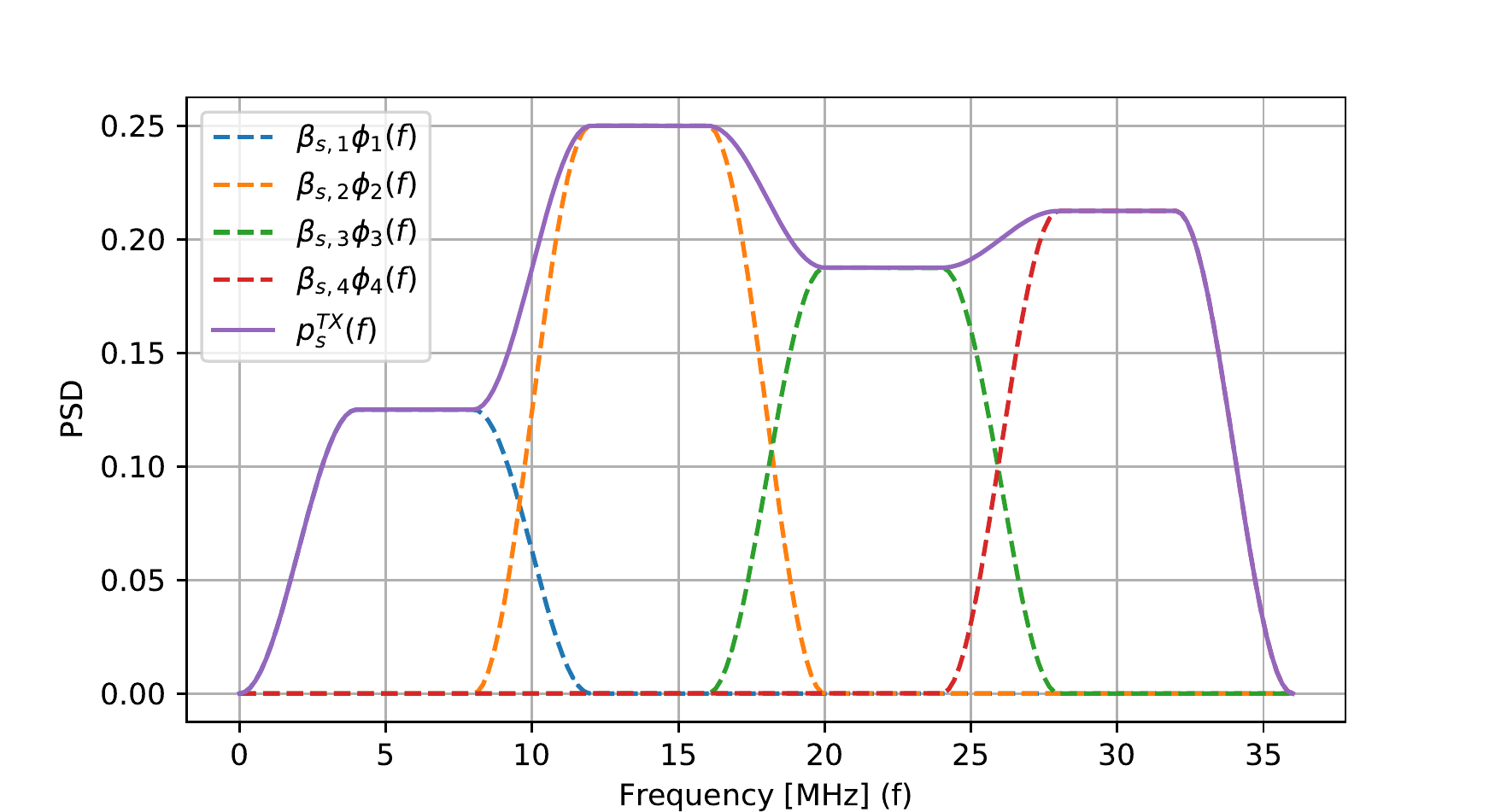}
  \caption{A basis expansion model can be used to decompose a PSD as a linear combination of functions in a basis. In this case, the basis functions are  raised cosine functions, each one corresponding to a transmission in a different band. This makes it possible to exploit prior information about bandwidths, central frequencies, and transmission pulse shapes.
  }
  \label{fig:bem}
\end{figure}
%%%%%%%%%%%%%%%%%%%%%%%%%%%%%%%%%

%%%%%%%%%%%%%%%%%%%%%%%%%%%%%%%%%%%%%%%%%%%%%%%%%%%%
\subsubsection{Estimation in Wideband Channels} 
\label{sssec:wbch}
%%%%%%%%%%%%%%%%%%%%%%%%%%%%%%%%%%%%%%%%%%%%%%%%%%%%
For a wideband channel, one cannot realistically assume that the channel response is flat. In order  to exploit the frequency domain structure, one can utilize  prior knowledge on the transmitter waveforms. %However, frequency-domain structure can still be exploited upon realizing that the transmit PSD within each subchannel is typically known up to a scaling factor, namely the power. 
Specifically, the PSDs of the transmitted waveforms are typically constrained by communication standards and spectrum regulations, which specify the bandwidth, carrier frequencies, transmission masks, roll-off factors, number of subcarriers, and so forth~\cite{romero2013wideband}. Therefore, the transmit PSD of a source can be approximated by a \emph{basis expansion model}~(BEM) as $\txpow_\txind(\freq)=\sum_{\fbasisind} \fcoef_{\txind,\fbasisind} \fbasisfun_\fbasisind(\freq)$, where $\fbasisfun_\fbasisind$ denotes the $\fbasisind$-th basis function and $\fcoef_{\txind,\fbasisind}$ is a nonnegative quantity.  This decomposition is illustrated in Fig.~\ref{fig:bem}.
    
If the signals transmitted by different sources are uncorrelated, the received PSD at a location $\loc$ can be  expressed as $\pow(\loc, \freq)=\sum_\txind \ch(\txloc_\txind,\loc,\freq)\txpow_\txind(\freq)$, where $\ch(\txloc_\txind,\loc,\freq)$ is the channel gain at frequency $\freq$. Then, using the BEM, one arrives at $\pow(\loc, \freq) =\sum_{\fbasisind}\sum_\txind\fcoef_{\txind,\fbasisind} \ch(\txloc_\txind,\loc,\freq) \fbasisfun_\fbasisind(\freq)$. If the bandwidths of the basis functions are small relative to the entire band, it is reasonable to assume that $\ch$ is approximately frequency-flat in the band of each basis function. This yields $\ch(\txloc_\txind,\loc,\freq) \fbasisfun_\fbasisind(\freq) \approx \ch(\txloc_\txind,\loc,\fbasisfreq_\fbasisind)     \fbasisfun_\fbasisind(\freq)$, where  $\fbasisfreq_\fbasisind$ is the central frequency of the $\fbasisind$-th basis function. With this approximation, one can write $\pow(\loc, \freq) =\sum_{\fbasisind}\pow_\fbasisind(\loc) \fbasisfun_\fbasisind(\freq)$, where $\pow_\fbasisind(\loc)= \sum_\txind \ch(\txloc_\txind,\loc,\fbasisfreq_\fbasisind) \fcoef_{\txind,\fbasisind}$ constitutes the power captured by the $\fbasisind$-th basis function at location~$\loc$.

Observe that introducing the BEM has reduced the problem of estimating $\freqnum$ power maps to the problem of estimating the $\fbasisnum\ll \freqnum $ power maps $\pow_1,\ldots,\pow_\fbasisnum$.  Clearly, the smaller $\fbasisnum$, the smaller the sensitivity to measurement noise. The approaches in the preceding subsections can be seen as the extreme cases of choosing $\fbasisnum=\freqnum$ and $\fbasisnum=1$, respectively.
To estimate a PSD map, the aforementioned technique can be used in combination with virtually any of the approaches for power map estimation discussed  earlier~\cite{bazerque2010sparsity,dallanese2012gslasso,bazerque2011splines,romero2017spectrummaps}. A recent example is \cite{teganya2020rme}, where a BEM is used in the last layer of a DNN for RME.
  
%   \blt[strengths]\acom{omit?}
%   which allows the estimation of the power of each
%   sub-channel and background noise as a byproduct. The resulting
%   estimates can be employed to construct signal-to-noise ratio
%   (SNR) maps, which reveal weak coverage areas, and alleviate
%   the well-known noise uncertainty problem in cognitive
%   radio\ncite{tandra2008snrwalls}.

\changed{
The main limitation of estimators that rely on a BEM is a manifestation of the well-known bias-variance trade-off. In particular, if the number of basis functions is small, the  approximation 
$\ch(\txloc_\txind,\loc,\freq) \fbasisfun_\fbasisind(\freq) \approx \ch(\txloc_\txind,\loc,\fbasisfreq_\fbasisind)     \fbasisfun_\fbasisind(\freq)$ may not hold, which will generally result in estimation bias. On the other hand, if the number of basis functions is large, the representation capacity of the BEM is large, which results in a small bias, but a larger variance must be expected as the result of the increase in the number of scalar maps to be learned for a fixed number of samples.
}

%%%%%%%%%%%%%%%%%%%%%%%%%%%%%%%%%%%%%%%%%%%%%
\section{Estimation of Propagation Maps}
\label{sec:propmapest}
%%%%%%%%%%%%%%%%%%%%%%%%%%%%%%%%%%%%%%%%%%%%%
% The goal here is to describe the main principles and the state of the art for estimating propagation maps. Both non-tomographic and tomographic approaches will be presented. To shed light on the challenges and principles, the radio tomography model will be described and illustrated graphically~\cite{patwari2008correlated,romero2018blind}. %\acom{Here, you might want to have separate subsections on channel gain maps and CSI maps, just like in Sec. 4.}

Propagation maps quantify  channel effects, such as  channel gains, for links between arbitrary pairs of locations where no sensors may have been deployed. The $\measind$-th measurement is collected by a pair of sensors, one at location $\loc_\measind$ and the other at $\locp_\measind$. The channel gain of the link between them can be measured possibly by employing pilot signals. The resulting measurement can be expressed as $\meas{\measind} = \ch(\loc_\measind, \locp_\measind)+\noise{\measind}$, where $\ch$ is the true map and $\noise{\measind}$ represents measurement noise. The RME problem is to obtain an estimate $\chest$ of $\ch$ given $\{(\loc_\measind,\locp_\measind,\meas{\measind})\}_{\measind=1}^{\measnum}$. A good RME algorithm should have good generalization properties, meaning that $\chest(\loc,\locp)\approx \ch(\loc,\locp)$ for all location pairs $(\loc,\locp)$, even those for which no measurements have been collected.

Like \signalstrength RME, propagation RME is a function estimation problem. Therefore, the techniques described in Sec.~\ref{sec:eststrength} can again be employed in principle. The key difference is that now the function to be estimated depends on two locations rather than one. If $\loc$ denotes a location in 3D space, it is clear that $\ch$ is a function of a $6$-dimensional input, namely the entries of $\loc$ and $\locp$. This means that the number of measurements necessary to attain a given accuracy may be considerably greater than for estimating a \signalstrength map --- a manifestation of the  curse of dimensionality. Thus, as explored next, a number of algorithms have been tailor-made for propagation RME to alleviate such a difficulty.

% \blt[more challenging] information about the source locations and tx. power is typically unreliable and costly to maintain. Measurements generally require orthogonality.

%%%%%%%%%%%%%%%%%%%%%%%%%%%%%%%%%%%%%%%%%%%%
\subsection{Non-tomographic Approaches}
\label{sec:nontomo}
%%%%%%%%%%%%%%%%%%%%%%%%%%%%%%%%%%%%%%%%%%%%

\def \bbB {{\boldsymbol{B}}}
\def \bbS {{\boldsymbol{S}}}

\def \bbs {{\boldsymbol{s}}}
\def \bbu {{\boldsymbol{u}}}
\def \bbx {{\boldsymbol{x}}}

\def \bbPsi {{\boldsymbol{\Psi}}}

\def \bbalpha {{\boldsymbol{\alpha}}}
\def \bbbeta {{\boldsymbol{\beta}}}
\def \bbpsi {{\boldsymbol{\psi}}}
\def \bbeta {{\boldsymbol{\eta}}}
\def \bbnu {{\boldsymbol{\nu}}}
\def \bbepsilon {{\boldsymbol{\epsilon}}}

\def \ccalN {{\mathcal{N}}}

\def \cov {{\mathrm{cov}}}

In the non-tomographic approaches, channel gains are directly modeled based on basic wireless propagation models without introducing any underlying auxiliary map. In order to maintain tractability, however, the RME problem is often simplified by fixing one end of a link. For example, one may consider estimating the maps $\{\ch_n(\bbx) := h(\bbx,\bbx_n)\}_n$ for fixed positions $\{\bbx_n\}_{n=1}^{N}$ where the  sensors are located. The individual functions $\{\ch_n(\bbx)\}_n$ can be estimated using  methods employed for  signal strength maps. \changed{Since  (static) signal strength map estimation techniques have been explained in the preceding sections, here we extend the RME problem to include the time domain to capture the temporal variation of channel gains. Needless to say, static channel gain maps can also be constructed in a non-tomograhpic fashion.}

Consider the channel gain $\ch_n(\bbx,t)$ between locations $\bbx$ and $\bbx_n$ at time $t$~\cite{kim2011kriged}. Suppose that the effect of small-scale fading  has been averaged out, allowing $\ch_n(\bbx,t)$ to be expressed in dB as
%\begin{align}
$\ch_{n,\dB}(\bbx,t) = %g_0 - 10 \gamma \log_{10}(\|\bbx - \bbx_n\|_2) 
\pathloss_n(\bbx) - \shad_n(\bbx,t)$, %\label{eq:gr}
%\end{align}
where $\pathloss_n(\bbx)$ is the known path loss from $\bbx_n$ to $\bbx$ and $\shad_n(\bbx,t)$ is the shadow fading  between $\bbx$ and $\bbx_n$ at time $t$. Note that $\pathloss_n(\bbx)$ can be assumed known whenever both $\bbx_n$ and $\bbx$ as well as the antenna gains are known. 
Thus, the problem becomes tracking the time-varying shadow fading map $\shad_n(\bbx,t)$. 

To do this, shadow fading measurements are needed, which can be obtained by subtracting the transmit-power and the path loss from the received power measurement. %; cf.~\eqref{eq:gr}. 
By letting $\ccalN_{-n} := \{1,\ldots,n-1,n+1,\ldots,N\}$, the noisy measurements $\{ \shadmeas_n(\bbx_j,t)\}_j$ of shadow fading obtained at time~$t$  by the sensor at $\bbx_n$ using the pilot signals sent from the radios at $\{\bbx_j\}_{j \in \ccalN_{-n}}$ can be expressed as
\begin{align}
\shadmeas_n(\bbx_j,t) = \shad_n(\bbx_j,t) + \noise{n}(\bbx_j,t), \ j \in \ccalN_{-n}, \label{eq:meas}
\end{align}
where $\noise{n}(\bbx_j,t)$ is zero-mean Gaussian measurement noise. Upon defining $\shadmeasvec_n(t) := [\shadmeas_n(\bbx_1,t), \ldots,$ $\shadmeas_n(\bbx_{n-1},t),\shadmeas_n(\bbx_{n+1},t),\cdots,\shadmeas_n(\bbx_N,t)]^\top$, the problem is to estimate $\ch_{n,\dB}(\bbx,t)$ for arbitrary $\bbx$ based on the  measurements $\shadmeasset_n(t) := \{\shadmeasvec_n(\tau)\}_{\tau=1}^t$ up to time $t$.

This problem can be tackled in the framework of kriged Kalman filtering, also known as space-time Kalman filtering~\cite{MGR98}. Employing the log-normal shadowing model, it is assumed that $\shad_n(\bbx,t)$ is a Gaussian process with spatio-temporal dynamics~\cite{MGR98,kim2011kriged}
\begin{align}
\shad_n(\bbx,t) &= \shadmean_n(\bbx,t) + \nu_n(\bbx,t) \label{eq:ss1}\\
\shadmean_n(\bbx,t) &= \int w_n(\bbx,\bbu) \shadmean_n (\bbu, t-1) d \bbu + \eta_n(\bbx,t), \label{eq:ss2}
\end{align} 
where $\shadmean_n(\bbx,t)$ is the spatio-temporally correlated component,  $w_n(\bbx,\bbu)$ captures the interaction of this component at location $\bbx$ at time $t$ and at location $\bbu$ at time $(t-1)$, and $\nu_n(\bbx,t)$ and $\eta_n(\bbx,t)$ are spatially correlated but temporally white zero-mean Gaussian processes. %with covariances $\cov \{\nu_r(\bbx,t),\nu_r(\bbu,\tau)\} = C_{\nu_r}(\bbx - \bbu)\delta(t-\tau)$ and $\cov \{\eta_r(\bbx,t),\eta_r(\bbu,\tau)\} = C_{\eta_r}(\bbx - \bbu)\delta(t-\tau)$, where $\delta(t) = 0$ if $t \neq 0$ and $\delta(0) = 1$. 
Process $\nu_n(\bbx,t)$ is uncorrelated with $\noise{n}(\bbu,\tau)$, and $\eta_n(\bbx,t)$ is uncorrelated with $\nu_n(\bbu,\tau)$ and $\noise{n}(\bbu,\tau)$ for all $\bbu$ and $\tau$. Moreover, $\mathbb{E}\{\nu_n(\bbx,t) \shadmean_n(\bbu,t)\} = \mathbb{E}\{\eta_n(\bbx,t) \shadmean_n(\bbu,t-1)\}=0$ for all $\bbx,\bbu$ and $t$. %For stability, it is also assumed that $|\int w_n(\bbx,\bbu) d\bbu| < 1$.

Since the state-space model in~\eqref{eq:ss1}--\eqref{eq:ss2} is infinite-dimensional, adopt a BEM for tractability, as in universal kriging. For a set of $K$ orthonormal basis functions $\{\psi_k(\bbx)\}_k$, $\shadmean_n$ and $w_n$ are respectively approximated as $\shadmean_n(\bbx,t) \approx \sum_{k=1}^K \alpha_{n,k}(t) \psi_k(\bbx)$ and $w_n(\bbx,t) \approx \sum_{k=1}^K \beta_{n,k}(t) \psi_k(\bbx)$ with expansion coefficients $\{\alpha_{n,k}(t)\}$ and $\{\beta_{n,k}(t)\}$. Substituting these expansions into~\eqref{eq:meas}--\eqref{eq:ss2} and evaluating the resulting equations at $\{\bbx_j\}_{j \in \ccalN_{-n}}$ yields the finite dimensional state-space model
\begin{align}
\shadmeasvec_n(t) &= \bbPsi_n \bbalpha_n(t) + \bbnu_n(t) + \noisevec_n(t) \label{eq:meas_eq}\\
\bbalpha_n(t) &= \bbPsi_n^\dagger \bbB_n \bbalpha_n (t-1) + \bbPsi_n^\dagger \bbeta_n(t). \label{eq:st_eq}
\end{align}
Here,  $\bbpsi(\bbx) := [\psi_1(\bbx),\ldots,\psi_K(\bbx)]^\top$, $\bbalpha_n(t) := [\alpha_{n,1}(t),\ldots,\alpha_{n,K}(t)]^\top$, and $\bbbeta_n(\bbx)$ is defined likewise. %:= [\beta_{r,1}(t),\ldots,\beta_{r,K}(t)]^\top$. 
 Vectors $\bbnu_n(t)$, $\noisevec_n(t)$, and $\bbeta_n(t)$ are  constructed in a similar fashion from $\{\nu_n(\bbx_j,t)\}_{j \in \ccalN_{-n}}$,  $\{\noise{n}(\bbx_j,t)\}_{j \in \ccalN_{-n}}$, and $\{\eta_n(\bbx_j,t)\}_{j \in \ccalN_{-n}}$, respectively. $\bbB_n$ and $\bbPsi_n$ are matrices constructed by respectively arranging $\bbbeta_n(\bbx_j)^\top$ and $\bbpsi(\bbx_j)^\top$ as rows for $j \in \ccalN_{-n}$.

Based on~\eqref{eq:meas_eq}--\eqref{eq:st_eq}, the MMSE estimate $\hat \bbalpha_n(t|t)$ of $\bbalpha_n(t)$ given $\shadmeasset_n(t)$ can be obtained via ordinary Kalman filtering, from which the temporally dynamic component $\shadmean_n(\bbx,t)$ can be estimated as $\mathbb{E}\{\shadmean_n(\bbx,t)|\shadmeasset_n(t)\} = \bbpsi(\bbx)^\top \hat \bbalpha_n(t|t)$. To capture $\nu_n(\bbx,t)$ as well, a kriging estimator is employed; cf. Sec.~\ref{sec:kriging}. Overall, the MMSE estimate $\shadest_n(\bbx,t) := \mathbb{E}\{\shad_n(\bbx,t) | \shadmeasset_n(t)\}$ can be obtained exploiting the covariance structure~\cite{kim2011kriged}. Once $\shadest_n(\bbx,t)$ is obtained, the channel gain map estimate $\chest_{n,\dB}(\bbx,t)$ can be constructed as $\chest_{n,\dB}(\bbx,t) = \pathloss_n(\bbx) - \shadest_n(\bbx,t)$.%; cf.~\eqref{eq:gr}. 

%%%%%%%%%%%%%%%%%%%%%%%%%%%%%%%%%%%%%%%%%%%
\subsection{Tomographic Approaches}
\label{sec:radiotomographicapproaches}
%%%%%%%%%%%%%%%%%%%%%%%%%%%%%%%%%%%%%%%%%%%

% \blt[overview]Rather than mapping
% power, other families of methods construct channel gain maps using
% \begin{bullets}%
%   \blt[RKHS]\acom{remove?}non-parametric regression in reproducing kernel
%   Hilbert spaces (RKHSs)~\cite{romero2016blindchannelgain},
% \end{bullets}

\begin{figure}[!t]
  \centering
  \includefig{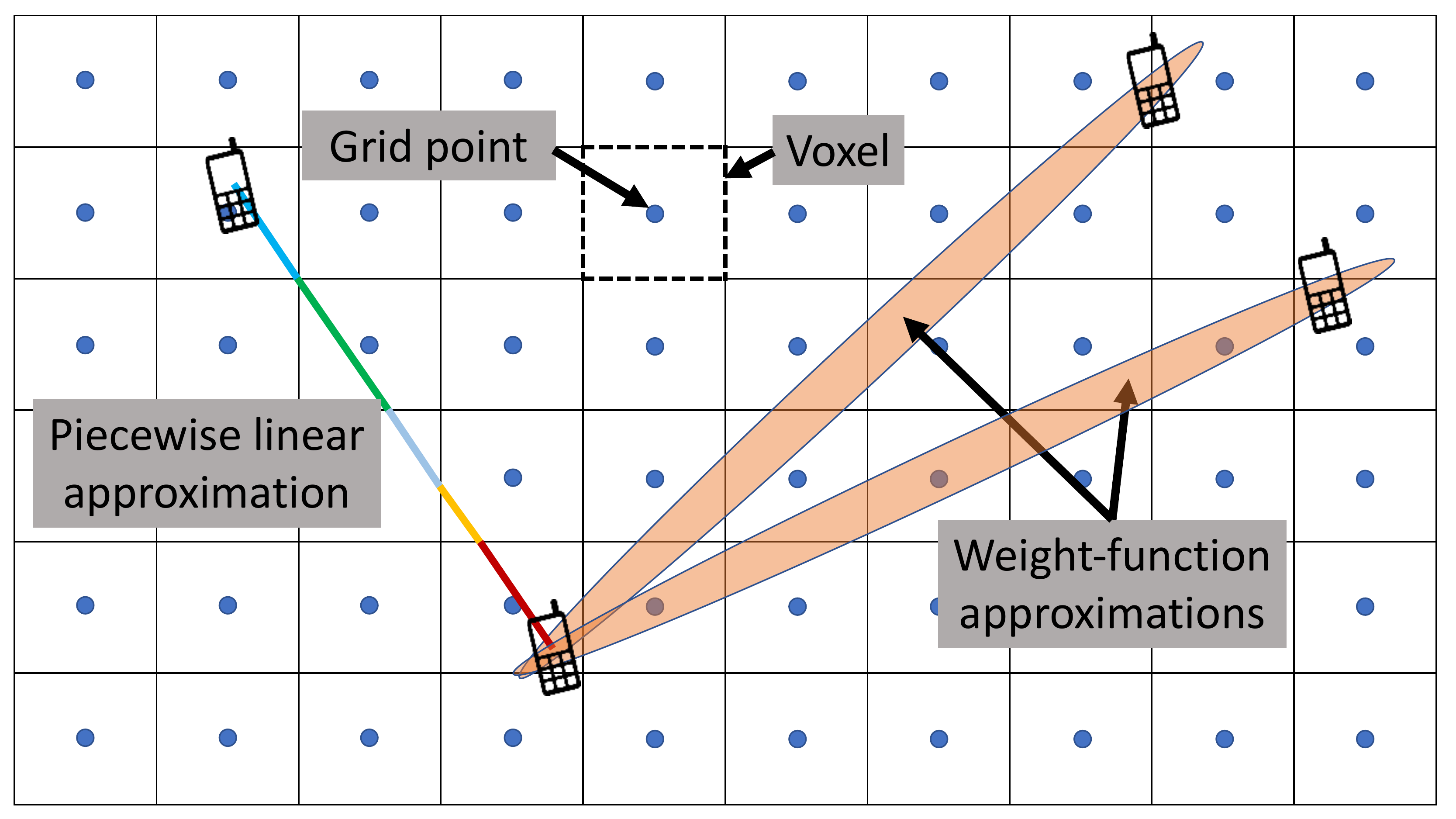}
  \caption{Illustration of possible approximations of the tomographic
    integral.  }
  \label{fig:tomography}
\end{figure}

\begin{bullets}
  \blt[Overview]%Most works on propagation RME rely on the so-called
  %\emph{radio tomographic} model~\cite{dallanese2011kriging,lee2016lowrank}.
%
  \blt[Model]
  \begin{bullets}    
    \blt[overview]%This model, whose name already suggests its
    %similarity with the models used for tomographic imaging in other
    %areas such as medical imaging, 
The radio tomographic model can be used to estimate shadow fading maps. It postulates that the attenuation due to shadowing can be expressed in terms of an underlying auxiliary map termed \emph{spatial loss field} (SLF)~\cite{dallanese2011kriging,lee2016lowrank}. The SLF characterizes how much  radio waves  attenuate when passing through each location and, hence, is specific to each propagation environment~\cite{patwari2008nesh}\ncite{agrawal2009correlated}.
    \blt[decomposition]%Specifically, decompose the channel gain (expressed in logarithmic units) as $\ch = \pathloss - \shad - \fad$, where $\pathloss$ is the path loss, $\shad$ is the so-called \emph{shadow fading} attenuation, caused by obstructions in the propagation path of the radio wave, and $\fad$ is the \emph{fast fading} attenuation, caused by the constructive and destructive     interference among multiple paths through which the radio signal travels.
    %\begin{bullets}
 %\blt[fading]The scale of spatial variations of the last term is in the order of the wavelength, which for contemporary communication systems ranges in the order of a few millimeters or centimeters. Since this is much smaller than the typical error of localization systems such as GPS used to determine the measurement locations, estimating $\fad$ may be challenging in practice. Thus, it can be regarded as a perturbation and absorbed into the measurement noise. 
      %
 %\blt[pathloss]On the other hand, the term $\pathloss$ can be considered known since it is determined by the endpoints of the communication link as well as by the antenna gains and       additional known factors. %\blt[shad]These observations mean that estimating $\ch$ amounts to estimating $\shad$. 
 Specifically, the radio tomographic model prescribes that the shadowing attenuation between locations $\loc$ and $\loc'$ is given by the line integral~\cite{patwari2008nesh} 
\begin{align}
\label{eq:tomoint}
\shad(\loc,\locp)=\frac{1}{\sqrt{\| \loc-\locp\|}} \int_\loc^{\locp} \slf(\locvar)d\locvar,
\end{align}
where $\slf:\region \rightarrow \rfield_+$ is the SLF.
This naturally captures the notion that nearby radio links generally experience similar shadowing due to the presence of common obstacles. \changed{Since the above integral provides the shadowing attenuation between two arbitrary  locations, one does not need to fix one end of a link, as in the non-tomographic approach explained earlier. Remarkably, time-varying maps can be readily accommodated in the tomograhpic approach~\cite{dallanese2011kriging}.} Furthermore, as the SLF can reveal the locations of obstacles, the SLF itself can be useful for various applications such as device-free passive localization~\ncite{woyach2005sensorless,patwari2008correlated}\cite{romero2018blind}, surveillance monitoring for intrusion detection~\cite{hamilton2014modeling}, and through-the-wall imaging for emergency or military operations~\cite{romero2018blind}\ncite{wilson2011throughwalls}\ncite{baranoski2008historical}.
  \end{bullets}

  \blt[Approximation]
  \begin{bullets}
    \blt[discretization]To approximate the line integral in
    \eqref{eq:tomoint}, $\slf$
    can be discretized on a regular grid of 2D or 3D spatial locations.
    \blt[line integral]%
    \begin{bullets}%
      \blt[weight fun]%
      \begin{bullets}%
        \blt[descr]One common approach is to  approximate this integral as a weighted sum of the SLF values on the
        grid points that lie inside an ellipse or an ellipsoid with foci at $\loc$ and $\locp$, as shown in Fig.~\ref{fig:tomography}.
        \blt[intuition] The intuition is
        that the attenuation between the two end points should be heavily
        affected by the obstacles around the line of sight or, more
        specifically, within the so-called \emph{Fresnel zone}, which
        is an ellipse whose geometry is dictated by the wavelength.
        \blt[weights] Several functions have been proposed  in the
        literature~\cite{hamilton2014modeling} to generate such weights, mainly based on heuristics.
        Alternatively, the weights can be learned from the data through blind schemes~\cite{romero2016channelgain}\ncite{gutierrezestevez2021hybrid}.

        \blt[limitations]While easy to
        implement, this  approximation yields  shadowing maps with discontinuities, as small changes in $\loc$ and $\locp$ may lead to a change in the set of grid points that lie in the ellipse. Even more, if the ellipse misses all the grid points, as shown by the left ellipse in Fig.~\ref{fig:tomography}, the approximation becomes $0$. Thus, to attain a good accuracy, the grid must be dense enough. 

        \begin{bullets}
          %\blt[continuous] First, the approximated $\shad(\loc,\locp)$
          %is not a continuous function of $\loc$ and $\locp$ since a
          %small change in either of these vectors may lead to a change
          %in the set of grid points that lie within the ellipse.  In
          %fact, it may well happen that the approximation is just 0,
          %as it occurs when the ellipse misses all grid points; see
          %the left ellipse on Fig.~\ref{fig:tomography}.
          %
          %\blt[resolution]To attain a reasonable accuracy, this
          %implies that the separation between grid points must be
          %small enough, which in turn may result in a prohibitively
          %large number of grid points. 
        \end{bullets}
      \end{bullets}

      \blt[piecewise constant]
      \begin{bullets}
        \blt[descr] This motivates an alternative approach, where
        the SLF is approximated as a piecewise constant function, taking a constant
        value within each grid cell~\ncite{kanso2009compressed}\cite{romero2022aerial}. The
        integral can then be computed as the weighted sum of the SLF values in the cells that the line of sight traverses. The weight simply corresponds to the
        distance traversed in each cell. This is illustrated by the
        colored line in Fig.~\ref{fig:tomography}.
        \blt[benefits]This approximation involves
        less computational burden than the one based on the ellipse,
        is continuous in $\loc$ and $\locp$, and does not vanish unless the SLF vanishes. Thus, the need for
        a dense grid is relaxed --- particularly attractive in 3D~\cite{romero2022aerial}.
       
      \end{bullets}
    \end{bullets}
  \end{bullets}

  \blt[Estimation]In either approach, the shadowing attenuation is a linear function of the SLF values at the
  grid points. Thus, the SLF can be estimated via (nonnegative) LS.
  \begin{bullets}
    \blt[LS]%This means that the measurements are affine functions of
    %these values, which means that the latter can be estimated, in
    %principle, via (possibly non-negative) LS. 
    However, this
    requires that the number of measurements is significantly
    larger than the number of grid points. %, which therefore poses a
    %practical constraint.
    %
    \blt[regularization]One can mitigate this through  appropriate regularizers~\cite{wilson2009regularization} or by specifying a prior distribution in a Bayesian framework~\cite{lee2018adaptive}. \changed{Another limitation of  tomographic approaches is that only the attenuation due to absorption (shadowing) is accounted for. Other propagation effects such as reflection, refraction, and diffraction are completely ignored.}
    %
    %\blt[bayesian]Yet another possibility is to pursue a Bayesian
    %approach, where the need for data is counteracted by the
    %introduction of prior distributions; see
    %e.g.~\cite{lee2018adaptive}.
    % \begin{bullets}
    %   \blt[Low rank and sparsity]low rank and
    %   sparsity~\cite{lee2016lowrank}

    %   \blt[Bayesian]or hidden
    %   Markov random fields~\cite{lee2018adaptive}.
    % \end{bullets}
    
  \end{bullets}

  \blt[limitations]
  
\end{bullets}

%%%%%%%%%%%%%%%%%%%%%%%%%%%%%%%%%
\section{Spectrum Surveying} 
\label{sec:ssurv}
%%%%%%%%%%%%%%%%%%%%%%%%%%%%%%%%%
In order to collect the measurements needed to build radio maps, traditionally technicians in a vehicle with measurement equipment would drive around the site. With the advances in mobile robotics of the last decade, it is now possible to employ an autonomous UAV with an on-board sensor to collect the desired measurements. This is clearly more efficient in terms of time and personnel cost.

An important task is to plan the path traversed by the autonomous UAV for acquiring measurements. A common approach is to define a grid and take measurements at each grid point. However, visiting each grid point can be very time-consuming and puts a strain on the limited battery capacity, especially when the grid is dense. A more efficient approach is to collect measurements at a small set of highly informative locations and apply the interpolation techniques discussed in previous sections to construct the entire map. To this end, besides the map estimate, RME algorithms need to provide an uncertainty map that indicates how informative a measurement would be at each location given the measurements collected so far~\cite{shrestha2022surveying}.
%To this end, the map estimators can not only provide the estimated map but also an uncertainty map that indicates how informative a measurement would be at each location given the measurements collected so far~\cite{shrestha2022surveying}.
Based on the uncertainty map, a route planning algorithm can produce a trajectory through areas of high uncertainty. This approach achieves a much higher estimation quality in the given surveying time (or requires much shorter time for a given quality) compared to the naive grid-based approach.

\begin{figure}[!t]
  \centering
  \includegraphics[width=1.\textwidth]{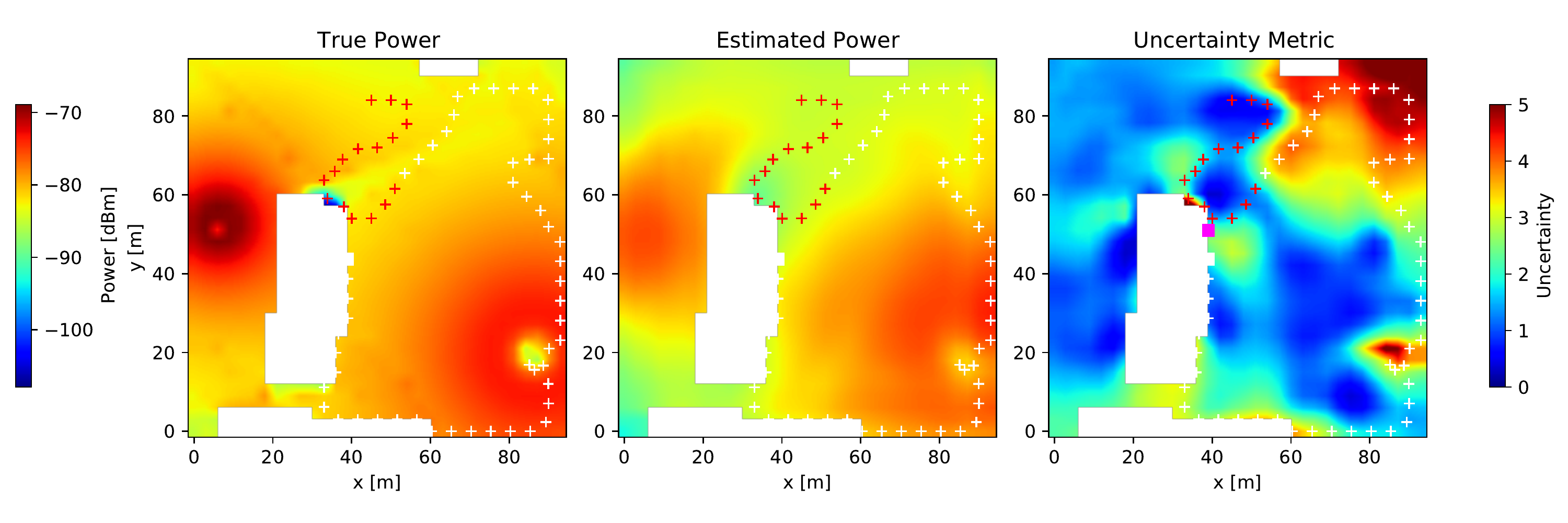}
  \caption{Example of a surveying operation with an autonomous UAV in an urban environment seen from above. White boxes denote buildings. Red and white crosses denote measurement locations.
  }
  \label{fig:surveying}
\end{figure}

Fig.~\ref{fig:surveying} illustrates an example of a surveying operation using a ray-tracing data set in a region of downtown Rosslyn, Virginia. The three panels show the UAV trajectory seen from above. White boxes correspond to space occupied by buildings, where no measurements can be taken. Red and white crosses denote measurement locations. The leftmost panel shows the ground truth power map in a setup with two transmitters. The middle and right panels respectively show the estimated power map and the uncertainty map when only the measurements marked by red crosses have been collected. At that point in time, the UAV plans a trajectory through areas of high uncertainty, represented by white crosses. The estimator in this case is a global DNN estimator capable of learning the nature of propagation phenomena from a data set; see Sec.~\ref{sec:deeplearning}.

%%%%%%%%%%%%%%%%%%%%%%%%%%%%%%
%%%%%%%%%%%%%%%%%%%%%%%%%%%%%%

%%%%%%%%%%%%%%%%%%%%%%%%%%%%%%%%%%%
\section{Practical Considerations}
%%%%%%%%%%%%%%%%%%%%%%%%%%%%%%%%%%%
In this section, some
challenges that arise when implementing  RME techniques in practical
setups are discussed. These include coping with  localization errors, non-isotropic antenna patterns, decentralized implementation, and reducing the bandwidth required to collect measurements. %Specifically, we will discuss the impact of the localization error and how to counter it via \emph{location-free cartography}, which comprises techniques that essentially construct radio maps that are not indexed by spatial coordinates but by \emph{channel-state information} (CSI)~~\cite{teganya2019locationfree,jiang2019structures}\ncite{chen2019remote}. Afterwards, we will explore decentralized implementations~\cite{bazerque2010sparsity,dallanese2011kriging} and how to reduce the bandwidth required by the sensors to report measurements~\cite{romero2017spectrummaps}.
%Finally, the issues
% regarding calibration will be
% outlined~\cite{patwari2008correlated,wilson2009regularization,wilson2010tomography}.

%%%%%%%%%%%%%%%%%%%%%%%%%%%%%%%%%%%
 \subsection{Localization Errors}
 \label{sec:localizationerrors}
%%%%%%%%%%%%%%%%%%%%%%%%%%%%%%%%%%%
\begin{bullets}
  
  \blt[Limitations of existing schemes]
  \begin{bullets}

    \blt[Need Locations]The RME schemes described earlier typically require accurate knowledge of the measurement locations.
    \blt[Location must be estimated]In practice, the sensor locations are themselves estimated based on localization systems such as GPS in the following way. 
    \blt[how loc works] A number of transmitters with known locations, such as 
    satellites or cellular base stations, regularly transmit signals
    termed \emph{localization pilots}. Each sensor then extracts
    certain \emph{features} from the pilots to estimate its
    location. For example, the received signal strength or
    the propagation delay, which contain information on the distance  to the
    transmitters, are used to produce location estimates based on geometric principles.
Thus, the quality of the estimates can be significantly degraded due to multipath propagation as in indoor and dense urban scenarios, where localization errors may reach tens of meters. 

% A sensor receives these signals
%     and determines their power, their time (difference) of arrival
%     (T(D)oA), or their direction of arrival (DoA). Using a geometric
%     model, an algorithm combines these values to produce a location
%     estimate. 
%  positioning pilot signals transmitted by satellites
%     (e.g. in GPS) or terrestrial base stations (e.g. in LTE or
%     WiFi~\cite{bshara2010fingerprinting})~\cite{naidu2017distributed,
%       bensky2016wireless}.
    \blt[Estimates not accurate in multipath env.]
    \begin{bullets}
      \blt[descr] %Unfortunately, this task may be challenged in
      %practice due to propagation phenomena affecting the pilot
      %signals. Specifically, multipath propagation, especially
      %important in indoor and dense urban scenarios, distorts the
      %pilot features and, consequently, the distance estimates
      %obtained by the sensors. This may result in localization errors
      %in the order of tens of meters.

      \begin{figure}[bhtp]
        \centering
        \subfloat[Free space.]{
          \label{fig:lf:fs}
          \includegraphics[width=.45\textwidth]{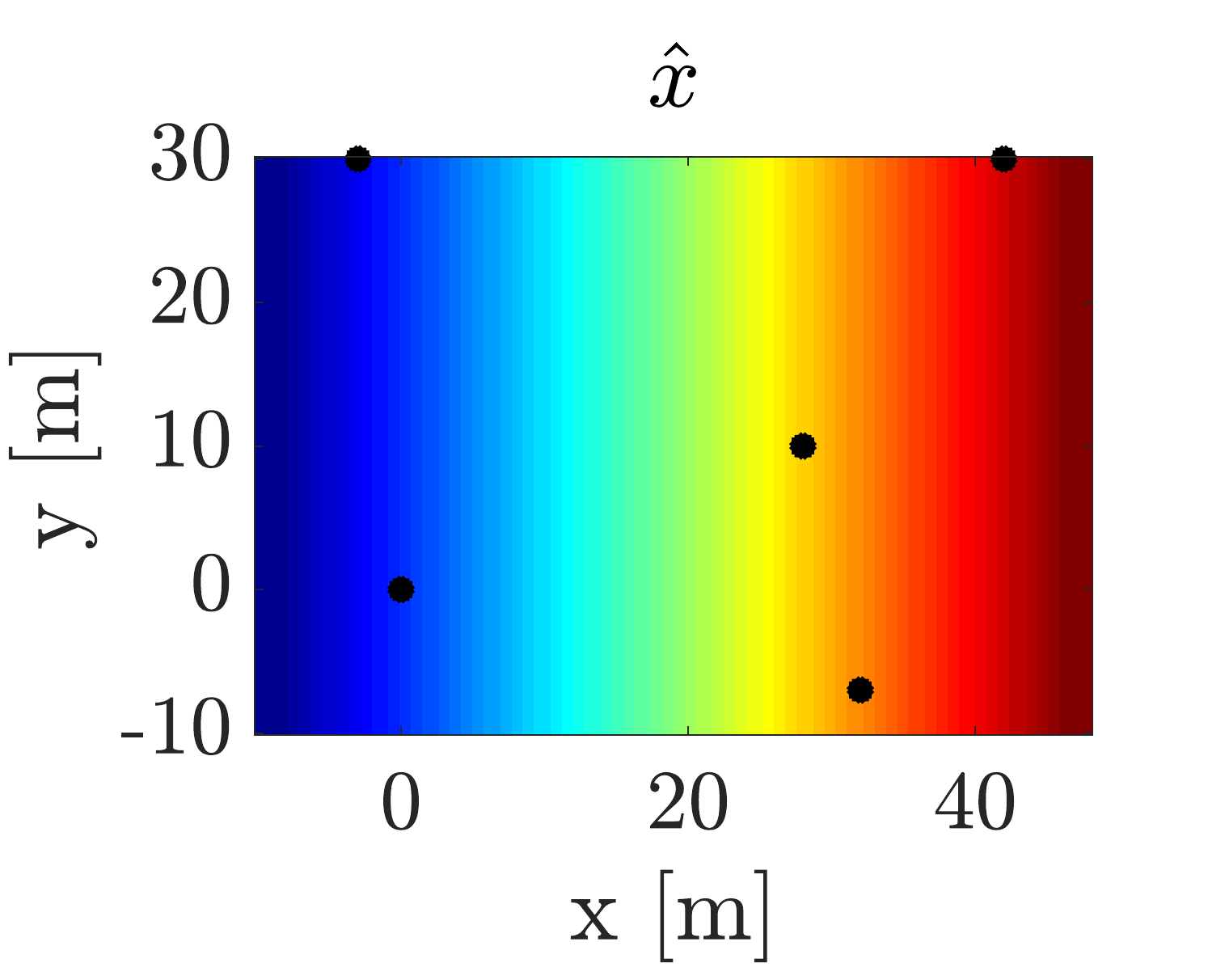}
        } %\\ -> to arrange vertically
        \subfloat[Scenario with 4 walls. ]{
          \label{fig:lf:mp}
          \includegraphics[width=.45\textwidth]{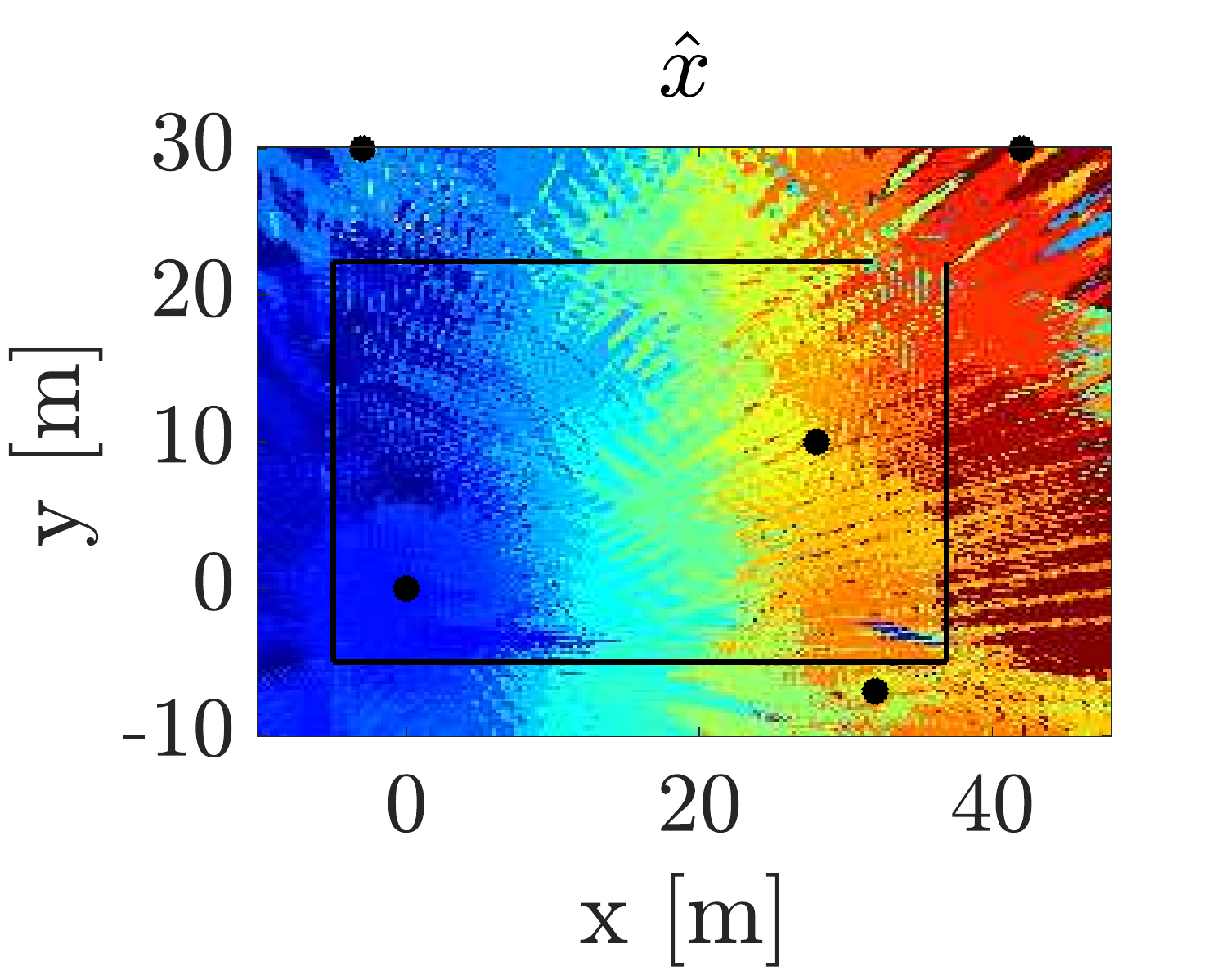}
        } % etc
        \caption{The color of each point indicates the x-coordinate of the
          location estimate obtained by a sensor at that location. The black
          circles indicate the positions of the transmitters.  The estimate
          accurately matches the true coordinate when there is no multipath;
          cf. Fig.~\ref{fig:lf:fs}. Thus, Fig.~\ref{fig:lf:fs} serves as a
          color bar. On the other hand, the estimation error is large in the
          presence of multipath; cf. Fig.~\ref{fig:lf:mp}. (Both figures were
          taken from~\cite{teganya2019locationfree}.)  }
        \label{fig:lf}
      \end{figure}

      \blt[Fig.]This phenomenon is illustrated in Fig.~\ref{fig:lf},
      where the x-coordinates of the location estimates are compared in a
      scenario without multipath (Fig.~\ref{fig:lf:fs}) and with
      multipath (Fig.~\ref{fig:lf:mp}). The localization algorithm is
      based on the time difference of arrival between the pilot
      signals arriving from each pair of transmitters; see \cite{teganya2019locationfree} for details. The poor quality of the location estimates in Fig.~\ref{fig:lf:mp} hinders the
      application of conventional RME techniques. This is because the maps are indexed by the locations (e.g., the input for the power map $p$ is $\loc$), and thus the localization error in $\loc$ propagates to the output $p(\loc)$. 
      
    \end{bullets}

  \end{bullets}

  \blt[Counteracting measures]%The localization error therefore limits the resolution achievable by RME. With typical errors in the order of a few meters, it is impossible to estimate fine details such as those created by fast fading.
  \begin{bullets}
    \blt[estimate only shadowing]%The RME can only capture phenomena with spatial variations in a scale larger than the localization error, such as shadow fading and path loss. In any case, it is worth noting that a relatively low localization error may not entail a loss of resolution since the latter can already be limited by the spatial density of the measurements. For example, if the separation between measurements is in the order of tens of meters, as may occur in practice, details in the order of meters cannot be reconstructed anyway regardless of how small the localization error is. For this reason, accurately mapping the fast-fading component in practice may be challenging: its coherence distance is much smaller than the distance between measurements.
    % \begin{bullets}
    %   \blt[Wavelength and position error]
    %   \blt[spatial averaging]
    % \end{bullets}

    \blt[Location-free] The key realization is, therefore, that $\loc$
      is not suitable as the ``index'' of the map. To mitigate this issue, one can resort to the
    so-called \emph{location-free (LocF) cartography}
    framework~\cite{teganya2019locationfree}.
    \begin{bullets}
      \blt[idea: bypass localization \ra paradigm shift]%Let us consider the power maps for concreteness. Recall
      %that a power map $\pow$ returns the power $\pow(\loc)$ received
      %at a location $\loc$. Since the location is the input variable,
      %it is expected that localization errors propagate to the output
      %of this function. 
%
      \blt[approaches]
      \begin{bullets}
        \blt[indexing via pilot features]
        \begin{bullets}
          \blt[how]To motivate this framework, it is
          worth stepping back and recalling that the location estimates
          are produced by a localization algorithm based on the
          pilot features. It is sensible, therefore, to bypass this
          step and directly use the pilot features to index the map, since
          these features evolve more smoothly across space than the
          location estimates.
          
          \blt[evaluation]Once such a map has been estimated, there are two approaches to evaluate it at a given location. If a terminal is present at that
          location, it can directly employ the features of the
          pilot signals. If no sensor is present, one
          can interpolate the features, e.g., based on the low rank prior.

        {\changedr
          \blt[limitations]Due to the larger
          input dimension of the map function,  location-free RME  requires a larger number of measurements than  location-based approaches in the absence of localization errors. Another difficulty is that the availability of the 
          features depends on the availability of the pilot
          signals. However, one can  reconstruct the missing features~\cite{teganya2019locationfree} or define features that can be extracted from regular
      communication signals, e.g. the ones broadcast by cellular base stations, rather than from dedicated localization pilots~\cite{jiang2019structures}.
      }
          % 
          % Remarkably, as a byproduct of
          % skipping the localization step,
          % \begin{bullets}%
          %   \blt[lower compl]the resulting cartography algorithm is
          %   typically computationally less expensive than its LocB
          %   counterparts \blt[cheap]and does not require
          %   additional localization infrastructure or the costly creation of
          %   fingerprinting datasets.
          % \end{bullets}%
          % \blt[Description]%
          % \begin{bullets}%
          %   \blt[Feature design]The key step is a design of pilot
          %   signal features tailored to multipath
          %   environments.\acom{explain} \blt[Accommodate missing
          %   features]special technique to accommodate scenarios where a
          %   sensor can only extract a subset of those features due to low
          %   signal-to-noise ratio (SNR).

          %   \blt[estimators]Although many algorithms can be
          %   devised within this framework,
          %   \begin{bullets}%
          %     \blt[kbl\ra simple]illustrates this
          %     approach using a kernel-based estimator. 
          %     \blt[frequency]
          %     carries over to other metrics such as PSD.
          %   \end{bullets}%

        % \end{bullets}%
        % \blt[caveat]As expected,
        % the proposed scheme outperforms LocB cartography in multipath
        % scenarios, but traditional LocB approaches are still preferable when
        % accurate location estimates are available.

      \end{bullets}

    \end{bullets}

    \end{bullets}

    \blt[Channel charting]
  \end{bullets}

\end{bullets}

\subsection{Antenna Patterns}
\label{sec:antennapatterns}

\def \bbtheta {{\boldsymbol{\theta}}}

{\changedr
So far, we assumed that a signal strength map $\pow(\loc)$ provides the power received by a sensor with an isotropic antenna at location $\loc$. If the   antenna pattern is not isotropic, the measured power will  depend on the sensor orientation. For this reason, it may be convenient to  estimate the \emph{angular spectrum map} $\pow(\loc,\bbtheta)$, which provides the angular power density received by a sensor at location $\loc$ from direction  $\bbtheta$. Here, $\bbtheta$ parameterizes the direction through, e.g. the azimuth and elevation angles. 

If $\pattern(\bbtheta-\bbtheta')$ denotes the antenna gain along direction $\bbtheta$ for a sensor with orientation $\bbtheta'$, it follows that the power received by such a sensor when placed at $\loc$ will be $\int \pattern(\bbtheta-\bbtheta')\pow(\loc,\bbtheta) d \bbtheta$. If the sensor orientations associated with all measurements are known, then each measurement is a noisy linear observation of $\pow(\loc,\bbtheta)$ and, therefore, the latter  can be estimated. The techniques described earlier for  PSD map estimation in the frequency domain can be adapted to this end, possibly upon discretizing the aforementioned integral. Specifically, $\pow(\loc,\bbtheta)$ can be estimated for a discrete set of angle bins separately or by parameterizing  $\pow(\loc,\bbtheta)$ by means of a BEM with standard or tailored basis functions along the lines of \cite[Sec.~III-A]{romero2017spectrummaps}, although the choice of  suitable basis functions seems to warrant further research. 

The challenges emerging in this approach are twofold. First, due to the curse of dimensionality and the fact that function $\pow(\loc,\bbtheta)$ takes the additional input $\bbtheta$, a significantly larger number of measurements may be required to estimate $\pow(\loc,\bbtheta)$ relative to $\pow(\loc)$. Second, sensors need to be able to measure their orientation, e.g. through accelerometers and magnetometers, which affects the cost and introduces additional error sources. 

A pragmatic alternative is to treat the sensor orientations as random variables with uniform distribution over orientations $\bbtheta$. This implies that the isotropic power map $\pow(\loc)$ equals the expectation of  $\pow(\loc,\bbtheta)$ and, thus, one can still estimate $\pow(\loc)$ using the procedures described in previous sections upon disregarding orientation. The uncertainty introduced by the directionality of the antennas translates into additional measurement noise, which therefore increases the number of measurements required to estimate $\pow(\loc)$ with a target accuracy. This is the price to be paid for circumventing the aforementioned limitations. 
}

\subsection{Decentralized Implementation}
%%%%%%%%%%%%%%%%%%%%%%%%%%%%%%%%%%
%\acom{Seung-Jun?}
%
%\begin{bullets}
%  \blt[relation to decentralized spectrum sensing]Different
%  from conventional spectrum sensing techniques, which assume a common
%  spectrum occupancy over the entire sensed
%  region~\cite{quan2008collaborativewideband,ariananda2014cooperative,mehanna2013frugal},
%  spectrum cartography accounts for variability across space. Benefit
%  \ra more aggressive spatial reuse in DSA.
%\end{bullets}

%\input{decentr.tex}

\def \bba {{\boldsymbol{a}}}
\def \bbu {{\boldsymbol{u}}}
\def \bbx {{\boldsymbol{x}}}
\def \bby {{\boldsymbol{y}}}

\def \bbI {{\boldsymbol{I}}}
\def \bbX {{\boldsymbol{X}}}

\def \ccalE {{\mathcal{E}}}
\def \ccalG {{\mathcal{G}}}
\def \ccalN {{\mathcal{N}}}

\def \bbgamma {{\boldsymbol{\gamma}}}
\def \bblambda {{\boldsymbol{\lambda}}}
\def \bbtheta {{\boldsymbol{\theta}}}

Unlike conventional spectrum sensing techniques, which often assume a common spectrum occupancy over the  entire area of interest~\cite{axell2010sensing}, spectrum cartography accounts for spatial variability. Thus, it is necessary that the measurements are obtained at various locations $\{\bbx_n\}_{n=1}^N$ within the region, which then must be processed jointly. While this can be achieved in theory by collecting the measurements at a fusion center (FC) for centralized processing, the feedback overhead and the associated delay can be significant in practice. Moreover, the FC must operate with higher resource and security requirements. An alternative is to employ distributed in-network processing, where all sensors collaboratively estimate the map via local interactions, i.e., the $n$-th sensor, $n \in \ccalN := \{1,\ldots,N\}$, exchanges information only with its set of single-hop neighbors $\ccalN_n\subset\ccalN$~\cite{kim2011link,bazerque2010sparsity,bazerque2011splines,kim2011kriged,dallanese2011kriging}. The key idea is that  RME tasks often boil down to a regression problem of the form
\begin{align}
\minimize_\bbtheta \frac{1}{2} \|\bby - \bbX \bbtheta\|_2^2 + \psi(\bbtheta), \label{eq:centr}
\end{align}
where $\bby \in \mathbb{R}^M$ and $\bbX \in \mathbb{R}^{M \times \Theta}$ represent the targets and the regressors, respectively, $\bbtheta \in \mathbb{R}^\Theta$ contains the regression coefficients, and $\psi(\cdot)$ is a convex regularizer that captures prior information; see, e.g.,~\eqref{eq:optalpha}. It is often the case that the data $\bbX$ and $\bby$ consist of the collection of the data $\{\bbX_n\}$ and $\{\bby_n\}$ from the individual sensors. That is, $\bby = [\bby_1^\top,\ldots,\bby_N^\top]^\top$, where $\bby_n \in \mathbb{R}^{M_n}$ for $n \in \ccalN$  and $\sum_{n=1}^N M_n = M$. Likewise, $\bbX = [\bbX_1^\top,\ldots,\bbX_N^\top]^\top$ with $\bbX_n \in \mathbb{R}^{M_n \times \Theta}$ for $n \in \ccalN$. %Thus, \eqref{eq:centr} represents the centralized problem solved by a FC. 

In order to solve~\eqref{eq:centr} in a decentralized manner, consider first an undirected graph $\ccalG := (\ccalN,\ccalE)$  with  vertex set $\ccalN$ and  edge set $\ccalE$, where vertices represent sensors and the edge $(n,n')$ is in $\ccalE$ whenever sensors $n$ and $n'$ can communicate in a single hop, i.e., $n'\in\ccalN_n$. If $\ccalG$ is connected, i.e., there is a (possibly multi-hop) path between every pair of sensors, it can be easily shown that \eqref{eq:centr} is equivalent to
\begin{subequations}
\begin{align}
\minimize_{\{\bbtheta_n,\bbgamma_n,\bbgamma_{(n,n')}\}} &\sum_{n=1}^N \left[\frac{1}{2} \|\bby_n - \bbX_n \bbgamma_n\|_2^2 + \frac{1}{N} \psi(\bbtheta_n) \right] \label{eq:distr}\\
\textrm{subject to } & \bbgamma_n = \bbtheta_n, \ n \in \ccalN \label{eq:constr1}\\
&  \bbtheta_n = \bbgamma_{(n,n')} = \bbtheta_{n'}, \ n' \in \ccalN_n, \ n \in \ccalN, \label{eq:constr2}
\end{align}
\end{subequations}
where $\{\bbgamma_n\}$ and $\{\bbgamma_{(n,n')}\}$ are auxiliary variables. Per~\eqref{eq:constr1}, $\bbgamma_n$ is just a copy of $\bbtheta_n$. $\{\bbgamma_{(n,n')}\}$ facilitate the derivation of simple update rules and are eventually eliminated. A decentralized algorithm can be derived by applying the alternating direction method of multipliers (ADMM) to~\eqref{eq:distr}--\eqref{eq:constr2}\ncite{BPC10}. Following steps similar to those in~\cite[App.~D]{bazerque2011splines}, one can obtain the decentralized update rules for iteration $k$ as
\begin{subequations}
\begin{align}
\bbu_n^{[k]} &= \bbu_n^{[k-1]} + \rho \sum_{n' \in \ccalN_n} \left(\bbtheta_n^{[k]} - \bbtheta_{n'}^{[k]} \right)  \label{eq:u} \\
\bblambda_n^{[k]} &= \bblambda_n^{[k-1]} + \rho \left(\bbtheta_n^{[k]} - \bbgamma_n^{[k]} \right) \label{eq:lamb} \\
\bbtheta_n^{[k+1]} &= \argmin_\bbtheta \frac{1}{Nc_n} \psi(\bbtheta) + \frac{1}{2} \|\bbtheta - \bba_n\|_2^2 \label{eq:theta}\\
\bbgamma_n^{[k+1]} &= \left(\rho \bbI_{\Theta} + \bbX_n^\top \bbX_n \right)^{-1} \left(\bbX_n^\top \bby_n + \rho \bbtheta_n^{[k+1]} + \bblambda_n^{[k]} \right), \label{eq:gamma}
\end{align}
where $\rho > 0$ is the step size, $c_n := \rho (1 + 2|\ccalN_n|)$, and
\begin{align}
\bba_n := \frac{1}{c_n} \left(\rho \sum_{n' \in \ccalN_n} \left(\bbtheta_n^{[k]} + \bbtheta_{n'}^{[k]} \right) + \rho \bbgamma_n^{[k]} - \bbu_n^{[k]} - \bblambda_n^{[k]} \right) \label{eq:a}
\end{align}
\end{subequations}
for $n \in \ccalN$. As can be seen in~\eqref{eq:u} and~\eqref{eq:a}, the updates involve only local communication with the neighbors. The \emph{proximal problem} in~\eqref{eq:theta}  admits a closed form solution for various common choices of $\psi(\cdot)$\ncite{Par13}. It can be proved that the iterate $\bbtheta_n^{[k]}$ for any $n \in \ccalN$ converges to the solution of~\eqref{eq:centr} as $k \rightarrow \infty$~\ncite{BPC10}\cite{bazerque2011splines}.

%%%%%%%%%%%%%%%%%%%%%%%%%%%%%%%%%%
\subsection{Rate Constraints}
%%%%%%%%%%%%%%%%%%%%%%%%%%%%%%%%%%
% Compression
% To alleviate the high power consumption and bandwidth needs that stem
% from obtaining and communicating these
% periodograms,~\cite{mehanna2013frugal} proposed a low-overhead sensing
% scheme based on single-bit data along the lines
% of~\cite{argg2006tsp}. However, this scheme assumes that the PSD is
% constant across space.

\begin{bullets}
  \blt[Motivation]To maintain up-to-date maps, every certain time interval,
  the sensors need to collect new measurements, which are then sent to a fusion center or shared with other nodes. For maps that change rapidly over time or require  high-dimensional measurements, such as  PSD maps, 
  the bandwidth required to report the measurements may be
  significant. %Since sensing and communication functionalities may
  %coexist in the same device, it is worth minimizing the bandwidth
  %used to report measurements so that more bandwidth can be used to
  %transmit user data.
  %
  \blt[Overview]To mitigate such an issue, compression and quantization can be employed~\cite{romero2017spectrummaps}.

  \blt[Measurements overview]The idea is twofold. First, instead of directly computing  PSD estimates at the sensors, each sensor measures
  the powers at the outputs of a filter bank acting on the received
  signal. Then, only  quantized versions of those measurements are reported; see Fig.~\ref{fig:quantized}.

\begin{figure}[!t]
  \centering
  \includegraphics[width=0.8\textwidth]{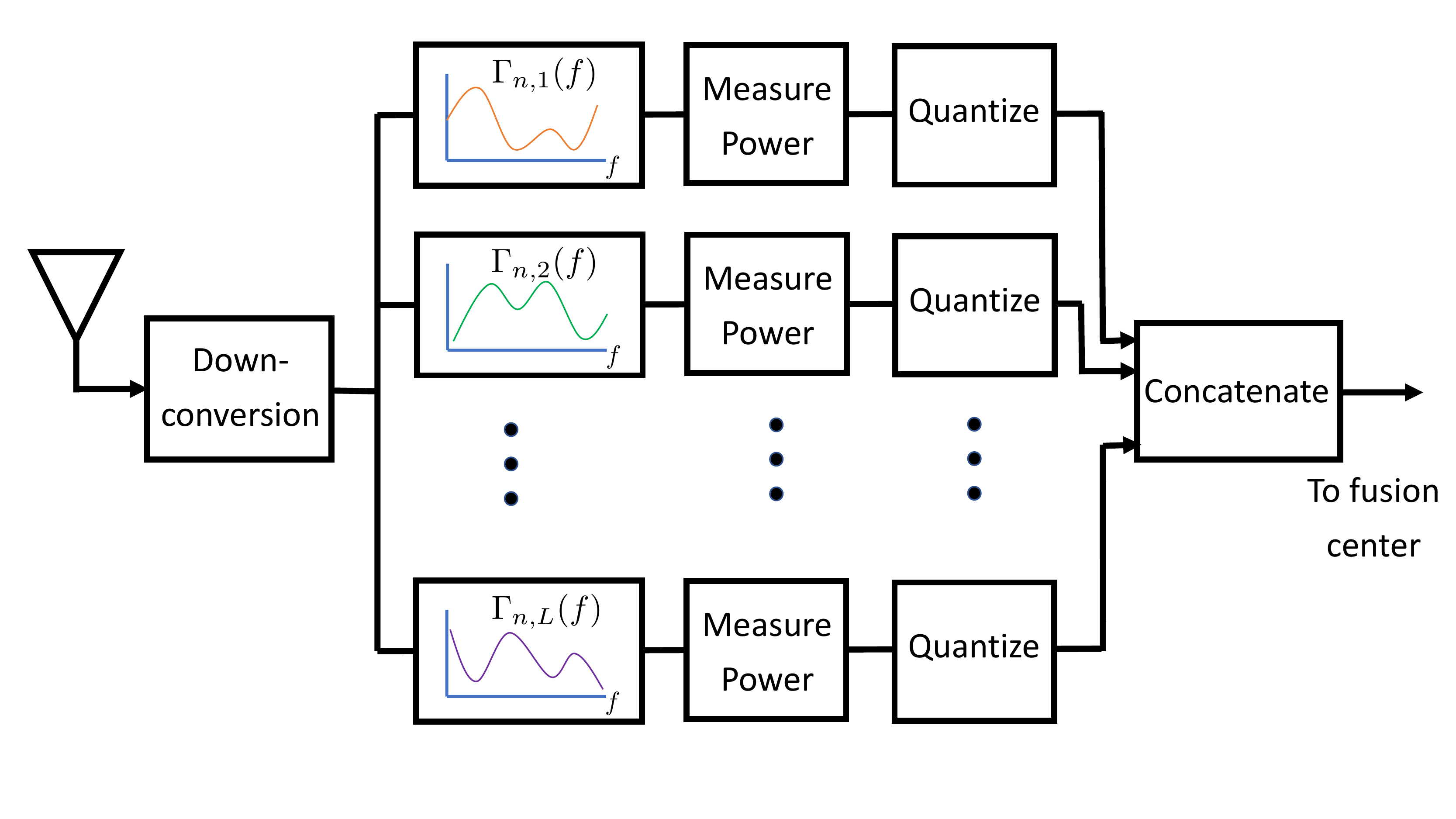}
  \caption{To reduce the rate necessary to report measurements, sensors may use a bank of random filters. The energy of each filter is measured, quantized, and sent to a fusion center that performs RME.
  }
  \label{fig:quantized}
\end{figure}

\def \qmeas {{\breve m}}

  \blt[single filter]
  \begin{bullets}
    \blt[measurements]
    To simplify the exposition, 
  assume for now that each sensor employs a single filter.
  %the extension to multiple filters is discussed later.
  Recall that $\pow(\loc, \freq)$ denotes the PSD at location
    $\loc$.  If the received signal at location
    $\loc_\measind$ is processed by a filter with frequency response $\freqresp_\measind(\freq)$, the output power is given by
    $\filtpow_\measind \define \int \pow(\loc_\measind, \freq)
    |\freqresp_\measind(\freq)|^2d\freq$. Due to the measurement noise, the measured value $\measpow_\measind$ will be generally different from the true $\filtpow_\measind $. Subsequently, $\measpow_\measind$ is
    quantized to $\qmeas_\measind$, which is then sent to the fusion center or other
    sensors. This clearly requires much smaller bandwidth than sending,
    e.g., the entire periodogram.

    \blt[Estimation]
    \begin{bullets}
      \blt[bem]
      \begin{bullets}
        \blt[review]
        To see how the map can be estimated from these
        linearly compressed and quantized measurements, recall the decomposition
        $\pow(\loc, \freq) =\sum_{\fbasisind}\pow_\fbasisind(\loc)
        \fbasisfun_\fbasisind(\freq)$ from Sec.~\ref{sssec:wbch}.
        \blt[implications]
        \begin{bullets}
          \blt[decompose problem] Since the basis functions
          $ \fbasisfun_\fbasisind$ are known, this decomposition reduces
          the problem of estimating $\pow$ to that of
          estimating $\fbasisnum$ functions
          $\pow_1,\ldots,\pow_\fbasisnum$.
          \blt[linear measurement] It also follows that $\filtpow_\measind$ can be
          written as
          $\filtpow_\measind =
          \sum_{\fbasisind}\pow_\fbasisind(\loc_\measind) \int
          \fbasisfun_\fbasisind(\freq)
          |\freqresp_\measind(\freq)|^2d\freq =
          [\pow_1(\loc_\measind),\ldots,\pow_\fbasisnum(\loc_\measind)] \fbasisvec_\measind $, where
          % $\powvec_\measind\define
          % [\pow_1(\loc_\measind),\ldots,\pow_\fbasisnum(\loc_\measind)]\transpose$
          % and
          the $\fbasisind$-th entry of vector $\fbasisvec_\measind$
          is
          $\int \fbasisfun_\fbasisind(\freq)
          |\freqresp_\measind(\freq)|^2d\freq$. In other words,
          $\filtpow_\measind$ is a linear combination of the values that
          the functions $\pow_1,\ldots,\pow_\fbasisnum$ take at
          $\loc_\measind$.
        \end{bullets}
      \end{bullets}
      \blt[$\filtpow_\measind$ known]If the true powers $\filtpow_1,\ldots,\filtpow_\measnum$ were known exactly, one could seek RKHS functions 
      $\hat \pow_1,\ldots,\hat \pow_\fbasisnum$ such that
      $[\hat \pow_1(\loc_\measind),\ldots,\hat \pow_\fbasisnum(\loc_\measind)]
      \fbasisvec_\measind = \filtpow_\measind \ \forall n$ using
      kernel-based learning. 

      \blt[no meas noise]Now consider the case where
      instead of $\filtpow_1,\ldots,\filtpow_\measnum$, one
      has the quantized measurements
      $\qmeas_1,\ldots,\qmeas_{\measnum}$, but it holds that
      $\measpow_\measind=\filtpow_\measind$ for all $\measind$, i.e., there is no measurement noise. 
      Each $\qmeas_{\measind}$ therefore indicates which quantization
      interval contains $\filtpow_\measind$. Upon denoting the endpoints
      of the interval that contains $\qmeas_{\measind}$ as $\powlow(\qmeas_{\measind})$ and
      $\powhigh(\qmeas_{\measind})$, it makes sense to now seek $\hat \pow_1,\ldots,\hat \pow_\fbasisnum$ that satisfy
      $[\hat \pow_1(\loc_\measind),\ldots,\hat \pow_\fbasisnum(\loc_\measind)]
      \fbasisvec_\measind \in [
      \powlow(\qmeas_{\measind}),\powhigh(\qmeas_{\measind})]$ for all
      $\measind$.

      \blt[no ass.]Finally, in the case where there is measurement
      noise, $\measpow_\measind$ is generally different
      from $\filtpow_\measind$. If the noise is small relative to the
      width of the quantization interval, the result of quantizing either
      values will be often the same, but not always. This means that one
      cannot impose that
      $[\rkhsfun_1(\loc_\measind),\ldots,\rkhsfun_\fbasisnum(\loc_\measind)]
      \fbasisvec_\measind$ \changed{necessarily} falls in the quantization interval
      $[ \powlow(\meas{\measind}),\powhigh(\meas{\measind})]$. Instead,
      the condition must be encouraged in a soft manner by
      penalizing deviations from the interval. Interestingly, by penalizing deviations in a linear fashion, it can be shown that the resulting
      estimates can be obtained through \emph{support vector regression}~\cite{romero2017spectrummaps}.
      
      % Following the
      % approach in \eqref{eq:pregr}, it makes sense to seek a map
      % estimate $\powest$ as a function that satisfies
      % $ \powvec_\measind\transpose\fbasisvec_\measind\in [
      % \powcent(\meas{\measind}) - \powwidth(\meas{\measind})/2,
      % \powcent(\meas{\measind}) + \powwidth(\meas{\measind})/2]$ for all
      % $\measind$.

      % Since this equality
      % will generally hold for infinitely many choices of
      % $\rkhsfun_1,\ldots,\rkhsfun_\fbasisnum$, one needs to adopt an
      % objective function

    \end{bullets}

  \end{bullets}

  \blt[Multiple filters]The previous considerations can be extended to
  the case where the filter bank at each sensor contains
  $\branchnum>1$ filters, as depicted in Fig.~\ref{fig:quantized}. Observe
  that now two subscripts are necessary to index each branch.
  \begin{bullets}
    \blt[impulse response] The power at the
    $\branchind$-th branch of the sensor at $\loc_\measind$ is given by
    $\filtpow_{\measind,\branchind} =
    [\pow_1(\loc_\measind),\ldots,\pow_\fbasisnum(\loc_\measind)]
    \fbasisvec_{\measind,\branchind} $. Since all the vectors
    $\fbasisvec_{\measind,1},\ldots, \fbasisvec_{\measind,\branchnum}$
    multiply the same
    $[\pow_1(\loc_\measind),\ldots,\pow_\fbasisnum(\loc_\measind)]$,
    the values $\filtpow_{\measind,\branchind}$ are not fully
    informative about $\pow_1,\ldots,\pow_\fbasisnum$ unless the vectors
    $\fbasisvec_{\measind,1},\ldots, \fbasisvec_{\measind,\branchnum}$
    are linearly independent. This imposes a design constraint on the
    filters. For example, filters with pseudorandom impulse
    responses may be utilized, which are expected to yield linearly
    independent vectors
    $\fbasisvec_{\measind,1},\ldots, \fbasisvec_{\measind,\branchnum}$
    so long as
    $\branchnum\leq \fbasisnum$.

  \end{bullets}

\end{bullets}

% \subsection{Calibration}
% \begin{bullets}

%   \blt[tomography]\begin{bullets}\blt[slf diff]Inspired by this
%     model,~\cite{patwari2008correlated,wilson2009regularization,wilson2010tomography}
%     proposed various techniques for radio tomographic imaging. Since
%     these techniques avoid calibration issues by estimating the
%     difference between the SLF at consecutive time instants instead of
%     the SLF itself, they reveal the location of changes in the
%     propagation medium but are unable to image static
%     structures. \blt[RSS variance]Similarly,
%     \cite{wilson2011throughwalls} builds on the arguments
%     in~\cite{patwari2011human} to replace the SLF with an indicator
%     function of the voxels that contain objects in motion and therefore
%     also suffers from this limitation.  \blt[indep. calibr]In
%     contrast, the scheme in~\cite{hamilton2014modeling} estimates the
%     SLF directly and therefore can image static structures, but involves
%     a separate calibration stage where these structures are
%     absent.  \end{bullets}

% \end{bullets}

\section{Future Directions}
\label{sec:future}
% \acom{This section will describe some key open problems in RME, which will include the following issues: i) Improving inference biases; ii) Uncertainty quantification; iii) Improving measurement collection; iv) Handling mobility; and v) Empirical validation.}
  
Although RME has been the subject of a sizable research body, a large
number of open issues still remain. First of all, the potential of
radio maps to endow applications with radio situational awareness is
yet to be fully exploited. A large part of the progress in this regard has
taken place in the context of device-free localization (see
e.g.~\cite{wilson2009regularization}) and UAV communications (see
e.g.~\cite{romero2022aerial} and references therein), but a number of
tasks arising in cellular networks such as resource allocation are yet
to be explored. Radio maps can also be used as priors for enhanced
channel estimation in mobile communications. 

Improving inference biases in data-driven radio map estimators is also
necessary. This can be achieved by collecting extensive data sets in
multiple bands, since most works so far rely on synthetic data
generated with ray-tracing software. Such data sets would also open
the door to devising improved uncertainty metrics for spectrum surveying. Remarkably, these
can be used for improving spectrum surveying
techniques~\cite{shrestha2022surveying}. \changed{Furthermore, hybrid
  model-based and data-driven approaches have the potential to combine the best of
  both worlds~\cite{thrane2020model}. The rationale is that radio propagation models may
  significantly reduce the amount of data required to train
  data-driven estimators, whereas learning from data can significantly
  improve the accuracy of model-based approaches. }

\changed{Methods for coping with various sources of error must also be
  devised. For example, time variations may be better predicted by
  exploiting side information on the mobility of terminals. In this context, trajectories of ground vehicles on the road or UAVs in aerial
  corridors may be instrumental to reduce the effective dimensionality
  of propagation maps.  One can also model how groups of
  persons or vehicles move to better predict \signalstrength maps as a
  whole. Other sources of error to counter include the use of  antennas with two polarizations and non-isotropic gain patterns.}

% \begin{bullets}
%   \blt[uplink]
%   \blt[mobility]
%   \blt Aerial, time-varying
%   \blt Data-driven + model-based
%   \blt Dataset collection and empirical validation
% \end{bullets}

% CSI prediction, channel charting
Recent developments adopt machine learning algorithms to predict the CSIs of desired multi-antenna channels based on pilot CSIs. This approach can capture the characteristics of small-scale fading, going beyond the channel gain maps. %In~\cite{AAV18}, the pilot CSIs are acquired from the {\em same} link as the target link, but using an omni-directional beampattern. On the other hand, 
In~\cite{jiang2019structures}, the pilot CSIs are obtained from a set of links that are {\em different} from the target link. The optimal transmit-beampattern of the desired link is predicted based on the acquired CSIs. When the source and the target links are not collocated, the traditional assumption is that the CSIs are statistically independent. In reality, there can be significant dependency between the CSIs and the  geometry of the propagation environment, transceiver locations, line-of-sight path, and other multipaths within the coherence time of the channels. Given sufficiently rich pilot CSI measurements that  capture the relevant geometry, an appropriate nonlinear mapping (e.g. via a DNN) can exploit this dependency. As a related idea, \emph{channel charting} obtains  in an unsupervised fashion  low-dimensional embeddings of the high-dimensional CSIs that approximately provide the spatial locations of the measurements~\cite{deng2021semisupervised}. %\cite{studer2018charting}. 

\changed{Finally, further types of radio maps may also be explored.
  For example, maps may be developed for massive MIMO and mmWave networks to benefit from reduced search time for beam selection. As
  another example, exploring delay-Doppler maps could be instrumental
  in the context of resource allocation for the emerging Orthogonal
  Time Frequency and Space (OTFS) modulation.  }

\changed{
\section{Related Work}
\label{sec:relatedwork}
The interested reader can delve deeper into RME through the surveys
\cite{hoyhtya2016survey,pesko2014survey}. In \cite{hoyhtya2016survey},
the focus is on \emph{occupancy maps}, which are  radio maps that
provide the fraction of time that a certain frequency channel is
used. On the other hand, the authors of \cite{pesko2014survey} focus on power map
estimation and review other methods that are not discussed here due
to space limitations. Relative to these works, the present tutorial is
more introductory in nature and considers more classes of maps as well
as more recent methods. }

\section{Conclusion}
Radio maps characterize important metrics of the RF spectrum landscape
across a geographical area. Two families of radio maps were considered
based on whether the  received signal strength or the
propagation channel effects are of interest, and a large number of
representative applications were discussed. Tutorial expositions of various
data-driven methods for RME have been presented, ranging from
parametric, non-parametric, and probabilistic approaches, to recent
powerful deep learning techniques, incorporating useful priors such as
sparsity, low rank, and union-of-subspace structures. Practical issues
related to  spectrum surveying, noisy location estimates,
decentralized implementation, and limited-rate measurements were also
discussed. With the advent of ultra-dense and ultra-dynamic deployment
scenarios often envisioned in  future wireless networking, the role
of data-driven spectrum cartography enabled via sophisticated RME
techniques will likely become even more relevant.

\printmybibliography
\end{document}